\documentclass{article}

\usepackage{custom_icml}

\usepackage[accepted]{icml2026}

\icmltitlerunning{General Synthetic-Powered Inference}

\begin{document}

\twocolumn[
  \icmltitle{General Synthetic-Powered Inference}

  \icmlsetsymbol{equal}{*}

  \begin{icmlauthorlist}
    \icmlauthor{Meshi Bashari}{equal,1}
    \icmlauthor{Yonghoon Lee}{equal,2}
    \icmlauthor{Roy Maor Lotan}{1}
    \icmlauthor{Edgar Dobriban}{2}
    \icmlauthor{Yaniv Romano}{1,3}
  \end{icmlauthorlist}

  \icmlaffiliation{1}{Department of Electrical and Computer Engineering, Technion IIT, Israel}
  \icmlaffiliation{2}{Department of Statistics and Data Science, The Wharton School, University of Pennsylvania, USA}
  \icmlaffiliation{3}{Department of Computer Science, Technion IIT, Israel}

  \icmlcorrespondingauthor{Meshi Bashari}{meshi.b@campus.technion.ac.il}
  \icmlcorrespondingauthor{Yonghoon Lee}{yhoony31@wharton.upenn.edu}

  \icmlkeywords{Machine Learning, ICML}

  \vskip 0.3in
]

\printAffiliationsAndNotice{\icmlEqualContribution}

\begin{abstract}

The rapid proliferation of high-quality synthetic data---generated by advanced AI models or collected as auxiliary data from related tasks---presents both opportunities and challenges for statistical inference. This paper introduces a \textbf{GE}neral \textbf{S}ynthetic-\textbf{P}owered \textbf{I}nference (\gespi) framework that wraps around a broad class of statistical inference procedures to safely enhance sample efficiency by combining synthetic and real data. Our framework leverages high-quality synthetic data to boost statistical power, yet adaptively defaults to the standard method using only real data when synthetic data are of low quality. 
The error rate of our method remains below a user-specified bound without any distributional assumptions on the synthetic data, and decreases as the quality of the synthetic data improves.
This flexibility enables seamless integration with conformal prediction, risk control, hypothesis testing, and multiple testing procedures, all without modifying the base inference method.
We demonstrate the benefits of our method on challenging tasks with limited labeled data, including AlphaFold protein structure prediction, and comparing large reasoning models on complex math problems.\footnote{Software for reproducing the experiments is available at \href{https://github.com/Meshiba/gespi}{https://github.com/Meshiba/gespi}.}
\end{abstract}

\section{Introduction}

Statistical inference lies at the core of data-driven decision-making, enabling researchers and practitioners to draw conclusions while rigorously controlling error rates or risks. 
The importance of such control cannot be overstated: uncontrolled errors can lead to misleading conclusions and costly mistakes. For example, in computational biology, researchers increasingly rely on AlphaFold's protein structure predictions, where substantial local errors can mislead downstream applications such as drug discovery or protein design. 
A/B testing can be used to assess whether a new large language model (LLM) outperforms the current one---yet falsely concluding that the new model is better can lead to revenue loss or customer dissatisfaction.

A fundamental limitation of statistical inference arises when data are scarce, leading to reduced utility (e.g., statistical power) and high variability in error rates. Yet limited data are almost unavoidable in domains where data acquisition is costly, difficult, or time-consuming. For example, in protein structure prediction, experimental validation is expensive and labor-intensive, resulting in relatively few labeled structures. Similarly, comparing LLMs on complex mathematical reasoning tasks requires curating high-quality problems with verified solutions: a process that is challenging and resource-intensive. 

At the same time, the availability of high-quality synthetic data presents new opportunities to mitigate data scarcity and enhance statistical utility. For example, such data can be generated by LLMs or diffusion models, or retrieved from related auxiliary databases. 
However, it is challenging to construct procedures that leverage synthetic data with provable theoretical error rate control; because synthetic data may not reflect the real-world distribution. 

To make this concrete, consider a practitioner with limited real data, abundant synthetic data, and an error-controlling algorithm, e.g., A/B testing procedure, conformal prediction, etc.
How can one perform statistical inference in this setting, with the goal of obtaining informative inference while controlling the error rate?

\begin{figure*}[!t]
    \centering
        \includegraphics[width=0.75\linewidth]{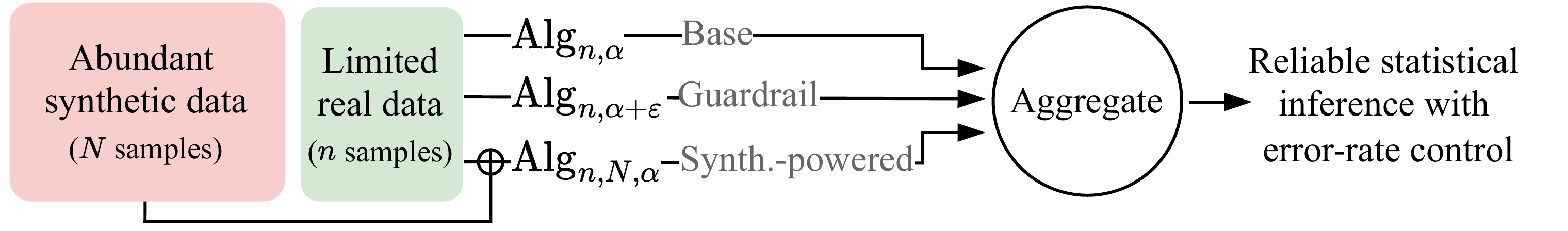}

    \caption{
    \textbf{Overview of \gespi framework.} 
    \gespi leverages a small real dataset and a large synthetic dataset. The procedure applies the base statistical method three times and aggregates the outputs in a way that guarantees error rate control while exploiting synthetic data when beneficial.
    }
    \vspace{-1em}
    \label{fig:gespi_illustration}
\end{figure*}

The first option is to completely ignore the synthetic data, even when they may be informative, and rely solely on the limited real data. This approach can lead to overly conservative inference, often prompting practitioners to increase the error level to increase power, at the cost of obtaining a higher error rate.
The second option is to blindly trust the synthetic data as if they were real, pooling them 
with the limited real data. However, this approach can in general result in an uncontrolled error rate, as the synthetic data distribution can be arbitrary.
Both options are insufficient. 

This fundamental challenge motivates the key question of this work: \emph{how can we utilize high-quality but technically unreliable synthetic data in a reliable way?}
We present a general approach that ``wraps'' around a broad class of statistical inference methods, providing them with the ability to harness synthetic data when beneficial, while attaining rigorous, distribution-free error rate control, even when the synthetic data quality is unknown \emph{a priori}.

\subsection{Key Contributions and Preview of Our Method}

As a first step, we introduce a novel theoretical formulation for leveraging synthetic data in the context of statistical decision theory, which is based on controlling a general loss function.
This formulation allows us to tackle a broad range of 
statistical problems in a unified framework, including hypothesis testing, multiple testing, confidence sets, etc.

Building on this formulation,
 our framework transforms the base statistical inference method to safely leverage synthetic data. 
As in \Cref{fig:gespi_illustration}, let $\Alg$ denote the base inference method, 
$\Dn$ the real observational dataset, and $\tDn$ the abundant synthetic 
dataset, with sizes $n$ and $N \gg n$, respectively. 
Ideally, the synthetic data distribution would match the real one, but we make no such assumption.

With these notations in place, \gespi\ invokes the base inference algorithm three times:

\begin{compactenum}
\item \textbf{Base} $\Alg_{n,\alpha}$, uses the real data $\Dn$, targeting a user-specified error rate level of $\alpha$ (e.g., 5\%).
\item \textbf{Guardrail} $\Alg_{n,\alpha+\ep}$, also uses only $\Dn$ but at a slightly higher user-specified level $\alpha+\ep$ (e.g., 7\%). 
\item \textbf{Synthetic-powered} $\Alg_{n,N,\alpha}$, which applies the base method to the pooled real and synthetic data $\Dn \cup \tDn$ at level $\alpha$.
\end{compactenum}
\gespi\ then aggregates (in a way we define) the outputs of these runs.
In terms of statistical validity, we prove in \Cref{thm:main,thm:main_2_extended} that this careful construction ensures \gespi's error rate never exceeds the guardrail bound $\alpha+\eps$, even when the synthetic data are of poor quality. 
Relaxing to $\alpha+\eps$ is essential to allow us to potentially benefit from additional synthetic data (\Cref{prop:impossibility}).
At the same time, as the quality of the synthetic data improves---better matching the distribution of the gold-standard real data---the error rate can approach the desired level $\alpha$. From the perspective of statistical efficiency, we prove in \Cref{thm:main_2} that \gespi's utility (e.g., power, set size) is always at least that of the base method and is bounded above by the guardrail. Our guarantees hold in finite samples and do not require any assumptions on the synthetic data distribution.

As a result, one can interpret the user-specified $\eps$ as the ``admission cost'' for the opportunity to benefit from synthetic data in statistical inference. In the worst case (poor synthetic data), the user ``pays'' an additional $\eps$ in error rate. In the best case (perfect synthetic data), the user enjoys enhanced sample efficiency and improved utility---where this gain increases with $\eps$.

An important feature of \gespi is its ability to automatically adapt to the quality of the synthetic data. This leads to various potential benefits in sample efficiency that go beyond utility improvements, depending on the application. In this work, we highlight two main benefits of our method:
(i) for tasks such as conformal prediction and risk control, \gespi reduces the variance of the empirical error rate compared to using only real data; and
(ii) for hypothesis testing problems, it improves the ROC curve relative to methods that ignore the synthetic data.

To illustrate these benefits, we evaluate \gespi across a variety of tasks in \Cref{sec:exp,app-sec:additional-exp},
including:
(i) image classification, where we control the coverage using conformal prediction;
(ii) AlphaFold protein structure prediction, where we apply conformal risk control to bound the average fraction of residues with large prediction error; 
(iii+iv) 
out-of-distribution detection, where we control the Type I error (in the single outlier case) and the family-wise error rate (for multiple outliers); 
(v+vi) comparison of large reasoning models on math datasets, where we use hypothesis testing for win rate, for both pairwise comparisons between models and hyperparameter selection;
(vii) mechanistic interpretability of a Vision Transformer model, where we use a two-sample test to provide evidence for a property-specific role of an attention head.
We also conduct ablation studies to evaluate the sensitivity of our method to $\ep$ and to the synthetic data quality.
These experiments support that \gespi adapts to the quality of synthetic data, providing meaningful improvements when possible while always maintaining error rate control.

\section{Related Work}

Our proposed \gespi framework is inspired by Synthetic-Powered Predictive Inference (SPI) \citep{bashari2025synthetic}, a recent procedure for incorporating synthetic data into conformal prediction. 
SPI constructs prediction sets with distribution-free, finite-sample coverage guarantees that hold regardless of synthetic data quality. 
Its core idea is to modify the construction of the prediction set by applying a novel transportation to the non-conformity scores from the real space to the synthetic one.
While we share the underlying motivation of leveraging synthetic data under rigorous error rate control, \textit{our GESPI approach offers a new perspective on how to leverage synthetic data, that serves as a foundation for addressing more general statistical inference problems}. 
Indeed, a key distinction between the two methods is that, instead of directly modifying the mechanism for constructing prediction sets---whose extension to broader statistical problems is unclear---\gespi treats the statistical inference method as a ``black-box'' procedure, without altering its inner workings. Accordingly, 
while SPI applies to conformal prediction, it does not extend to the broader statistical inference tasks considered in this paper.
Additional review of works broadly related to our work is provided in~\Cref{app-sec:related-work}.
Notably, these methods 
do not provide distribution-free control of risk when leveraging arbitrary synthetic data, which is the goal of this work. 

\section{General Synthetic-Powered Inference}\label{sec:gespi_appl}

For illustration, we first present applications of \gespi to different inference problems, and then introduce the general framework and theory.

\subsection{\gespi for Conformal Prediction and Risk Control}\label{sec:gespi-pi}

Consider a predictive inference task where we are given $n$ i.i.d.~(real) data points $\smash{\Dn = {(X_i, Y_i)}_{i=1}^n \iidsim P}$, where $X_i \in \mathcal{X}$ and $Y_i \in \mathcal{Y}$ represent the features (e.g., an image) and labels (e.g., pedestrian), respectively. 
Given a new test input $X_{n+1}$, we aim to construct a prediction set $\smash{\ch_n(X_{n+1})}$ such that the expected loss (i.e., risk) of $\ch_n$ does not exceed a user-specified level $\alpha$, for a loss $\ell$ of interest:
\begin{center}
$
    \mathbb{E}_{\scriptsize\hspace{-0.5em}\begin{array}{l}\Dn \iidsim P_{X,Y},\\ (X_{n+1},Y_{n+1}) \iidsim P_{X,Y}\end{array}}\hspace{-3em}\left[\ell({ \ch_n(X_{n+1}), Y_{n+1}})\right] \le \alpha.
    $
\end{center}

Conformal prediction~\citep{saunders1999transduction,vovk1999machine,vovk2005algorithmic, papadopoulos2002inductive} is a framework that takes a holdout dataset $\Dn$ and transforms the output of any machine learning model into a prediction set $\ch_n(X_{n+1})$ that satisfies the finite-sample guarantee above for the 0-1 miscoverage loss, $\mathbb{I}\{Y_{n+1} \notin \ch_n(X_{n+1})\}$. Conformal risk control 
extends this framework to any monotone loss function, such as the false negative rate or the F1 score.

\begin{figure}[!h]
    \centering
    \begin{subfigure}[t]{\linewidth}
    \centering
    \includegraphics[width=\linewidth]{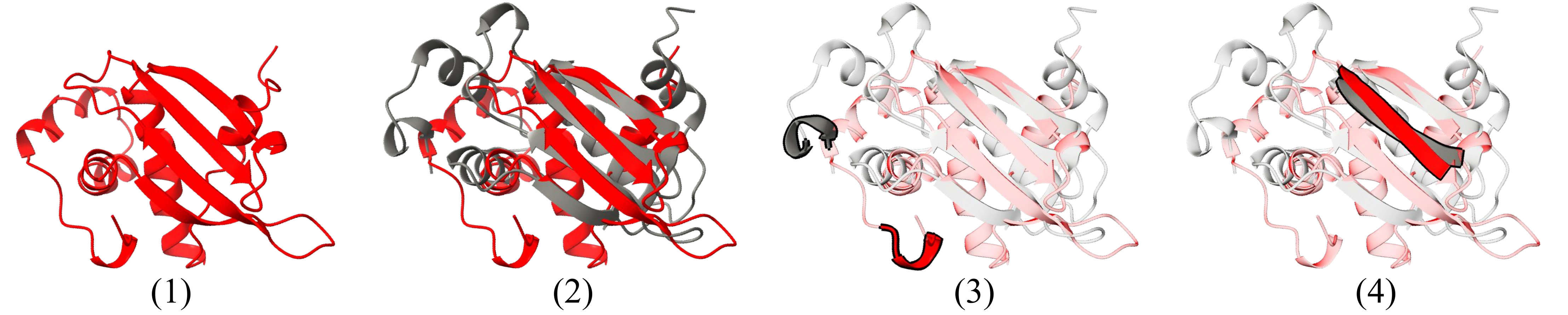}
    \caption{
    \texttt{OnlyReal}: Conformal risk control using only real data, applied at level $\alpha=5\%$}
    \label{fig:protein-illus-crc}
    \end{subfigure}
    \begin{subfigure}[t]{\linewidth}
    \centering
    \includegraphics[width=\linewidth]{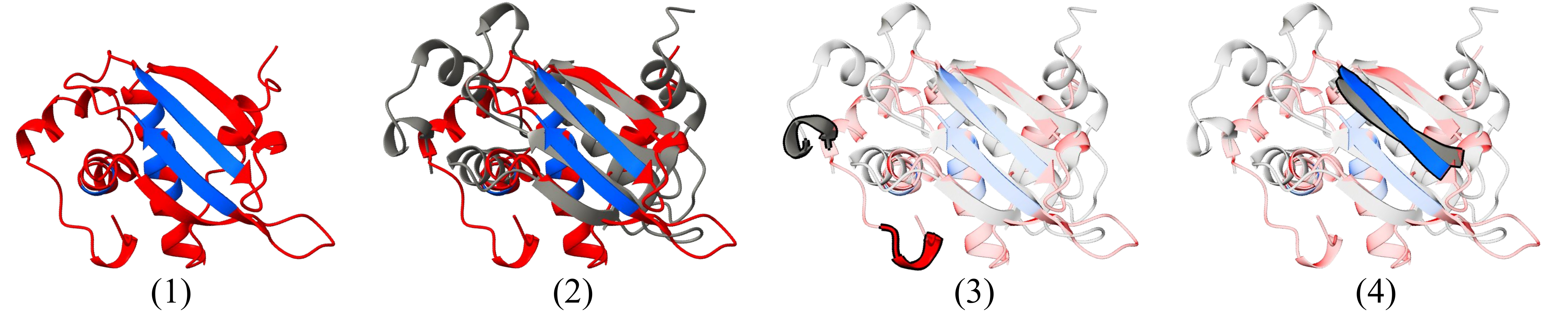}
    \caption{
    \texttt{\gespi}: Our method applying conformal risk control at level $\alpha=5\%$ and guardrail $\alpha+\ep=10\%$}
    \label{fig:protein-illus-scrc}
    \end{subfigure}
    \caption{
    \textbf{Visualization of protein structure prediction with error rate control.} Panels show protein T1029 predictions with residues abstained on by (a) \texttt{OnlyReal} and (b) \texttt{\gespi} methods. {\color{red}Red}: residues abstained on; {\color{blue}Blue}: accepted residues. {\color{gray}Gray}: real experimental structure, aligned with AlphaFold2 predicted structure. Quantitative results \{abstention ratio, risk\}: \texttt{OnlyReal} -- \{$100\%$,  $0\%$\}; \texttt{\gespi} -- \{$85.6\%$,  $\approx7\%$\}. See text in \Cref{sec:exp-protein}.
    }
    \label{fig:protein-illus}
\end{figure}

A fundamental limitation of these methods, however, is that when the sample size $n$ is small, they often produce excessively large and uninformative prediction sets, or exhibit unstable empirical risk with high variability---limiting their practical applicability.

Motivated by this small sample-size limitation, our \gespi procedure utilizes synthetic data to enhance sample efficiency. Let $\tDn = (\tilde{X}_j, \tilde{Y}_j )_{j=1}^N \iidsim Q_{X,Y}$ denote the synthetic data. Let $\ch_{n,\alpha}(\cdot)$ denote the prediction set function obtained by applying conformal prediction to the real data $\Dn$ at level $\alpha$, and let $\ch_{n,N,\alpha}(\cdot)$ denote the corresponding function obtained from the pooled data $\Dn \cup \tDn$. Then, given an additional error tolerance level $\ep>0$, the \gespi prediction set is given as follows.

\begin{tightbox}
\begingroup
\setlength{\abovedisplayskip}{4pt}%
\setlength{\belowdisplayskip}{4pt}%
\setlength{\abovedisplayshortskip}{2pt}%
\setlength{\belowdisplayshortskip}{2pt}%
The  \gespi conformal prediction set when the test datapoint $X_{n+1}$ takes value $x$ is given by
\begin{equation}\label{eq:gespi-cp}\small
\hat{C}^{\gespi}(x)
:= \ch_{n,\alpha}(x)\cap\!\bigl(\ch_{n,N,\alpha}(x)\cup \ch_{n,\alpha+\ep}(x)\bigr).
\end{equation}
\endgroup
\end{tightbox}
The intuition behind our \gespi construction is as follows. If the synthetic and real data distributions are identical, the second term $\ch_{n,N,\alpha}(x)$ amounts to applying conformal prediction on a larger real dataset, achieving the target risk level $\alpha$ while attaining tighter and more stable prediction sets.\footnote{We usually have $\ch_{n,\alpha+\ep} \subseteq \ch_{n,\alpha}$, as sets get wider when a tighter error guarantee (smaller $\alpha$) is required. In this case, we have 
$\ch_{n,\alpha+\ep} \subseteq \hat{C}^{\gespi} \subseteq \ch_{n,\alpha}$. When the synthetic data has high quality, we expect that $\ch_{n,N,\alpha}$ is small, and does not increase $\ch_{n,\alpha+\ep}$ by much. In such a setting, we will have that $\hat{C}^{\gespi}$ is close to $\ch_{n,\alpha+\ep}$, which can be a much tighter set than the original 
set $\ch_{n,\alpha}$. This explains how \gespi can lead to tighter sets.} 
At the same time, if the synthetic data are of poor quality, there are two guardrail bounds: (i) $\ch_{n,\alpha+\ep}$ ensures the risk level of $\hat{C}^{\gespi}$ does not exceed $\alpha + \ep$; and (ii) $\ch_{n,\alpha}$ prevents the \gespi prediction set from being even wider than the base method $\ch_{n,\alpha}$.

\subsection{\gespi for One-Sided Hypothesis Testing}\label{sec:gespi-ht-parameter}
We now turn to describe how \gespi can 
enhance sample efficiency in the canonical problem of one-sided hypothesis testing; see e.g., \cite{lehmann2005testing} for an overview of hypothesis testing. Consider a parameter $\theta$ of interest. Our goal is to test the following hypothesis
\[
\mathcal{H}_0: \theta \leq \theta_0\text{ versus }\mathcal{H}_1: \theta > \theta_0,
\]
where $\mathcal{H}_0$ is the null hypothesis and $\mathcal{H}_1$ the alternative. 
The parameter $\theta$ could represent, for example, 
the prediction error of a model,
or the win rate of model A compared to model B. In the latter, rejecting the null that $\theta_0 \leq 0.5$ provides evidence that model A outperforms model B, an application we revisit in the experiments (\Cref{app-sec:exp-win-rate}).

Let $\Dn=(X_i)_{i=1}^n$ denote the real dataset, with $\Dn \iidsim P_\theta$, where $P_\theta$ 
depends on\footnote{This distribution could also depend on other parameters, which are omitted here for clarity.} $\theta\in\Theta$. When only a small dataset is available, statistical tests may suffer from low power; for example, the empirical mean estimate $\hat{\theta}$ can be noisy when evaluated on limited data, making it difficult to detect whether $\theta > \theta_0$ when using $\hat{\theta}$ as a test statistic.

Now suppose we also have access to a large synthetic dataset $\tDn$. To effectively leverage this additional data in a statistically valid way, we construct the \gespi testing procedure as follows. Given a testing procedure with Type~I error rate control, let $\phi_{n,\alpha} \in \{0,1\}$ and $\phi_{n,N,\alpha} \in \{0,1\}$ denote the output of the tests applied to the real data $\Dn$ and the pooled data $\Dn \cup \tDn$, respectively, at level $\alpha$. By convention, we say that the test $\phi_{n,\alpha}$ rejects the null hypothesis if $\phi_{n,\alpha} = 1$, and fails to reject otherwise. Consider an additional error tolerance level $\ep>0$. 

\begin{tightbox}
\begingroup
\setlength{\abovedisplayskip}{4pt}%
\setlength{\belowdisplayskip}{4pt}%
\setlength{\abovedisplayshortskip}{2pt}%
\setlength{\belowdisplayshortskip}{2pt}%
The \gespi\ hypothesis test $\phi^{\gespi}\in\{0,1\}$ is given by
\begin{equation}\label{eq:gespi-hypothesis-testing}\small
\phi^{\gespi}
:= \phi_{n,\alpha}\; \text{ OR } \;\bigl(\phi_{n,N,\alpha} \text{ AND } \phi_{n,\alpha+\ep}\bigr).
\end{equation}
\endgroup
\end{tightbox}

Intuitively, \gespi for hypothesis testing works as follows. We first apply the test $\phi_{n,N,\alpha}$ to the pooled dataset at level $\alpha$. If it rejects the null, we do not immediately reject, since the synthetic data may come from a distribution that differs significantly from the real one. To account for this, we also run the test on the real dataset $\phi_{n,\alpha+\ep}$ at a slightly relaxed level $\alpha+\ep$, and reject the null only if this test also rejects. In any case, if the base test $\phi_{n,\alpha}$ on the real dataset 
rejects the null, we reject it immediately. This ensures that \gespi never loses power compared to the base test at level $\alpha$.

Importantly, for both Type I error rate control and power, the synthetic data do not need to follow the exact distribution of the real data. This flexibility stems from the structure of one-sided tests.
To see this, consider for illustration a simple setting where power is increasing\footnote{This holds generally, for one-dimensional families of probability distributions with the monotone likelihood ratio property, 
including exponential families such as 
the normal mean 
; see \cite{lehmann2005testing}.} 
in the true parameter $\theta$ of the real distribution. 
Suppose that the pooled distribution can be described by the parameter $\theta^\text{pool}$.
To control the Type I error when the null is true, it suffices that $\theta^{\text{pool}} \leq \theta \leq \theta_0$, even if the synthetic distribution differs from the real one (i.e., $\theta^{\text{pool}} \neq \theta$). Analogously, under the alternative $\theta > \theta_0$, it suffices that $\theta^{\text{pool}} > \theta_0$, without requiring $\theta^{\text{pool}} = \theta$. This property greatly expands the range of useful synthetic data that \gespi\ can leverage to improve power in one-sided hypothesis testing.

\begin{remark}\label{rmk:od}
The formulation in \eqref{eq:gespi-hypothesis-testing} is general and can be applied to additional hypothesis testing problems. One important problem that we revisit later in the experiments is outlier detection. Formally, given a set of {\em inliers} $\Dn=(X_i)_{i=1}^n$, with $X_i \iidsim P$, and a test point $X_{n+1}$, the task is to determine whether $X_{n+1}$ is an inlier sampled from $P$---or an outlier sampled from a different distribution. This can be framed as testing the null hypothesis:
\(\mathcal{H}_0: X_{n+1}\sim P\).
 Any test for this null---e.g. those by \citet{vovk2005algorithmic,bates2023testing,laxhammar2011sequential}---can be used within our \gespi framework.
\end{remark}

\subsection{\gespi for Multiple Hypothesis Testing}\label{sec:gespi-ht-exch-multiple-ht}

\gespi can be used for a number of additional statistical inference problems.
Due to space limitations, we are only able to present here a high-level overview of the multiple testing problem \cite{lehmann2005testing}\footnote{Multiple hypothesis testing has a broad range of applications across science and engineering, see for instance \cite{benjamini1995controlling,efron2012large,bretz2016multiple}, etc.}, and defer details to the appendix (See \Cref{sec:examples,app-sec:applications}).

Suppose we want to simultaneously test $m$ null hypotheses: $\mathcal{H}_{0,j}$, for $j = 1, \ldots, m$. 
When testing many hypotheses at once---such as identifying outliers among a batch of test points---the probability of incorrectly labeling at least one inlier as an outlier can increase rapidly if each hypothesis is tested separately at level $\alpha$. This motivates the goal of simultaneously testing all $m$ nulls while controlling the family-wise error rate (FWER), or more generally, the $k$-FWER \citep{lehmann2005generalizations} at level $\alpha$:
\(\PP{\sum_{j=1}^m \One{\text{$\mathcal{H}_{0,j}$ is true but rejected}} \ge k } \le \alpha\),
for some predetermined $k > 0$. 

Suppose now that we have an FWER-controlling procedure which, given appropriate data, 
outputs a candidate set of rejections $\hat{\mathcal{S}}_{n, \alpha} = \{j : \mathcal{H}_{0,j} \text{ is rejected}\}$. 
Similarly, we define $\hat{\mathcal{S}}_{n,N,\alpha}$ as the rejection set obtained by applying the same FWER procedure using the real and synthetic data together. With this notation in place, we can now state the \gespi\ procedure.

\begin{tightbox}
\begingroup
\setlength{\abovedisplayskip}{4pt}%
\setlength{\belowdisplayskip}{4pt}%
\setlength{\abovedisplayshortskip}{2pt}%
\setlength{\belowdisplayshortskip}{2pt}%
The \gespi\ rejection set $\hat{\mathcal{S}}^{\gespi}\subseteq \{1,\ldots,m\}$ for multiple testing is given by
\begin{equation}\label{eq:gespi-m-hypothesis-testing}\small
\hat{\mathcal{S}}^{\gespi}
:= \hat{\mathcal{S}}_{n,\alpha}\;\cup\;\bigl(\hat{\mathcal{S}}_{n,N,\alpha} \cap \hat{\mathcal{S}}_{n,\alpha+\ep}\bigr).
\end{equation}
\endgroup
\end{tightbox}

\subsection{The Proposed GESPI Framework}\label{sec:gespi-theory}

In this section, we describe our general framework for synthetic-powered inference that covers the applications in the previous section and extends beyond them.

{\bf Problem setup.}
Suppose we have a dataset $\Dn = (Z_1,Z_2,\ldots,Z_n) \in \Z^n$---e.g., in the setting of supervised learning, each $Z_i$ represents a (feature, outcome) pair $(X_i,Y_i)$. 
Consider a general statistical inference problem where the goal is to construct an algorithm\footnote{We let the input of the algorithm be $\Z^\infty$ so that the same algorithm can be used for different sample sizes.} 
$\Alg : \Z^\infty \rightarrow \A$ that maps the data to an action in the action space $\A$---with $Z^\infty = \Z \cup \Z^2 \cup \Z^3 \cup \ldots$---and controls a risk:

\begin{equation}\label{eqn:target}
\begin{split}
    \mathcal{R}(\Alg,P) = \Ep{\Dn \iidsim P, V \sim \T(P)}{\ell(\Alg(\Dn),V)} \le \alpha,\\ \text{ for all } P \in \mathcal{P} \text{ and } n \in \N.
\end{split}
\end{equation}
or equivalently,
\(\sup_{P \in \mathcal{P}} \mathcal{R}(\Alg,P) \le \alpha, \text{ for all } n \in \N,\)
for a predetermined target level $\alpha$. 

Here, $\mathcal{P}$ is a set of distributions on $\Z$, $V \in \V$ denotes a quantity used for evaluating of the algorithm (e.g., a new test point in a predictive inference task, the target parameter in a confidence interval task, etc), and $\T : \mathcal{P} \rightarrow \mathcal{P}_{\V}$ is a function that maps the data distribution $P \in \mathcal{P}$ to a distribution on $\V$---where $\mathcal{P}_{\V}$ denotes the set of all distributions on $\V$---so that $\T(P)$ defines the distribution\footnote{In some cases, $V$ is deterministic, such as for confidence intervals, when it is the parameter of interest. In that case, the distribution of $V$ simplifies to a point mass.} of $V$. The function $\ell : \A \times \V \rightarrow \R^+$ is a loss function that evaluates the quality of the procedure $\Alg(\Dn)$ with respect to $V$.
See Table~\ref{table:examples} for a non-exhaustive set of examples.

\begin{table}[htbp]
\centering
\scalebox{0.8}{
\renewcommand{\arraystretch}{1.4}
\begin{tabular}{lcccc}
\noalign{\hrule height 1pt}
 & $\substack{\text{Predictive}\\\text{inference}}$ & $\substack{\text{Hypothesis}\\\text{testing}}$ & $\substack{\text{Multiple hypothesis}\\\text{testing}}$ \\
\hline
$\Z$ & $\X \times \Y$ & $\X$ & $\X$ \\
$\Alg(\Dn)$ & Prediction set & Rejection indicator & Rejection set\\ 
$V$ & New test point & None & None \\
Risk & Miscoverage rate & Type I error & FWER\\
\noalign{\hrule height 1pt}

\end{tabular}
}
\caption{
Examples of problems covered by our framework.
}
\label{table:examples}
\vspace{-1em}
\end{table}

Now suppose we also have access to a synthetic/auxiliary dataset $\tDn = (\tilde{Z}_1,\ldots,\tilde{Z}_N) \in \Z^N$. Given a family of algorithms $\Alg_{\alpha}$ for each $\alpha>0$ that attains the guarantee~\eqref{eqn:target}, we aim to construct an algorithm $\tAlg : \Z^\infty \times \Z^\infty \rightarrow \A$ that takes both $\Dn$ and $\tDn$ as input, 
such that the synthetic-leveraging procedure $\tAlg(\Dn, \tDn)$ improves upon the standard procedure $\Alg_{\cdot}(\Dn)$. 

\subsubsection{General Algorithm and Theory}

To introduce our method, we begin with a simpler procedure, which only aim to upper bound the risk, and not to lower bound it; 
later we will consider the formulation presented in the examples. 

\begin{condition}[informal]\label{con:simplified}
The action space $\A$ is partially ordered by $\preceq$, and for any $a_1, a_2 \in \A$ the minimum $a_1 \wedge a_2$ and the maximum $a_1 \vee a_2$ are well defined. 
In addition, the loss $\ell$ is bounded by a constant $c$, and 
monotone with respect to $\preceq$.
\end{condition}

A formalized statement of Condition~\ref{con:simplified} is given in~\ref{con:target_order_formal}. Intuitively, this condition allows us to robustly aggregate real and synthetic-powered actions by (i) comparing them, and (ii) choosing the better action in the view of the risk we aim to control.
We note here that this is a mild condition, satisfied by all the inference problems discussed in this work; see \Cref{table:order_examples}.
For example, in hypothesis testing, the action space consists of accept/reject $\{0,1\}$ and the partial order $\preceq$ is defined by $\leq$.
Recall that in the examples, our algorithm relies on taking unions and intersections or alternatively performing OR/AND operations. Generalizing our examples, our \gespi method takes the minimum ($\wedge$) of the output of two carefully chosen algorithms:
\begin{equation}\label{eqn:alg_spi}
    \tAlg(\Dn, \tDn) = \Alg_{\alpha}(\Dn \cup \tDn) \wedge \Alg_{\alpha+\ep}(\Dn),
\end{equation}
where $\Dn \cup \tDn$ is the vector $(Z_1,\ldots,Z_n,\tilde{Z}_1,\ldots,\tilde{Z}_N)$.

Intuitively, the first component of~\eqref{eqn:alg_spi}, $\Alg_{\alpha}(\Dn \cup \tDn)$, serves as the main part that incorporates the synthetic data, thereby producing a procedure based on a larger sample. 
The second component, $\Alg_{\alpha+\ep}(\Dn)$, at a relaxed level $\alpha+\ep$, serves as a guardrail that does not depend on the synthetic data, and thus provides reliable statistical inference. 
The resulting procedure $\tAlg(\Dn, \tDn)$ tightly controls the risk when the synthetic distribution closely resembles the real one, while still guaranteeing risk control at level $\alpha + \ep$ even when the synthetic data are of low quality.

\begin{theorem}\label{thm:main}
Given $\alpha,\ep >0$, suppose that algorithm $\Alg$
satisfies~\eqref{eqn:target} for
   $\alpha$ and $\alpha+\ep$, and that Condition~\ref{con:target_order_formal} holds.
   Then the algorithm $\tAlg$ defined in~\eqref{eqn:alg_spi} satisfies
   \begin{align*}
       &\Ep{\Dn \iidsim P, \tDn \iidsim Q, V \sim \T(P)}{\ell(\tAlg(\Dn, \tDn), V)} 
    \\&  \qquad \le\alpha + 
   \min\{\ep,c \cdot \dell(P,Q)\}
   \end{align*}
    for all $P,Q \in \mathcal{P}$,
    where\footnote{Here, $\dtv$ denotes the total variation distance.} 
    $\dell(P,Q) = \dtv(P_{\ell,\Alg}(P,Q), P_{\ell,\Alg}(Q,Q))$, 
    and $P_{\ell,\Alg}(P,Q)$ denotes the distribution of $\ell(\Alg_{\alpha}(\Dn \cup \tDn), V)$ under $\Dn \iidsim P$, $\tDn \iidsim Q$, $\V \sim \T(P)$.
\end{theorem}

The above result provides a general upper bound that depends on the synthetic data quality, as measured by 
$\dell(P,Q)$. If the synthetic data are of high quality, this term is small and the risk is controlled close to level $\alpha$. 
However, even if the synthetic data are of arbitrary poor quality and $\dell(P,Q)\to\infty$, the guardrail ensures risk control at $\alpha + \ep$. 
Tighter and more interpretable bounds for specific applications are detailed in \Cref{app-sec:applications}.

{\bf Inference with two-sided guardrails.}
For this setting, we have a similar version of \gespi, but one that also takes the maximum ($\vee$) with the output of the base algorithm:
\begin{equation}\label{eqn:alg_spi_2}
\begin{split}
    \tAlg & (\Dn, \tDn) = \\ &\Alg_{\alpha}(\Dn) \vee (\Alg_{\alpha}(\Dn \cup \tDn) \wedge \Alg_{\alpha+\ep}(\Dn)) ,
\end{split}
\end{equation}
where $\ep \ge 0$ is a predetermined level.

Similarly to procedure~\eqref{eqn:alg_spi}, a TV-distance-type bound can be derived for~\eqref{eqn:alg_spi_2}; we defer it to \Cref{app-sec:general-theory} due to space limitations.
 Here, we state a simpler observation that codifies that the two-sided guardrail version of \gespi is sandwiched between the base algorithm at levels $\alpha$ and $\alpha + \ep$. 

\begin{theorem}\label{thm:main_2}
Suppose that Condition~\ref{con:target_order_formal} holds. Then given $\alpha,\ep >0$, the algorithm
defined in~\eqref{eqn:alg_spi_2} deterministically satisfies
\(\Alg_{\alpha}(\Dn) \preceq \tAlg(\Dn, \tDn) \preceq \Alg_{\alpha+\ep}(\Dn)\).
\end{theorem}

This result implies that the component $\Alg_{\alpha+\ep}(\Dn)$ serves as a guardrail for the validity of the synthetic-leveraged procedure $\smash{\tAlg(\Dn,\tDn)}$, while $\Alg_{\alpha}(\Dn)$ ensures that $\smash{\tAlg(\Dn,\tDn)}$'s utility is at least that of the base procedure $\Alg_{\alpha}(\Dn)$ (e.g., prediction interval width, test power, etc).

{\bf Practical choice of $\eps$.}
We suggest viewing the user-specified $\eps$ as the additional worst-case risk the user is willing to tolerate when the synthetic data are of poor quality. That is, 
$\eps$ allows the user to control the trade-off between the worst-case risk (low quality synthetic data) versus best-case gain (high quality synthetic data). Similar to $\alpha$, the choice of $\eps$ is application-dependent; in practice, a reasonable heuristic is to set $\eps$ proportional to $\alpha$ ($0.2\cdot\alpha$, $0.5\cdot\alpha$, etc), 
where a larger value reflects the practitioner's belief that the synthetic data are of high quality or a willingness to accept higher risk in order to potentially obtain greater benefit from the synthetic data.

To further guide this choice, one may first decide whether to use the synthetic data at all. A practical approach is to apply a goodness-of-fit test. Specifically, the test compares the empirical risk of the synthetic data to that of the real data.
If the two are significantly different, so that a suitable null hypothesis is rejected, this indicates that the synthetic data differ substantially from the real data and should not be used. Importantly, although it may seem that this requires data-splitting, in practice the test can be applied as a part of the synthetic-powered algorithm, with a fallback mechanism of running the only-real method at level $\alpha$.

\section{Experiments}\label{sec:exp}
We now demonstrate the performance of \texttt{\gespi} across several applications. Additional results and applications---including comparison of LLMs and mechanistic interpretability of a Vision Transformer model---are provided in \Cref{app-sec:additional-exp}. This section also includes controlled experiments on simulated data
to provide further insight into \gespi’s performance. 
Full experimental details are provided in \Cref{app-sec:exp-details}.

\textbf{Methods.} For each application, we compare the following methods: \texttt{OnlyReal}---the base inference method using only real data; 
\texttt{Guardrail}---same as \texttt{OnlyReal}, but with a higher error level $\alpha+\eps$; \texttt{OnlySynth}---the same inference method using only synthetic data; this method lacks error rate control guarantees and is included solely to illustrate the unknown quality of the synthetic data; \texttt{\gespi}---the proposed method that leverages both real and synthetic data, and supported by error rate control guarantees.

In \Cref{app-sec:all-exp-full} we include an additional baseline---denoted \texttt{Synth+Real}---that pools the real and synthetic data and then applies the base inference method at level~$\alpha$. 
This baseline does not have any error rate control guarantees, similarly to \texttt{OnlySynth}, and therefore does not constitute a competitive method; it is included only to illustrate the unknown quality of the pooled data.
Moreover, in the regime $N \gg n$---which is our main setting---\texttt{OnlySynth} and \texttt{Synth+Real} perform similarly.

\subsection{Conformal Prediction for Image Classification}\label{sec:exp-imagenet}

We begin with an image classification task, where the goal is to obtain class-conditional coverage. Specifically, given a test image $X_{n+1}$, we aim to construct a prediction set $\hat{C}(X_{n+1})$ such that
$
    \PP{Y_{n+1}\in \hat{C}(X_{n+1}) \mid Y_{n+1} = y}\geq 1-\alpha, \text{for each } y \in \mathcal{Y}.
$

The purpose of this section is to better understand and visualize how \gespi adapts to the quality of the synthetic data, and how this quality affects the performance compared to using only the real data.

\textbf{Data and experimental setup.} We follow the setup introduced in \citet{bashari2025synthetic}. We use a subset of 30 classes from ImageNet as the real data and their corresponding FLUX-generated images as the synthetic data~\cite{flux2024}. The target coverage level is $1-\alpha = 95\%$ with $\eps=2\%$. 
Additional details are in \Cref{app-sec:cp-exp-details}.

For this particular setting of the hyperparameters,
SPI~\cite{bashari2025synthetic} performs similarly to our method, as $N \gg n$; so we omit the results.
\emph{Note, however, that this is the only application among those considered in this paper where SPI can be applied.} That said, we include a comparison with SPI for different values of $N$ in \Cref{app-sec:cp}. There, we show that for smaller values of $N$, \gespi outperforms SPI. This is because our synthetic-powered prediction sets are constructed using the pooled data and not merely using the synthetic data as in SPI.  

\begin{figure}[!h]
    \includegraphics[width=0.7\linewidth]{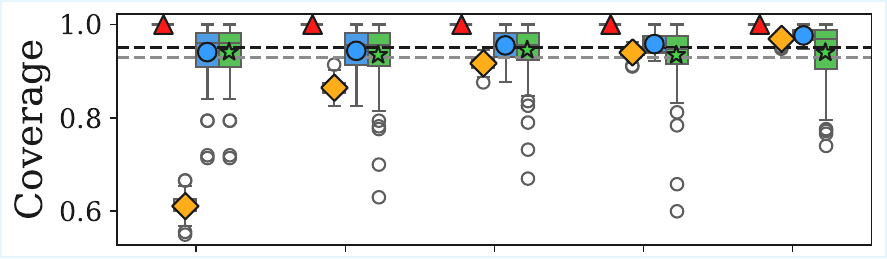}
\includegraphics[width=0.7\linewidth, valign=t]{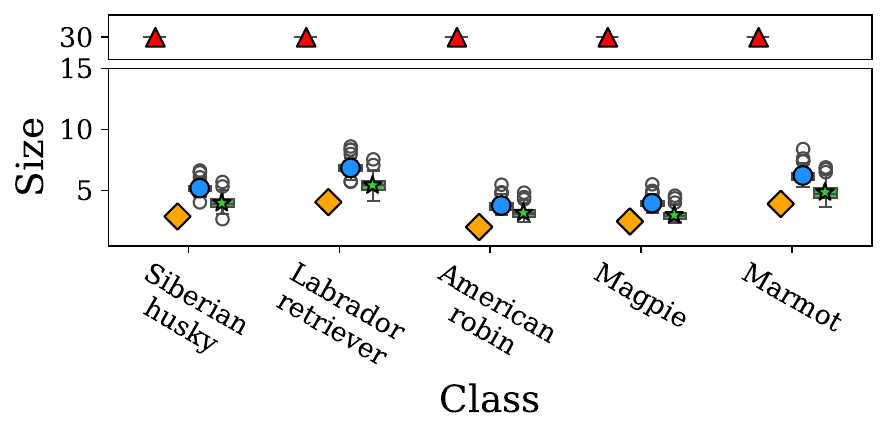}
\includegraphics[width=0.22\linewidth, valign=t]{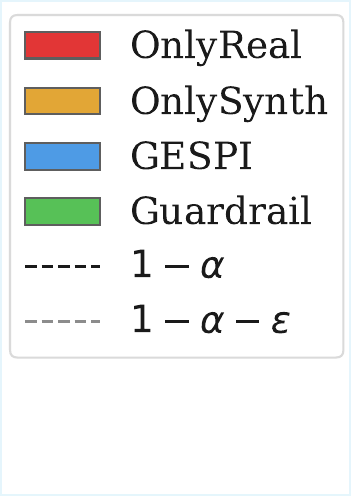}
    
    \caption{\textbf{Performance comparisons for image classification with class-conditional coverage on ImageNet}. Conformal prediction methods applied at level $\alpha = 5\%$ and $\varepsilon=2\%$. FLUX-generated images serve as the synthetic data. Results are shown for selected classes; see \Cref{app-tab:img-coverage,app-tab:img-size} for all classes.}
    \label{fig:imagenet}
\end{figure}

\Cref{fig:imagenet} presents the performance of various conformal methods for selected classes. \texttt{OnlyReal} conservatively controls the coverage at level $1-\alpha$, producing trivial, non-informative prediction sets due to the limited sample size. 
In contrast, \texttt{Guardrail} controls the coverage at level $1-\alpha-\eps$, though it exhibits high variability for the same reason. \texttt{OnlySynth} obtains arbitrary coverage levels with low variability, depending on the (unknown) synthetic data quality. The deviation of the observed coverage from the nominal level $1-\alpha$ serves as a proxy for synthetic data quality: the smaller the deviation, the higher the quality.

In turn, \texttt{\gespi} controls the coverage in the worst-case at level $1-\alpha-\eps$, while achieving coverage close to the nominal level $1-\alpha$ with low variability when the synthetic data are of high quality. By design, \texttt{\gespi} handles the full spectrum of synthetic data quality---from worst to best---without requiring any a-priori knowledge of that quality.

This adaptivity allows \texttt{\gespi} to achieve coverage close to the nominal level with low variability, rather than always attaining $1-\alpha-\eps$ coverage as \texttt{Guardrail} does. 
The variance reduction is practically important, as practitioners typically have access to only a single dataset. A lower variance ensures that the holdout-data-conditional coverage is more likely to be close to the target level $1-\alpha$.

For example, for the \textit{Magpie class}, the variance of \texttt{Guardrail} is extremely large due to the small sample size, resulting in realized coverage as low as 60\%, despite the target level being 95\%. In contrast, \texttt{\gespi} achieves coverage close to 95\% with substantially lower variability. 

To further illustrate that the quality of synthetic data is non-trivial to assess---and specifically, unknown in practice---\Cref{app-fig:husky} shows example FLUX-generated images for the \textit{Siberian Husky} class. 
Interestingly, although these synthetic images appear
to be of a high quality, 
they do not prove useful for our current task; 
further emphasizing the importance of safely leveraging synthetic data. 
Indeed, as shown in \Cref{fig:imagenet}, \texttt{OnlySynth} achieves only 60\% 
coverage, well below the target level of $95\%$, 
indicating that the synthetic data are of low quality.

We include additional results in \Cref{app-sec:cp}, including a comparison of \texttt{\gespi} and \texttt{OnlyReal} as a function of empirical coverage. 
In this experiment, we artificially align the two approaches at the same mean empirical coverage to isolate the effect of using additional high-quality synthetic data. This illustrate the variance reduction benefit of \gespi compared to using only real data. Of course, such a comparison is only illustrative, as it requires knowing the empirical coverage, which is not feasible in practice.

\subsection{Conformal Risk Control for Protein Structure Prediction}\label{sec:exp-protein}
We consider the protein structure prediction problem, where the input $X$ is the amino-acid (residue) sequence and the target $Y$ is the corresponding 3D structure (coordinates per residue). 
The goal is to control the proportion of residues whose prediction error exceeds a threshold (e.g., 3\AA).
We achieve this by abstaining predicted coordinates of residues that are likely to have such large prediction errors.
Formally, we define a prediction set $C_{\lambda}(X)\subseteq X$ as the subset of the residues abstained on. We employ the conformal risk control framework \citep{angelopoulos2022conformal,Bates2021} and utilize real data to tune a threshold $\hat{\lambda}$ such that 
$\EE{\frac{1}{|X|} \sum_{i\in X} \I{\text{err}_i > 3\text{\AA}}\cdot \I{i\notin C_{\hat{\lambda}}(X)}} \leq \alpha.$
Here, $\text{err}_i$ is the prediction error for residue $i$, which is formally defined in \Cref{app-sec:risk-control-exp-details}. Note that the choice of 3\text{\AA} is a standard scale for error, see, e.g., \cite{jumper2021highly}.

\textbf{Real data and prediction set formulation.}
We use the CASP-14 dataset, focusing on monomer protein structure prediction using AlphaFold2~\citep{jumper2021highly}. In addition to predicted structures, AlphaFold provides per-residue confidence scores (pLDDT, 0–100), which we use to construct the prediction set of residues abstained on: \(
C_{\lambda}(X) = \{i\in X: \text{pLDDT}_i < \lambda\}.
\)

\textbf{Synthetic data.} A key component of AlphaFold2 is the use of multi-sequence alignments (MSAs), where the model searches a terabyte-scale database for related sequences to improve predictions. Inspired by this, we treat the same MSAs 
as synthetic data ($\tilde{X}$) and generate
predicted structures to approximate the prediction error, since true structures for the synthetic data are unavailable. 
\textit{The appeal of this construction is that we show how the powerful MSA component of AlphaFold2 can be utilized beyond its original purpose of improving predictions: we harness the MSA sequences to form high-quality synthetic data that boost statistical inference.} Further details on the construction of the synthetic data are provided in \Cref{app-sec:risk-control-exp-details}. 

\textbf{Experimental setup and metrics.}
We use $n = 10$ out of $38$ proteins to form the real dataset $\mathcal{D}_n$, and reserve the remaining proteins for the test set; the synthetic dataset contains $N = 1{,}000$ proteins.
We apply \texttt{\gespi} with  $\ep = 5\%$, chosen relative to the $\alpha$ levels ($5-15\%$) used in the experiments.
Results are averaged over 10 repeated trials, including the average risk (average fraction of residues with error $>$ 3\AA), the average fraction of residues abstained on, and the selected pLDDT threshold $\hat{\lambda}$.

We begin by visualizing the differences between the base \texttt{OnlyReal} method and our \texttt{\gespi} procedure. To illustrate how our method performs, we select protein T1029, where AlphaFold's prediction is only partly accurate, and show the resulting prediction sets obtained by \texttt{OnlyReal} (\Cref{fig:protein-illus-crc}) and \texttt{\gespi} (\Cref{fig:protein-illus-scrc}).
Panel (1) shows the predicted structure, with residues abstained on highlighted in red and accepted residues in blue. Observe how \texttt{OnlyReal} conservatively abstains from all residues; this is a consequence of the limited real data used to tune $\lambda$. By contrast, \texttt{\gespi} abstains less, demonstrating the advantage of using synthetic data.
Panel (2) shows the predicted structure (red and blue) aligned with the true structure (gray).
Panel (3) highlights a small subset of residues for which the prediction is clearly inaccurate and where both methods abstain.
Lastly, panel (4) shows a well-aligned region where \texttt{OnlyReal} abstains unnecessarily, while \texttt{\gespi} correctly accepts these residues.
For completeness, we also visualize protein T1078 in \Cref{app-sec:risk-control}, where AlphaFold achieves relatively high accuracy.

\begin{figure}[!h]
\includegraphics[width=0.16\textwidth, valign=t]{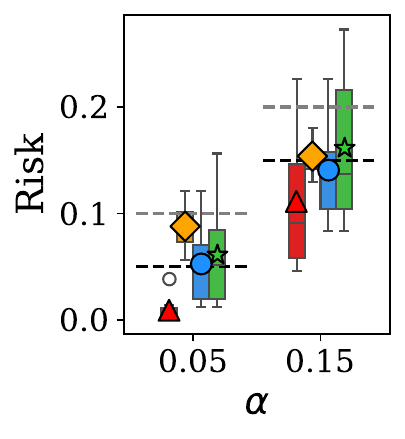}
\includegraphics[width=0.16\textwidth, valign=t]{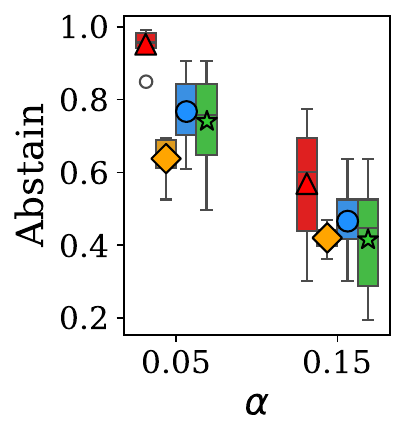}
\includegraphics[width=0.11\textwidth, valign=t]{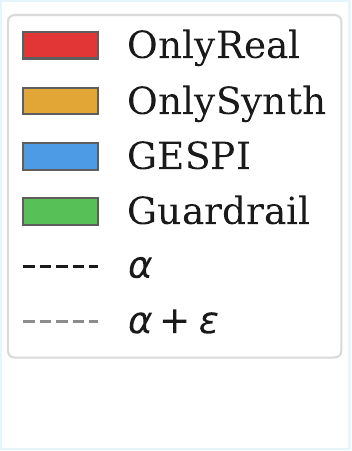}
    
    \caption{
    \textbf{Performance comparisons for protein structure prediction with error rate control}. Conformal risk control methods at
     $\alpha=5\%$ and $15\%$. Left: average risk (fraction of residues with error $>3$\AA). Right: average abstention rate. }
    \label{fig:protein-alpha--0.05-0.15}
\end{figure}

\Cref{fig:protein-alpha--0.05-0.15} presents quantitative results for two $\alpha$ levels, showing a consistent trend: \texttt{OnlyReal} conservatively controls the risk but at the cost of a high abstention rate. 
\texttt{Guardrail} controls the risk at a higher level $\alpha+\eps$ but exhibits high variability due to limited real data.
In contrast, \texttt{\gespi} achieves risk close to the nominal $\alpha$ level with lower variability, while reducing the abstention rate compared to \texttt{OnlyReal}. Crucially, \texttt{OnlySynth} serves only as a heuristic baseline and does not provide risk control guarantees. Additional results for $\alpha = 10\%$, along with the selected thresholds for all $\alpha$ levels, are in~\Cref{app-sec:risk-control}.

\subsection{Multiple Hypothesis Testing for Outlier Detection}\label{sec:exp-od}

We now consider the task of conformal outlier detection (introduced in \Cref{rmk:od}) for multiple testing. 
Conformal outlier detection guarantees FWER control given a reference set of pure inliers. In practice, however, one often has access to only a small inlier dataset $\Dn$---which can make conformal methods conservative \citep{bashari2025robust}---as well as a larger, unlabeled dataset, $\tDn$, that is contaminated with a small percentage of outliers, say $5\%$. An ideal, but infeasible, \texttt{Oracle} would annotate $\tDn$ and use only the inliers from both datasets as reference data. 
As a cheap, annotation-free alternative, we use an ML model to trim the top $q\%$ of samples from $\tDn$ that are suspected to be outliers by the model. We then treat the remainder as synthetic data consisting of ``pseudo-inlier'' points. Notably, this trimming can make \texttt{OnlySynth} less conservative, but does not guarantee error rate control at the desired level.

\textbf{Data.} We compare the performance of conformal outlier detection methods on three benchmark tabular datasets for outlier detection: {\em shuttle} \citep{shuttle}, {\em credit card} \citep{creditcard}, and {\em KDDCup99} \citep{KDDCup99}.  See \Cref{sec:exp-od-details} for additional details on the experiments.

\begin{figure}[!h]
    \centering
        \includegraphics[width=0.9\linewidth, valign=t]{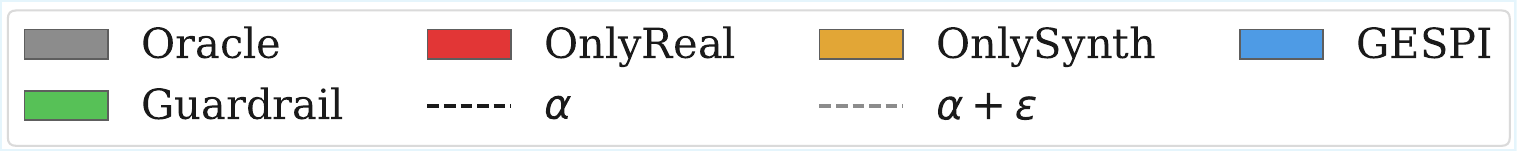}

    \includegraphics[width=0.42\linewidth, valign=t]{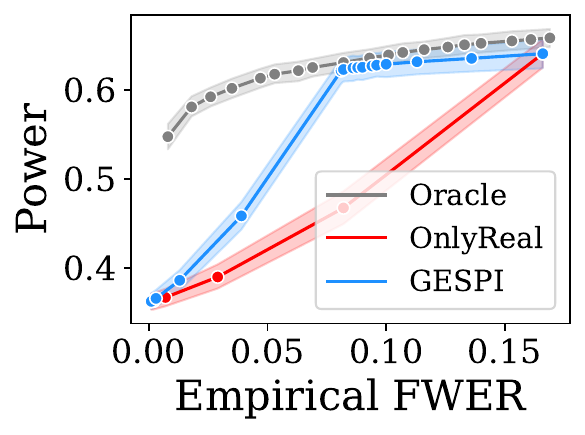}
    \includegraphics[height=0.31\linewidth, valign=t]{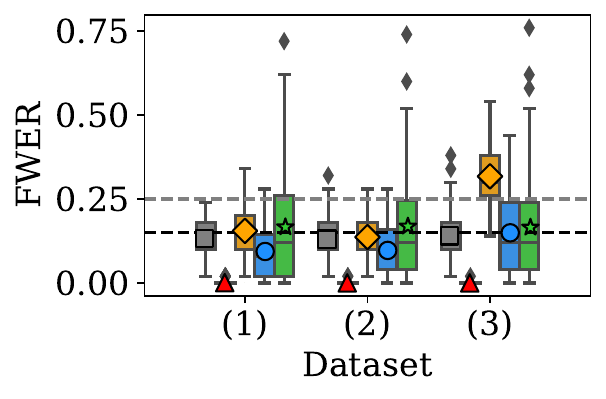}
        \caption{
        \textbf{Performance comparisons for outlier detection}. Right: average FWER evaluated on three datasets: (1) Shuttle, (2) Credit-card, (3) KDDCup99, with $\alpha=15\%$. Left: empirical power versus empirical FWER on the \textit{Shuttle} dataset. For both panels, the trimming proportion is $q=2.5\%$ and $\eps=10\%$.
        }
    \label{fig:od-fwer-roc}
\end{figure}

\Cref{fig:od-fwer-roc} shows the performance of multiple testing methods. In the right panel, we report the FWER to verify the validity of \texttt{\gespi}. The results show trends similar to those observed across all applications: \texttt{\gespi} achieves FWER close to the target $\alpha$ with lower variability than \texttt{Guardrail}.
To provide a more detailed comparison of power, we present ROC curves (left panel) that compare the methods when they achieve the same empirical error. As portrayed, \texttt{\gespi} achieves higher power compared to using only real data, with performance close to that of \texttt{Oracle}. 
This shows that \texttt{\gespi} not only attains empirical error near the target $\alpha$ (as in the right panel) but also improves power compared to using only real data when synthetic data are of high quality.
Additional results---power under FWER, single hypothesis testing, and effect of trimming proportions---are in \Cref{app-sec:od}.

\section{Discussion}
This work introduces \gespi, a new paradigm for leveraging synthetic data in statistical inference while preserving distribution-free, finite-sample error rate control guarantees.
Extensive experiments across different applications 
show that \gespi adapts automatically to the synthetic data quality: it yields substantial gains when synthetic data are useful, but never underperforms the base method. 
One limitation is that the power gain depends on the quality of the synthetic data. A promising future direction is to develop adaptive methods for selecting synthetic data to further enhance statistical power.

\section*{Acknowledgments}
M.~B., R.~M-L., and Y.~R. were supported by the European Union (ERC, SafetyBounds, 101163414). Views and opinions expressed are however those of the authors only and do not necessarily reflect those of the European Union or the European Research Council Executive Agency. Neither the European Union nor the granting authority can be held responsible for them. This research was also partially supported by the Israel Science Foundation (ISF grant 729/21). 
E.~D. and Y.~L. were partially supported by the US NSF, NIH, ARO, AFOSR, ONR, and the Sloan Foundation.
Y.~R. acknowledges additional support from the Career Advancement Fellowship at the Technion.

\section*{Impact Statement}
This work aims to advance the field of statistical inference. As with much research in this field, it may have potential societal implications, none of which are considered important enough to highlight here.


\bibliographystyle{icml2026}

\FloatBarrier
\clearpage
\appendix

\renewcommand{\thefigure}{S\arabic{figure}}
\setcounter{figure}{0}

\renewcommand{\thetable}{S\arabic{table}}
\setcounter{table}{0}

\onecolumn
\section{Additional Related Work}\label{app-sec:related-work}

Our formulation can be viewed through the lens of statistical decision theory.
In particular, our work is connected to 
robust statistics,
statistical estimation under distribution shift.
The guarantee \eqref{eqn:target} can be viewed as saying that the minimax optimal risk over the class $\mathcal{P}$ of probability distributions is upper bounded by $\alpha$.
The algorithm $\Alg$ certifies this.

Then, \Cref{thm:main} can be viewed as an upper bound on the minimax risk for the partial distribution shift problem where we observe $n$ datapoints from the original distribution, and $N$ datapoints from the shifted distribution.
The algorithm $\tAlg$ certifies this.
There has been work on 
statistical learning under a variety of distribution 
shifts, including 
the Huber contamination model
\citep[e.g.,][etc]{huberLocation,huberRatio,huber2004robust,hampel2005robust,chen2016general, chen2018robust,zhu2020robuststat}
and Wasserstein shifts
\cite{zhu2020robuststat,Liu2021RobustWE,chao2023statistical}.
However, these works typically
focus on the scenario where either 
(1) some random or adversarial subset of the data (that is not known to the analyst) is corrupted, or
(2) all datapoints are potentially corrupted.
We are not aware of studies of
the scenario where a \emph{known} subset is from the ground truth distribution, while the remaining subset is potentially shifted.

Our work is also related to 
transfer learning
\citep{Pan2010,weiss2016survey,zhuang2020comprehensive}, 
semi-supervised learning
\cite{BlumMitchell1998, BenDavid2008,chapelle2006semi}
and other forms of structured distribution shift, where a known part of the data is from the target distribution, while another part of the data shares some similarities with the target
\citep[see, e.g.,][]{Storkey2013,shimodaira2000improving,Sugiyama2012,kouw2018introduction,qiu2024efficient}.
For instance, in semi-supervised learning, we have additional unlabeled data from the target.
The question is then how to use the additional data.
However, most work in this area concerns certain known forms of relations between the auxiliary data and the original data (e.g., in semi-supervised learning, the distribution of the features is informative for the conditional distribution of the outcome given the features), and we are not aware of studies where using arbitrarily shifted data has been provably used to benefit in transfer learning.

A related line of work within semi-supervised, initiated by prediction-powered inference \citep{angelopoulos2023prediction}, studies how to combine a small labeled dataset with a larger set of synthetically generated labels to obtain unbiased inference and valid confidence intervals for parameters of interest. This framework has since been extended to more efficient procedures \citep{angelopoulos2023ppi++,ji2025predictions}, to unbiased model evaluation using synthetic labels \citep{boyeau2024autoeval,fisch2024stratified,oosterhuis2024reliable}, and to ranking problems \citep{chatzi2024prediction}. 
Notably, as in semi-supervised settings, this line of work assumes that the covariates of the unlabeled and labeled data are i.i.d., in striking contrast to \gespi.
A related direction, moment-based inference \citep{byun2026valid}, studies valid inference with imperfect synthetic data and assumes access to an unlabeled corpus sampled i.i.d. from the same underlying distribution, which similarly differs from our setting.

Our work is also related to work in causal inference 
that develops methods to pool unbiased estimators from real (experimental) data
with biased but more accurate estimators from another (usually observational) distribution \citep[see, e.g.,][etc]{cheng2021adaptive,rosenman2023combining,rosenman2025methods,de2025efficient}. 
Unlike these works, we do not focus on causal inference.

There has been a large amount of work on using prior data and information in hypothesis testing, see e.g., \cite{spjotvoll1972optimality,roeder2009genome,bourgon2010independent,dobriban2015optimal}, etc.
In this line of work, the question is: How to use data from prior studies on the same hypotheses (at the simplest level, p-values for the same null hypotheses)
to improve power in multiple testing. 
Strategies have been developed that rely on choosing a class of methods, such as based on p-value weighting (e.g., the weighted Bonferroni method), and then characterizing the optimal choice of weights as well as how to estimate them based on the available data. 
Our work is different because we do not assume an explicit statistical model that connects the prior and current data sets, but instead try to be adaptively useful when their distributions are close. 


Other recent related works include
\cite{decruyenaere2023real}, who discuss how to use synthetic tabular data in statistical inference problems, arguing that such synthetic data cannot be used as if it were real data. 
\cite{keret2025glm} discuss using synthetic data in generalized linear models, proposing to use mis-specified linear regression estimators that they argue can have a faster speed of convergence. 
An important prior work is by\cite{mccaw2024synthetic}, which develops methods for improved confidence interval construction in mixed linear models using synthetic data. 
The crucial difference between this approach and ours is that we do not make any explicit modeling assumptions
 on the synthetic data.

More broadly, our methodology enables the application of statistical methods to the analysis of a variety of generative AI models. 
In our work, we illustrate this by studying the evaluation of large reasoning models as well as the identification of internal components of vision transformer models. 
Both of these areas of application (evaluation and identification of internal components of black-box models) 
has been discussed as promising avenues where statistical methods can be used \citep{dobriban2025statistical}, and our work supports that thesis.

\section{Experimental Details}\label{app-sec:exp-details}

\subsection{Conformal prediction for image classification}\label{app-sec:cp-exp-details}

\paragraph{Data.} As mentioned in the main manuscript, we adopt the experimental setup of \citet{bashari2025synthetic}, along with the FLUX-generated images used therein. We work with a subset of 30 ImageNet classes and their corresponding synthetic images generated by FLUX \citep{flux2024}; see \citet[Appendix H.4.2]{bashari2025synthetic} for details on the image-generation procedure. The 30 classes are listed in \Cref{app-tab:img-coverage,app-tab:img-size}.

\paragraph{Model.} Following~\citep{bashari2025synthetic}, we use a CLIP model~\citep{Radford2021LearningTV} as the predictive model and the adaptive prediction sets (APS) score function~\citep{romano2020classification}; see \citet[Appendices C.1 and H.3]{bashari2025synthetic} for additional details on the score function and predictive model.

\paragraph{Experimental setup.} Unless otherwise specified, for each class the real dataset contains 15 ImageNet images, and the synthetic dataset contains 1,000 images. The test set includes 500 real images per class. The target coverage level is $1-\alpha=95\%$ with $\varepsilon=2\%$. All results are averaged over 100 runs with different data splits.

\subsection{Conformal Risk Control for Protein Structure Prediction}\label{app-sec:risk-control-exp-details}

\paragraph{Model.} We use AlphaFold2 \citep{jumper2021highly} through ColabFold \citep{mirdita2022colabfold} with the MMseqs2 search strategy over the UniRef and Environmental databases for MSA construction. Each prediction is run with five models, three recycles, and an early stopping criterion at a confidence score of 97.

\paragraph{Data.} The real dataset is taken from CASP-14. For each CASP-14 protein, we retrieved the corresponding MSA files generated during the AlphaFold run. For every protein sequence appearing in the UniRef MSAs, we queried the AlphaFold Database \citep{varadi2024alphafold} to collect predicted structures, per-residue pLDDT scores, and PAE matrices (an additional AlphaFold output representing the predicted alignment error for each residue pair). CASP-14 proteins for which predictions were unavailable for any of their MSAs were excluded, leaving a total of 38 proteins. In each experiment, the CASP-14 proteins are split into real ($\Dn$) and test sets. For a given realization of the real dataset $\Dn$, the synthetic dataset is constructed using only the MSAs of proteins in this set. If more than 1,000 synthetic samples are available, we randomly sample a balanced subset of 1,000, ensuring roughly equal representation from each protein; otherwise, we include all available samples.

\paragraph{Prediction error.} Let $\text{pred}$ and $\text{real}$ denote the predicted and real structures of a given sequence $X$, respectively, where $\text{pred}[i]$ and $\text{real}[i]$ are the 3D coordinates of the $i$-th residue.
The per-residue prediction error is defined as the average absolute difference between the pairwise distances from residue $i$ to all other residues:
\[
\text{err}_i = \frac{1}{|X|} \sum_{j=1}^{|X|} \big| \|\text{pred}[i]-\text{pred}[j]\|_2 - \|\text{real}[i]-\text{real}[j]\|_2 \big|.
\]
Intuitively, $\text{err}_i$ quantifies how well the local spatial geometry relative to residue $i$ is preserved in the predicted structure compared to the real one.

Building on this, the risk from \Cref{sec:exp-protein} 
$$\EE{\frac{1}{|X|} \sum_{i\in X} \I{\text{err}_i > 3\text{\AA}}\cdot \I{i\notin C_{\hat{\lambda}}(X)}} \leq \alpha,$$
which is the proportion of residues with error greater than 3\AA, is bounded by 1 by definition. In practice, however, AlphaFold2 predictions are fairly accurate, so the observed risk is far below this maximum. Conformal risk control uses an upper bound on the risk, denoted by $B$, to account for the unknown risk of a test point. Given AlphaFold2’s performance, we set $B=0.5$, meaning that in the worst case, at most half of the residues may have a prediction error exceeding 3\AA.

For synthetic data, the true structures are unavailable. Instead, we approximate the per-residue error using the predicted alignment error (PAE) matrix. Specifically, the synthetic error for residue $i$ is taken as the mean of the $i$-th row of the PAE matrix, and we discard any synthetic protein for which more than 50\% of residues exceed the 3\AA\ threshold according to this proxy.

\paragraph{Visualization.} All protein visualizations were performed using UCSF ChimeraX~\citep{pettersen2021ucsf}, developed by the Resource for Biocomputing, Visualization, and Informatics at the University of California, San Francisco, with support from National Institutes of Health R01-GM129325 and the Office of Cyber Infrastructure and Computational Biology, National Institute of Allergy and Infectious Diseases.

\subsection{Hypothesis Testing for Win Rate Comparison between LLMs}\label{app-sec:win-rate-exp-details}

\paragraph{Testing the null hypothesis.}
In this setting, we aim to test whether \texttt{model A} performs better than \texttt{model B} in terms of win rate. Formally, we test the null hypothesis that the win rate of \texttt{model A} over \texttt{model B} is at most $0.5$, rejecting it when we have sufficient evidence that \texttt{model A} wins more frequently.

In more detail, one observation corresponds to a trinomial random variable
\(
Z \sim \text{Trinomial}(p_{\text{win}}, p_{\text{equal}}, p_{\text{loss}}),
\)
where one coordinate of
\(
Z = (Z_{\text{win}}, Z_{\text{equal}}, Z_{\text{loss}})
\)
is equal to unity, and all others are equal to zero.
After observing $n$ independent trials/observations, we summarize them into
\(
N \sim \text{Trinomial}\!\left(n;\, p_{\text{win}}, p_{\text{equal}}, p_{\text{loss}}\right),
\)
where
\(
N = (N_{\text{win}}, N_{\text{equal}}, N_{\text{loss}})
\)
is the vector of corresponding counts.

Now, for any given observed value $N_{\text{equal}} = n_{\text{equal}}$, the conditional distribution of $N_{\text{win}}$ is
\[
N_{\text{win}} \,\big|\, N_{\text{equal}} = n_{\text{equal}}
\;\sim\; \text{Binomial}\!\left(n - n_{\text{equal}}, p_{\text{win}}/({p_{\text{win}} + p_{\text{loss}}})\right).
\]
Under the null hypothesis, this is a
\(
\text{Binomial}\!\left(n - n_{\text{equal}}, q\right), \text{with } q\leq \tfrac{1}{2}.
\)
distribution. 
Since the binomial distribution has the monotone likelihood ratio property, it suffices to perform the test at $q=0.5$.
Hence, we can apply a randomized binomial (or sign) test conditionally on $n_{\text{equal}}$, using $q=0.5$, at level $\alpha$.
This test will maintain level $\alpha$ even unconditionally.

\paragraph{Models.} We use the following models from the vLLM library for the comparisons:
\begin{compactitem}
    \item \texttt{deepseek-ai/DeepSeek-R1-Distill-Qwen-1.5B}
    \item \texttt{deepseek-ai/DeepSeek-R1-Distill-Qwen-7B}
    \item \texttt{Qwen/Qwen3-1.7B}
\end{compactitem}

For the K-way majority vote experiment, we use \texttt{deepseek-ai/DeepSeek-R1-Distill-Qwen-7B} with different values of $K\in \{1,4,8,16,64\}$.

Unless specified otherwise, models are run with the following settings: temperature 0.6, top-p = $0.95$, top-k = $20$, and min-p = $0$. For each run, we generate $K = 64$ answers per question and take the majority vote as the final answer. The maximum token limit for all runs is 32,768.

\paragraph{Data and answer verification.} We use datasets from the Hugging Face datasets library:
\begin{compactitem}
    \item Real data: AIME25 test split (\texttt{math-ai/aime25}), containing 30 challenging math reasoning questions in English.
    \item Synthetic data: a subset of OlympiadBench (\texttt{Hothan/OlympiadBench OE\_TO\_maths\_en\_COMP}), containing English-language math reasoning questions without full proofs.
\end{compactitem}

For each question, we use the following system prompt:
\begin{verbatim}
You are a helpful AI Assistant. First, think through the reasoning 
inside <think/>...</think>. Then, always present the final answer 
in \boxed{}.
\end{verbatim}
The question itself is provided as the user input.

To determine whether a model’s answer is correct, we first extract the answer from the \texttt{\textbackslash boxed\{\}} in the model’s response and apply \texttt{math\_verify.verify}. 
If the model fails to produce a complete answer within the token limit, the response is counted as incorrect and no further evaluation is performed. 
In cases where an answer is flagged as incorrect by \texttt{math\_verify.verify} (e.g., due to notation mismatches), we re-evaluate it using the LLM \texttt{deepseek-ai/deepseek-math-7b-instruct}. This LLM returns ``Yes'' if it considers the answer correct and ``No'' otherwise, and this output is used as the final evaluation. Specifically, we use the following prompt for the LLM:
\begin{verbatim}
You are given a math problem, a reference solution, and a generated 
answer. Determine if the generated answer is equivalent to the so-
lution.  Answer "Yes" or "No".

Problem:
{prompt}

Reference Solution:
{solution}

Generated Answer:
{generated_answer}

Are they equivalent? Answer Yes or No:
\end{verbatim}

\paragraph{Experimental setup and metrics.} For each experiment, we consider two experimental schemes:
\begin{compactitem}
\item \textbf{Original answers:} We compare the two models using their original responses. If \texttt{model A} outperforms \texttt{model B}, the null hypothesis should be false, and---given sufficient evidence against the null---we expect the rejection rate to exceed the nominal level $\alpha$.
\item \textbf{Shuffled answers:} As a complementary baseline, we randomly shuffle the responses of the two models. In this setting, the null hypothesis holds by design. This scheme allows us to estimate the Type I error rate.
\end{compactitem}

For each scheme, we estimate the power and Type I error by running 50 independent trials. 
In each trial, we randomly sample the real and synthetic datasets and test the null hypothesis on the resulting subset. To quantify variability, we repeat this entire procedure independently 50 times. For the shuffled-answers scheme, the random reassignment of responses is performed once at the start of each outer replicate and kept fixed across its 50 inner trials. 

Note that for these experiments, the variability of the error depends only on the number of trials, which is the same across all methods. Therefore, unlike in the other applications presented in this paper, the variability in the error does not differ between methods.

\subsection{Single and Multiple Hypothesis Testing for Outlier Detection}\label{sec:exp-od-details}

\paragraph{Experimental setup and metrics.} For single hypothesis testing (Type I error rate control), we use the following setup:
For Shuttle and KDDCup99, we use 5,000 training datapoints; for Credit Card, 2,000. Both training and reference sets are contaminated at a 5\% rate. The contaminated reference set contains 2,500 datapoints, while the clean reference set, $\Dn$, contains 40 inlier points. We use Isolation Forest \citep{liu2008isolation}, implemented using \texttt{scikit-learn} with 100 estimators, as the outlier detection model used by the conformal outlier detection framework.
Specifically, the trained model is used as a score function for computing conformal p-values. For each datapoint, it assigns a score, where a larger score indicates stronger evidence that this point may be an outlier. After computing the scores on the contaminated set, we trim the top $p\%$ of scores to attain the synthetic data.
Test sets contain 950 inliers and 50 outliers.
We report the average detection power and the average Type I error over 100 independent trials.

For multiple hypothesis testing (FWER control), the setup is similar with two key differences. First, the clean reference set consists of 100 inlier points. Second, the test set contains 1,000 datapoints with 5\% outliers, randomly partitioned into 50 batches. For each batch, every method produces a rejection set; we record whether it contains at least one false rejection (indicator 0/1). Averaging these indicators across all batches yields the empirical FWER, computed over 100 independent trials. All methods use the Simes-Hochberg procedure \citep{hochberg1988sharper} for testing.

In both experiments, we apply \texttt{\gespi} with $\eps$ chosen proportional to the target level $\alpha$: we set $\eps=1\%$ for single-hypothesis testing and $\eps=10\%$ for multiple-hypothesis testing.

\clearpage
\section{Additional Experiments}\label{app-sec:additional-exp}

\subsection{Hypothesis Testing with Simulated Data}\label{app-sec:exp-synt}
In this section, we present controlled experiments on simulated data to systematically study the performance of \texttt{\gespi}.

We focus on hypothesis testing for a single parameter. Let $\Dn=(X_i)_{i=1}^n$ denote real datapoints drawn i.i.d. from $\text{Binomial}(n,\rho)$, and let $\tDn=(\tilde{X}_i)_{i=1}^N$ denote synthetic (auxiliary) datapoints drawn from a related but potentially different distribution $\text{Binomial}(N, \rho_{\text{synt}})$. We test the null hypothesis
\[
\mathcal{H}_0: \rho \leq 0.5 \text{ versus } \mathcal{H}_1: \rho > 0.5.
\]
We use the randomized binomial test and report both power (under the alternative $\mathcal{H}_1$) and Type I error (under the null $\mathcal{H}_0$). 

\paragraph{Experimental setup and metrics.} Unless stated otherwise, the real dataset contains 50 datapoints and the synthetic dataset 500 datapoints. The target Type I error level is $\alpha = 5\%$, and \texttt{\gespi} is applied with $\ep = 2\%$. Under the alternative, the real data parameter is set to $\rho = 0.6$, while the synthetic data parameter is $\rho_{\text{synt}} = 0.55$, illustrating a scenario where the synthetic data have a lower signal yet are still useful due to the larger sample size. Each experiment is repeated 100 times to estimate the power and Type I error, and the entire procedure is repeated 100 times to evaluate variability. 

\begin{figure}[!h]
    \centering
    \hspace{-7em}
    \begin{subfigure}[t]{0.3\linewidth}
    \centering
    \includegraphics[width=\linewidth, valign=t]{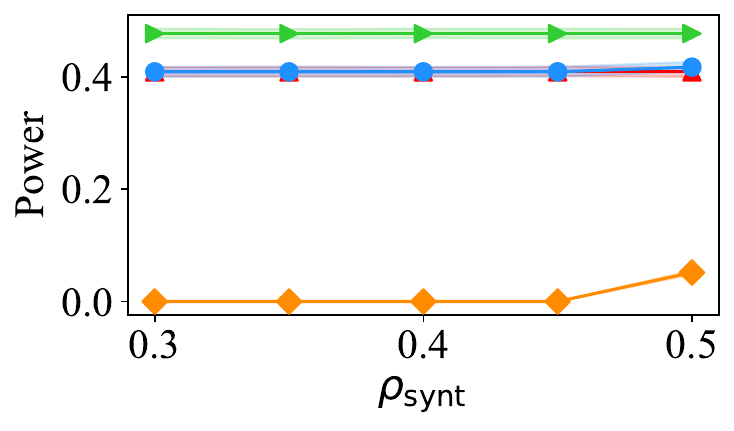}
    \caption{
    real under the alternative,\\ synthetic under the null}
    \label{app-fig:synt-p_B-1}
    \end{subfigure}
    \begin{subfigure}[t]{0.3\linewidth}
    \centering
    \includegraphics[width=\linewidth, valign=t]{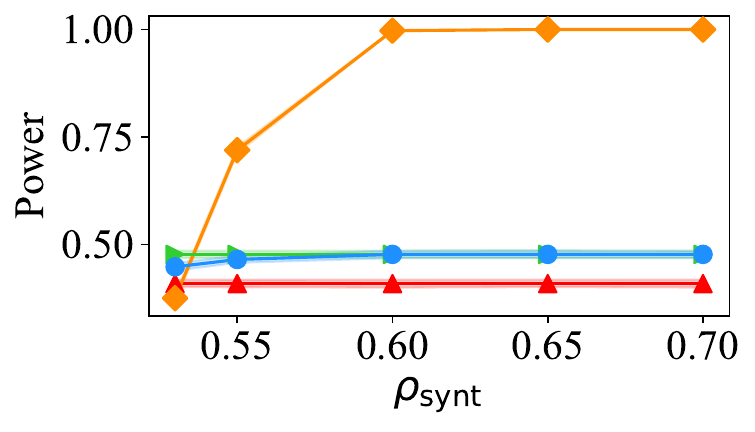}
    \caption{
    both real and synthetic under the alternative}
    \label{app-fig:synt-p_B-2}
    \end{subfigure}
    \\
    \begin{subfigure}[t]{0.3\linewidth}
    \centering
    \includegraphics[width=\linewidth, valign=t]{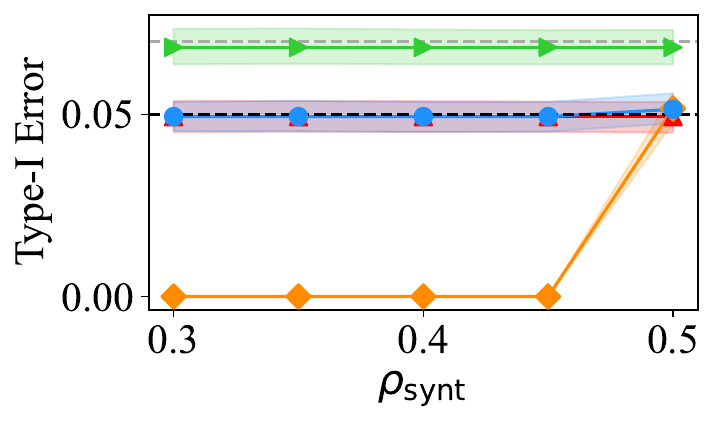}
    \caption{
    both under the null}
    \label{app-fig:synt-p_B-3}
    \end{subfigure}
    \begin{subfigure}[t]{0.3\linewidth}
    \centering
    \includegraphics[width=\linewidth, valign=t]{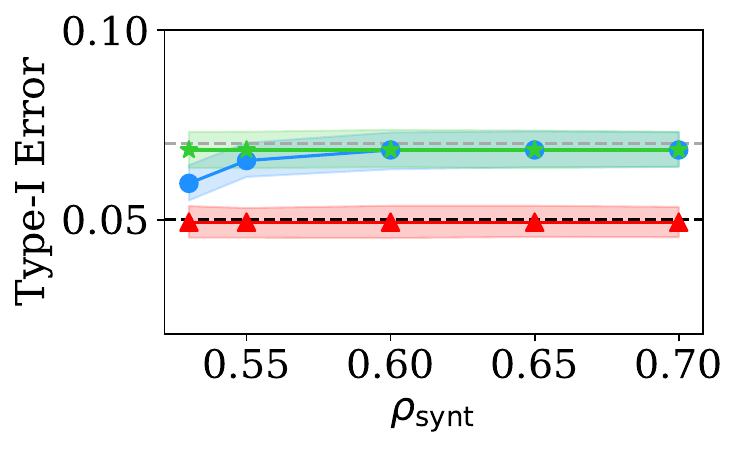}
    \caption{
    real under the null, synthetic under the alternative}
    \label{app-fig:synt-p_B-4}
    \end{subfigure}
    \includegraphics[width=0.14\linewidth, valign=t]{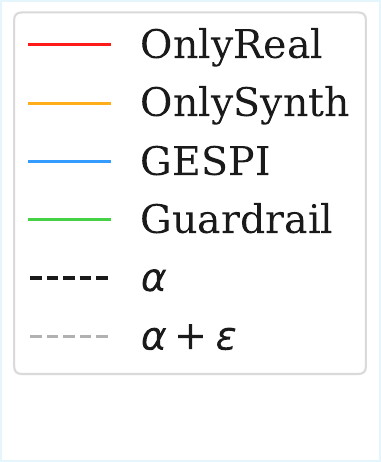}
    \caption{
    \textbf{Performance comparison as a function of $\rho_{\text{synt}}$}.
    Hypothesis testing methods across different values of $\rho$ applied at level $\alpha=5\%$ and $\ep=2\%$. Top row: $\rho = 0.6$ (alternative). Bottom row: $\rho = 0.5$ (null). }
    \label{app-fig:synt-p_B}
\end{figure}

\paragraph{The effect of the distance between real and synthetic distributions.}
We begin by examining how performance varies with different choices of $\rho_{\text{synt}}$. \Cref{app-fig:synt-p_B} summarizes results for two regimes: $\rho = 0.6$ (alternative, first row) and $\rho = 0.5$ (null, bottom row).

\Cref{app-fig:synt-p_B-1} considers a setting where the alternative holds for the real data, while the synthetic data correspond to the null. Here, the synthetic data do not provide useful information for inference, and as a result, both \texttt{OnlyReal} and \texttt{\gespi} achieve comparable power.

In contrast, \Cref{app-fig:synt-p_B-2} presents the case where the alternative holds for both the real and synthetic data. In this setting, \texttt{\gespi} achieves higher power than \texttt{OnlyReal}. Its power is comparable to \texttt{Guardrail}, which represents the maximum power \gespi can attain, as stated in \Cref{thm:main_2,thm:main_2_extended}. Meanwhile, \texttt{OnlySynth} attains even higher power. This is because OnlySynth naively treats the synthetic data as if it were real---a strategy that is invalid in this distribution-free setting, where the synthetic data may differ arbitrarily from the real distribution. Importantly, even when $\rho_{\text{synt}}\neq \rho$, we still observe a clear gain in power. This highlights that synthetic data do not need to perfectly match the real distribution; it suffices that the synthetic data support the same hypothesis as the real one (the alternative, in this case).

\Cref{app-fig:synt-p_B-3} presents the setting where the null holds for both the real and the synthetic data. All methods control the Type I error at level $\alpha$, 
except \texttt{Guardrail}, which controls it at a higher level $\alpha+\eps$. This is expected but not desired, since the goal is to control the Type I error at level $\alpha$.
Note that \texttt{OnlySynth} exhibits Type I error approaching zero when $\rho_{\text{synt}} < 0.5$. Importantly, \texttt{\gespi} still benefits from the synthetic data in this setting, which highlights the point made above: the synthetic data need not perfectly match the real distribution, as long as they support the same hypothesis (here, the null).

Finally, \Cref{app-fig:synt-p_B-4} shows the case where the null holds for the real data, but the synthetic data follow the alternative. As in \Cref{app-fig:synt-p_B-1}, the synthetic data are uninformative for inference. \texttt{OnlySynth} obtains very high Type I error and is therefore omitted from the plot (its Type I error matches the power reported in \Cref{app-fig:synt-p_B-2}). In contrast, \texttt{\gespi} controls the Type I error at most $\alpha + \ep$, as guaranteed by \Cref{thm:main_2_extended}.

\begin{figure}[!h]
    \centering
    \hspace{-7em}
    \begin{subfigure}[t]{0.3\linewidth}
    \includegraphics[width=\linewidth, valign=t]{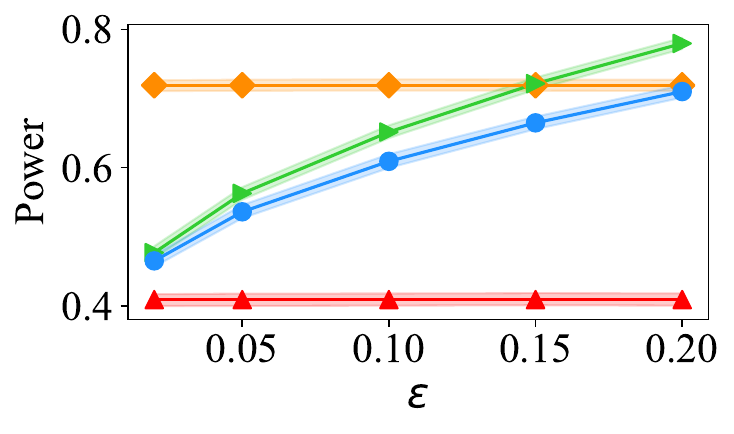}
    \caption{
    $\rho=0.6, \rho_{\text{synt}}=0.55$}
    \label{app-fig:synt-epsilon-1}
    \end{subfigure}
    \begin{subfigure}[t]{0.3\linewidth}
    \includegraphics[width=\linewidth, valign=t]{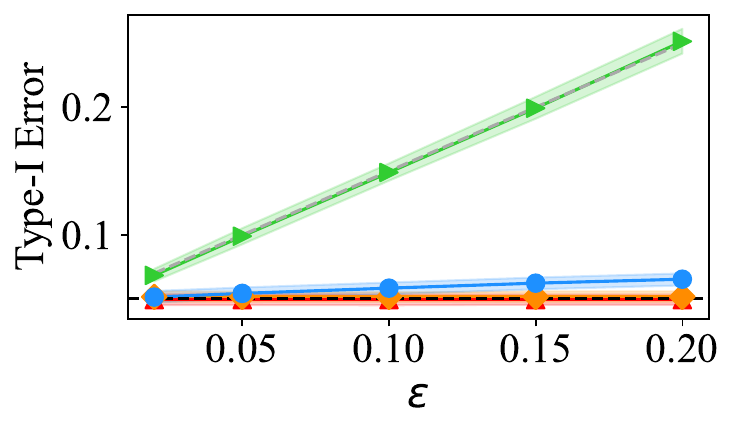}
    \caption{
    $\rho=0.5, \rho_{\text{synt}}=0.5$}
    \label{app-fig:synt-epsilon-2}
    \end{subfigure}
    \\
    \begin{subfigure}[t]{0.3\linewidth}
    \includegraphics[width=\linewidth, valign=t]{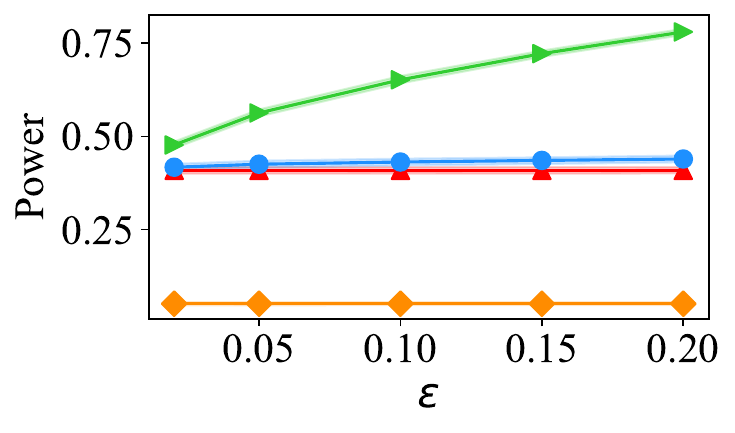}
    \caption{
    $\rho=0.6, \rho_{\text{synt}}=0.5$}
    \label{app-fig:synt-epsilon-3}
    \end{subfigure}
    \begin{subfigure}[t]{0.3\linewidth}
    \includegraphics[width=\linewidth, valign=t]{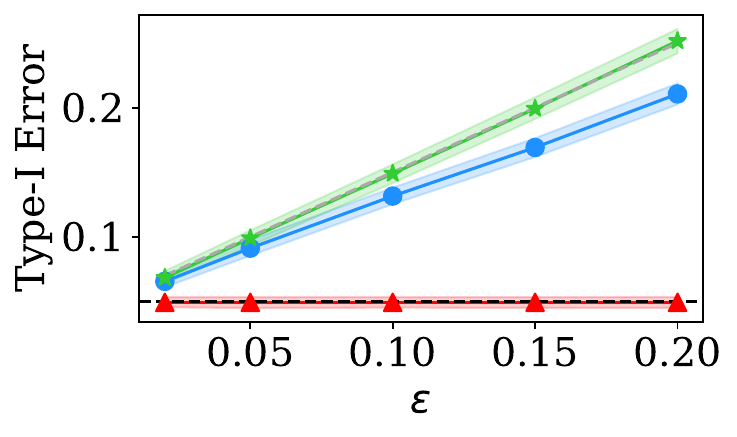}
    \caption{
    $\rho=0.5, \rho_{\text{synt}}=0.55$}
    \label{app-fig:synt-epsilon-4}
    \end{subfigure}
    \includegraphics[width=0.14\linewidth, valign=t]{figures/simulated/Type-I-Error_legend.pdf}
    \caption{
    \textbf{Performance comparison as a function of $\ep$}.
    Hypothesis testing methods across different values of $\rho$ and $\rho_\text{synt}$ applied at level $\alpha=5\%$. }
    \label{app-fig:synt-epsilon}
\end{figure}

\paragraph{The effect of $\ep$.} \Cref{app-fig:synt-epsilon} investigates how different values of $\ep$ affects the performance of \texttt{\gespi}. The experiments follow a similar setup to the one in~\Cref{app-fig:synt-p_B}, with results shown as a function of $\ep$ across four different scenarios:
(a) the alternative holds for both real and synthetic data ($\rho=0.6$, $\rho_{\text{synt}}=0.55$), where the synthetic data provide a weaker signal against the null;
(b) the null holds for both datasets ($\rho = \rho_\text{synt}=0.5$);
(c) the alternative holds for the real data ($\rho=0.6$) and the null for the synthetic data ($\rho_{\text{synt}}=0.5$);
(d) the null holds for the real data ($\rho = 0.5$) and the alternative for the synthetic data ($\rho_\text{synt} = 0.55$).

In scenario (a), shown in \Cref{app-fig:synt-epsilon-1}, \texttt{\gespi} achieves higher power than \texttt{OnlyReal}, and its power increases with $\ep$. 
Intuitively, this occurs because larger values of $\ep$ make the test on the real dataset at level $\alpha+\ep$ more liberal (as reflected by the higher power of \texttt{Guardrail}), which in turn allows \texttt{\gespi} to rely more on the pooled-data decision: the method rejects the null if both the pooled-data test at level $\alpha$ and the real-data test at level $\alpha+\ep$ reject. In scenario (b) from \Cref{app-fig:synt-epsilon-2}, where both datasets follow the null, all methods achieve Type I error close to the nominal level $\alpha$, except for \texttt{Guardrail}, which by design attains the higher level $\alpha+\eps$.

For scenarios (c) and (d), where the real and synthetic datasets correspond to opposing hypotheses, the synthetic data do not provide useful information for inference. In scenario (c), shown in~\Cref{app-fig:synt-epsilon-3}, \texttt{\gespi} and \texttt{OnlyReal} achieve comparable power, as expected. In scenario (d), \Cref{app-fig:synt-epsilon-4}, \texttt{OnlyReal} controls the Type I error at level $\alpha$, while \texttt{\gespi} exhibits a higher Type I error, but it remains controlled at level $\alpha + \ep$, as guaranteed by \Cref{thm:main_2_extended}.

\begin{figure}[!h]
    \centering
    \includegraphics[width=0.3\linewidth, valign=t]{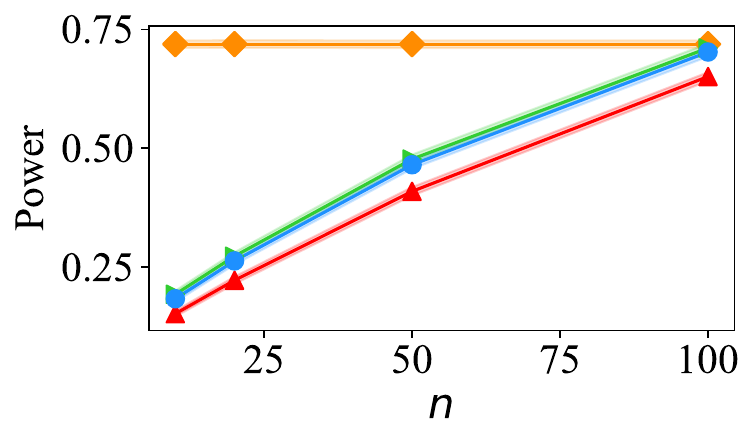}
    \includegraphics[width=0.3\linewidth, valign=t]{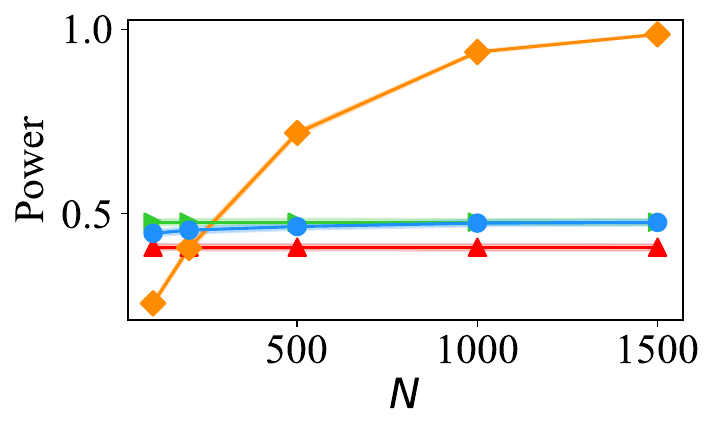}
    \includegraphics[width=0.14\linewidth, valign=t]{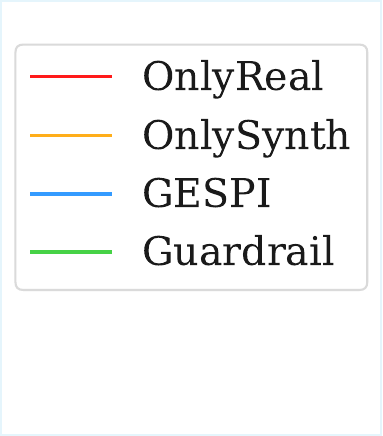}
    \caption{
    \textbf{Performance comparison as a function of the real dataset size $n$ and the synthetic dataset size $N$}.
    Hypothesis testing methods under the alternative ($\rho=0.6$ and $\rho_\text{synt}=0.55$) applied at level $\alpha=5\%$ and $\ep=2\%$. }
    \label{app-fig:synt-n}
\end{figure}

\paragraph{The effect of sample size.} \Cref{app-fig:synt-n} compares the performance of various hypothesis testing methods as a function of the real dataset size $n$ and the synthetic dataset size $N$, under the alternative hypothesis for both real and synthetic data. Following the left panel in that figure, we can see that \texttt{\gespi} consistently improves power compared to the baseline, \texttt{OnlyReal}, across all values of $n$. The right panel demonstrates a similar trend with respect to $N$: \texttt{\gespi} outperforms \texttt{OnlyReal}, and for sufficiently large synthetic datasets ($N \geq 500$), its power remains relatively unchanged under these conditions.

\begin{figure}[!h]
    \centering
    \hspace{-7em}
    \begin{subfigure}[t]{0.3\linewidth}
    \includegraphics[width=\linewidth, valign=t]{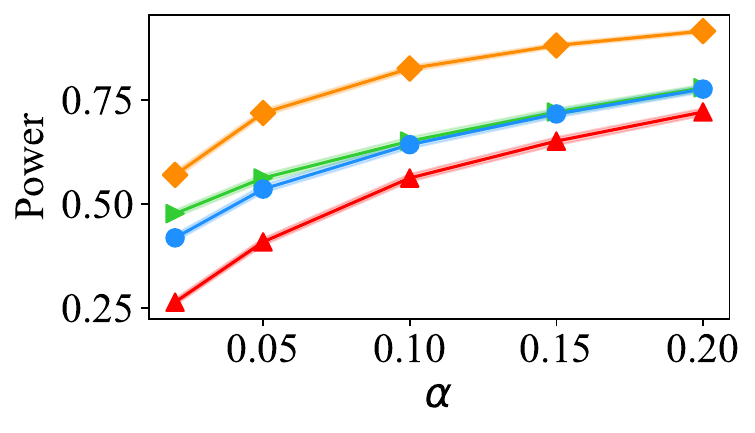}
    \caption{
    $\rho=0.6, \rho_{\text{synt}}=0.55$}
    \label{app-fig:synt-alpha-eps-05-1}
    \end{subfigure}
    \begin{subfigure}[t]{0.3\linewidth}
    \includegraphics[width=\linewidth, valign=t]{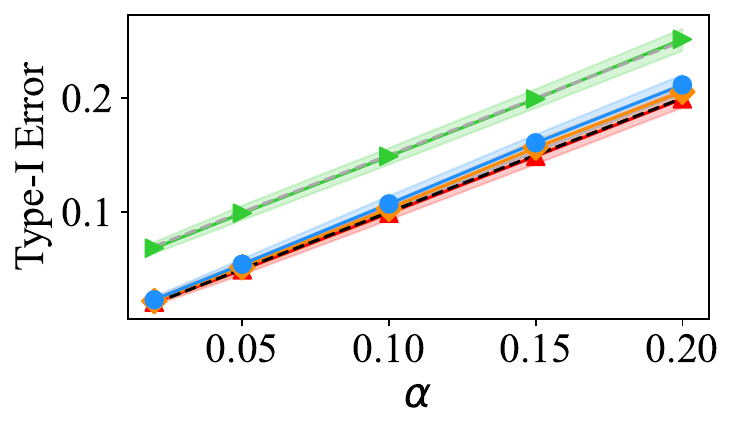}
    \caption{
    $\rho=0.5, \rho_{\text{synt}}=0.5$}
    \label{app-fig:synt-alpha-eps-05-2}
    \end{subfigure}
    \\
    \begin{subfigure}[t]{0.3\linewidth}
    \includegraphics[width=\linewidth, valign=t]{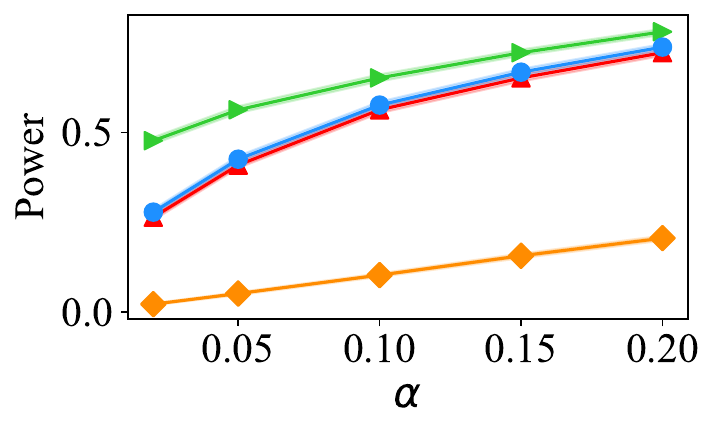}
    \caption{
    $\rho=0.6, \rho_{\text{synt}}=0.5$}
    \label{app-fig:synt-alpha-eps-05-3}
    \end{subfigure}
    \begin{subfigure}[t]{0.3\linewidth}
    \includegraphics[width=\linewidth, valign=t]{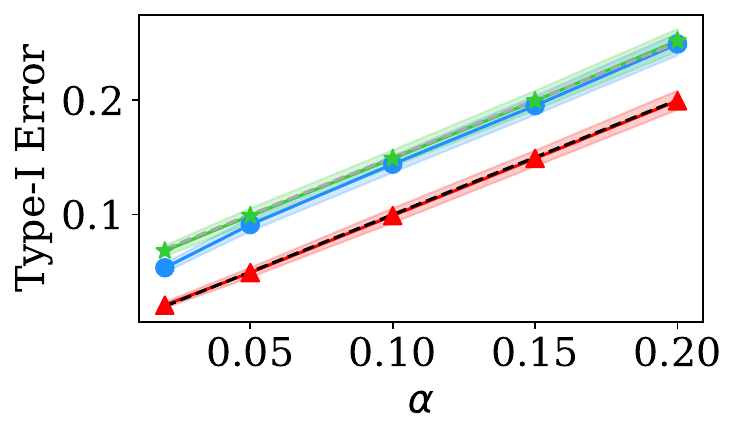}
    \caption{
    $\rho=0.5, \rho_{\text{synt}}=0.55$}
    \label{app-fig:synt-alpha-eps-05-4}
    \end{subfigure}
    \includegraphics[width=0.14\linewidth, valign=t]{figures/simulated/Type-I-Error_legend.pdf}
    \caption{
    \textbf{Performance comparison as a function of the target Type I error level $\alpha$}.
    Hypothesis testing methods across different values of $\rho$ and $\rho_\text{synt}$. \gespi applied with $\ep=5\%$.}
    \label{app-fig:synt-alpha-eps-05}
\end{figure}

\paragraph{The effect of the target error rate $\alpha$.} \Cref{app-fig:synt-alpha-eps-05} reports performance as a function of the target Type I error $\alpha$ across the four settings considered in \Cref{app-fig:synt-epsilon}, with \texttt{\gespi} applied using $\ep=5\%$.

In \Cref{app-fig:synt-alpha-eps-05-1}, where the alternative holds for both real and synthetic data, \texttt{\gespi} consistently achieves higher power than \texttt{OnlyReal} across all values of $\alpha$, with power comparable to \texttt{Guardrail}. In \Cref{app-fig:synt-alpha-eps-05-2}, where the null holds for both datasets, all methods control the Type I error close to the nominal level, except for \texttt{Guardrail}, which controls the error at level $\alpha+\eps$. 
Together, these figures illustrate a setting in which the synthetic data are of high quality: \texttt{OnlyReal} and \texttt{\gespi} achieve the same Type I error, while \texttt{\gespi} attains higher power.

We now turn to a scenario in which the synthetic data are of poor quality.
As before, when the real and synthetic data correspond to opposing hypotheses (\Cref{app-fig:synt-alpha-eps-05-3,app-fig:synt-alpha-eps-05-4}), the synthetic data provide no useful information for inference. In the former case, \texttt{\gespi} and \texttt{OnlyReal} attain comparable power, while in the latter, \texttt{OnlyReal} controls the Type I error at level $\alpha$ and \texttt{\gespi} at level $\alpha+\eps$, as guaranteed.

\begin{figure}[!h]
    \centering
    \hspace{-7em}
    \begin{subfigure}[t]{0.3\linewidth}
    \includegraphics[width=\linewidth, valign=t]{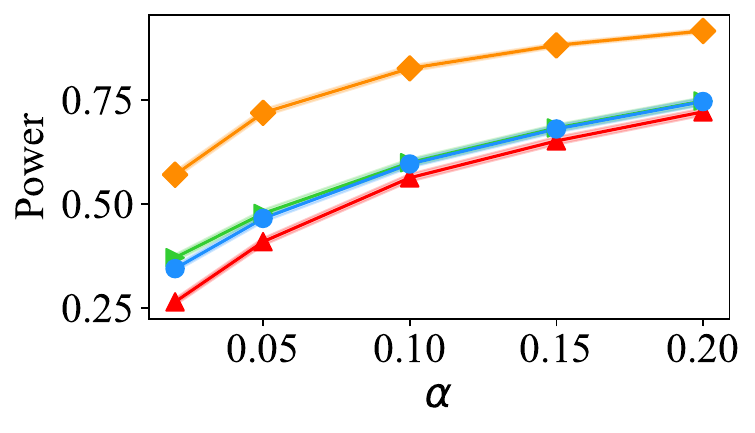}
    \caption{
    $\rho=0.6, \rho_{\text{synt}}=0.55$}
    \label{app-fig:synt-alpha-1}
    \end{subfigure}
    \begin{subfigure}[t]{0.3\linewidth}
    \includegraphics[width=\linewidth, valign=t]{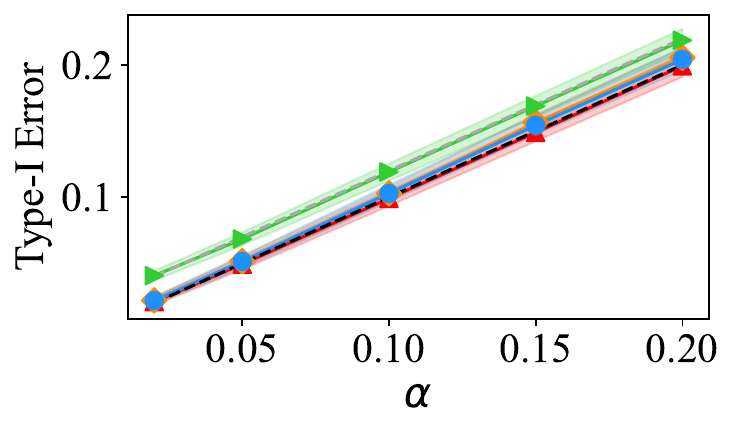}
    \caption{
    $\rho=0.5, \rho_{\text{synt}}=0.5$}
    \label{app-fig:synt-alpha-2}
    \end{subfigure}
    \\
    \begin{subfigure}[t]{0.3\linewidth}
    \includegraphics[width=\linewidth, valign=t]{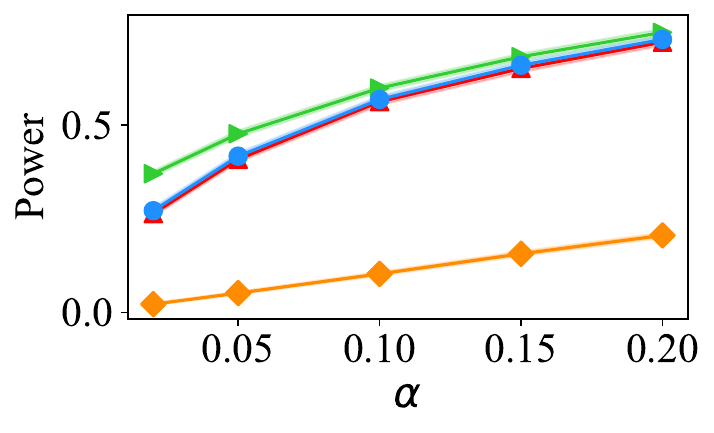}
    \caption{
    $\rho=0.6, \rho_{\text{synt}}=0.5$}
    \label{app-fig:synt-alpha-3}
    \end{subfigure}
    \begin{subfigure}[t]{0.3\linewidth}
    \includegraphics[width=\linewidth, valign=t]{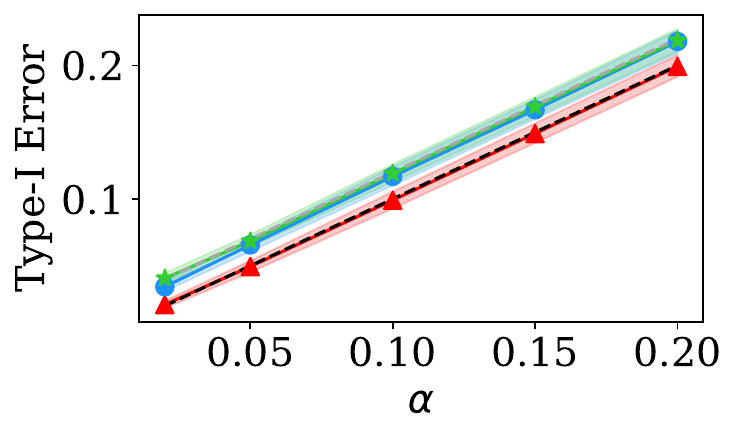}
    \caption{
    $\rho=0.5, \rho_{\text{synt}}=0.55$}
    \label{app-fig:synt-alpha-4}
    \end{subfigure}
    \includegraphics[width=0.14\linewidth, valign=t]{figures/simulated/Type-I-Error_legend.pdf}
    \caption{
    \textbf{Performance comparison as a function of the target Type I error level $\alpha$}.
    Hypothesis testing methods across different values of $\rho$ and $\rho_\text{synt}$. \gespi applied with $\ep=2\%$.}
    \label{app-fig:synt-alpha}
\end{figure}

Finally, \Cref{app-fig:synt-alpha} presents the same analysis for $\ep=2\%$, showing the same overall trends.

\FloatBarrier

\subsection{Conformal Prediction for Image Classification}\label{app-sec:cp}

\begin{figure}[!h]
    \centering
    \includegraphics[width=0.15\linewidth]{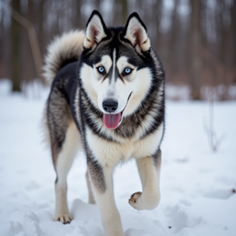}
    \includegraphics[width=0.15\linewidth]{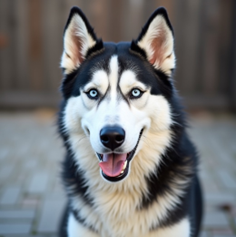}
    \includegraphics[width=0.15\linewidth]{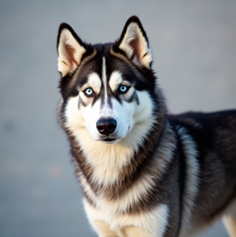}
    \includegraphics[width=0.15\linewidth]{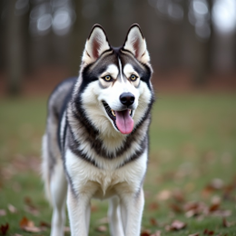}
    \includegraphics[width=0.15\linewidth]{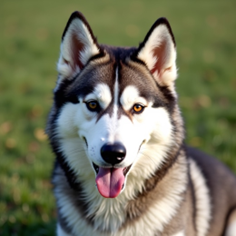}
    \caption{Examples of images generated using FLUX~\citep{flux2024} for the Siberian Husky class.}
    \label{app-fig:husky}
\end{figure}

\Cref{app-fig:husky} shows examples of images generated using FLUX\citep{flux2024} for the Siberian Husky class.
The images are visually high-quality and appear realistic. However, as shown in \Cref{fig:imagenet}, these images constitute low-quality synthetic data for statistical purposes, as their distribution deviates significantly from the real data. This is evidenced by the actual coverage of \texttt{OnlySynth} being $\approx 60\%$, whereas the target coverage is 95\%.

\subsubsection{Comparing \texttt{\gespi} and \texttt{OnlyReal}}

\begin{figure}[!h]
    \centering
    \includegraphics[height=0.16\linewidth]{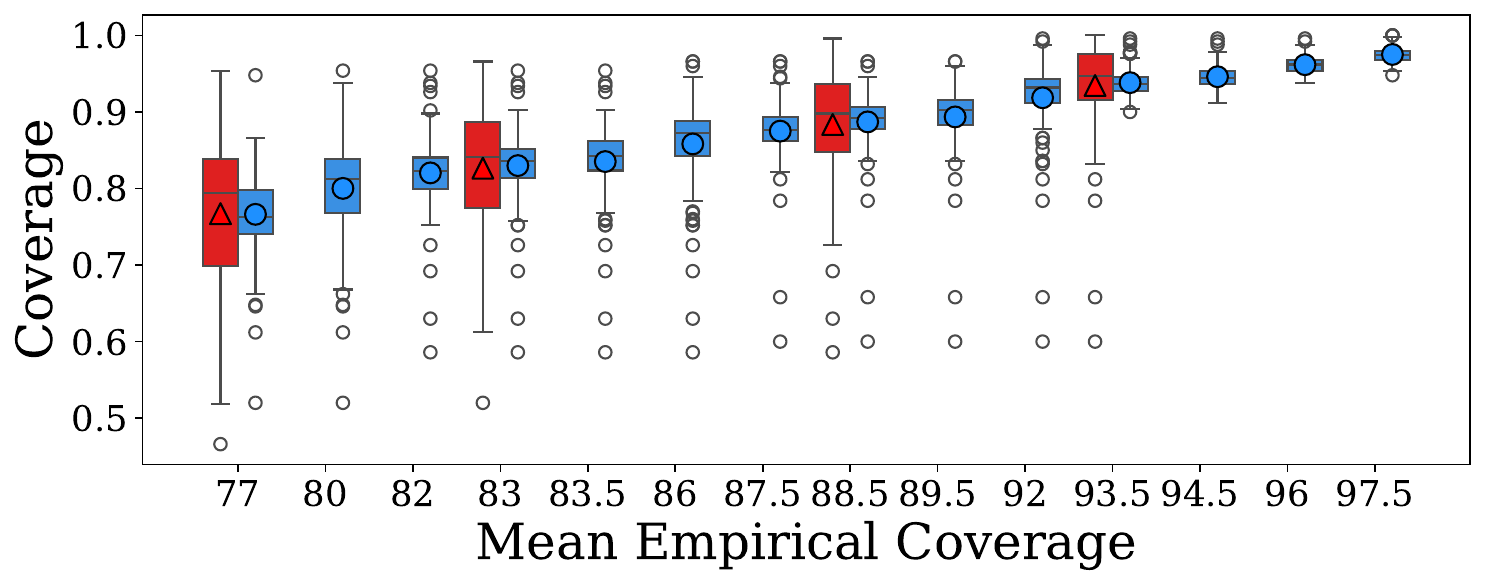}
    \includegraphics[height=0.16\linewidth]{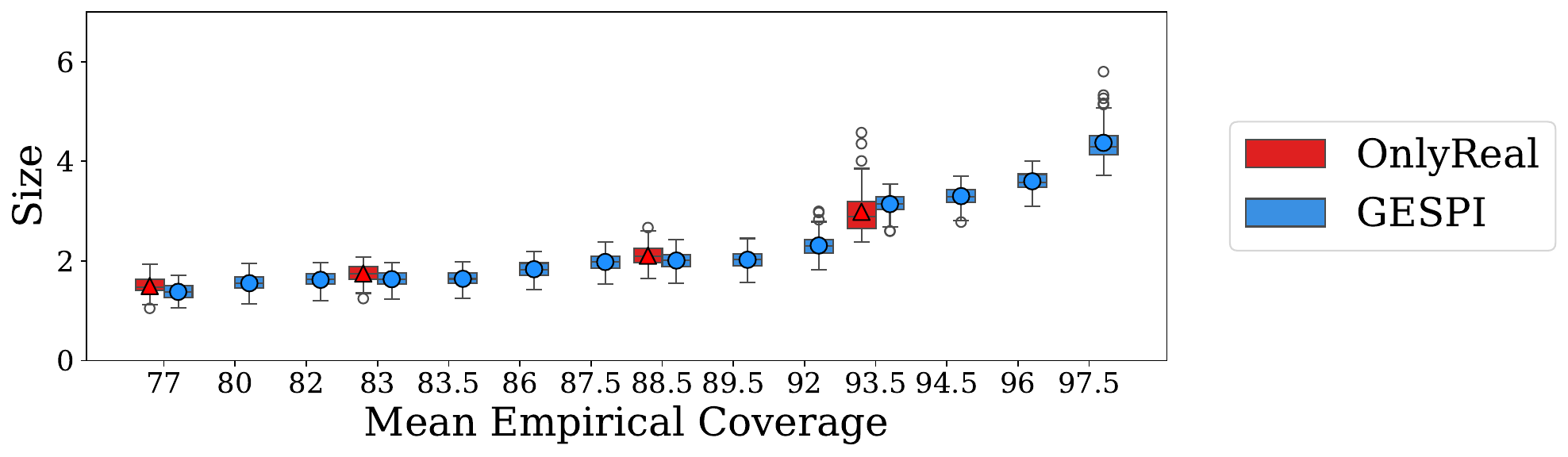}
    \caption{\textbf{Performance of GESPI and OnlyReal as a function of the mean empirical coverage for image classification on ImageNet}. 
    Results shown for the \textit{Magpie} class for varying $\alpha$ levels. GESPI applied with $\eps=10\%$.}
    \label{app-fig:cp-two-sided}
\end{figure}

\begin{figure}[!h]
    \centering
    \includegraphics[height=0.16\linewidth]{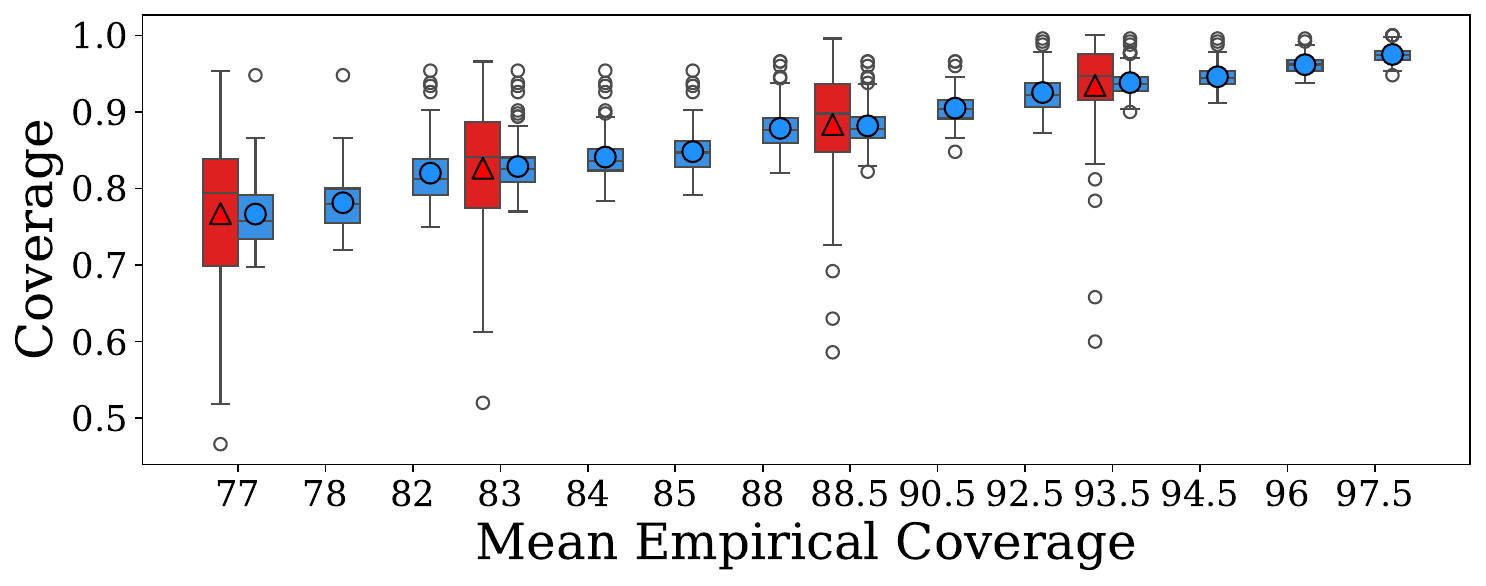}
    \includegraphics[height=0.16\linewidth]{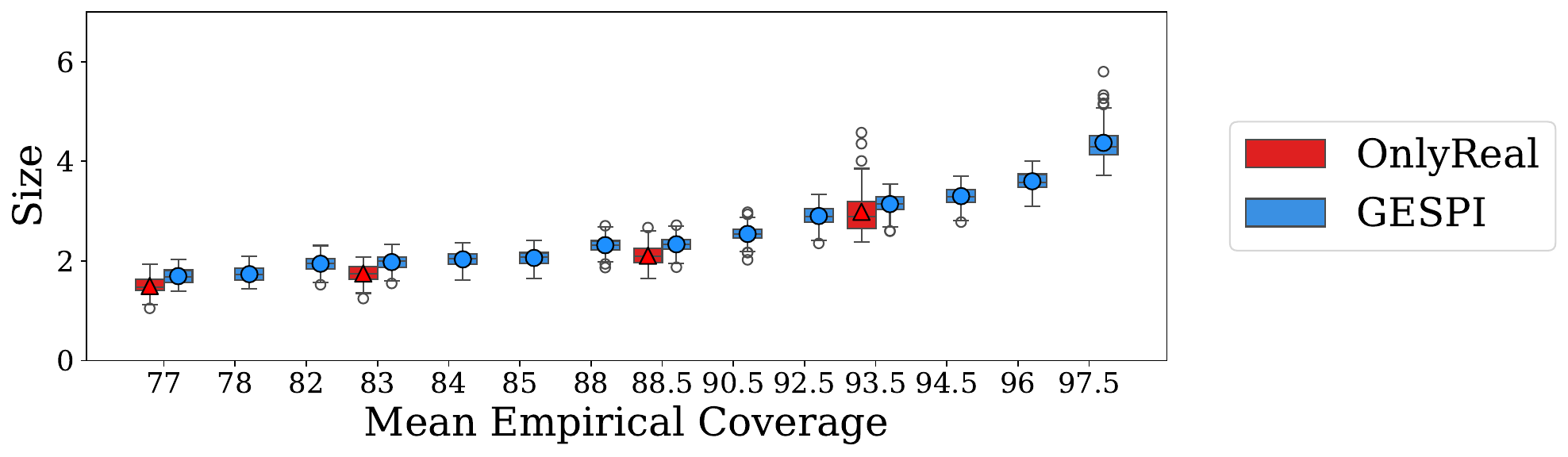}
    \caption{\textbf{Performance of one-sided GESPI and OnlyReal as a function of the mean empirical coverage}. Same experiment as \Cref{app-fig:cp-two-sided}, but using the one-sided guardrail version of GESPI.
    }
    \label{app-fig:cp-one-sided}
\end{figure}

\Cref{app-fig:cp-two-sided} revisits the class-conditional coverage experiment presented in~\Cref{fig:imagenet}, comparing \texttt{\gespi} and \texttt{OnlyReal} when they achieve the same mean empirical coverage. 
Specifically, we consider multiple target levels $\alpha$ and plot performance as a function of the observed empirical coverage.
\Cref{app-fig:cp-one-sided} shows the same setting but for the one-sided version of \texttt{\gespi}. Since both \texttt{\gespi} and \texttt{OnlyReal} use the same machine-learning model and the same prediction-set construction base algorithm, and obtain the same mean empirical coverage, the meaningful quantity to examine is the variance.

The variance reflects the scenario in which a practitioner would run the procedure multiple times with different realizations of the data. Accordingly, each point in the figures represents the holdout-data-conditional coverage observed from a single run using one data realization.

We observe that \texttt{\gespi} obtains lower variance than \texttt{OnlyReal}. In addition, \texttt{\gespi} attains coverage levels that \texttt{OnlyReal} cannot due to the small sample size. The achieved variance reduction demonstrates another benefit of \texttt{\gespi}, which goes beyond the theoretical guarantees on risk control and power. When synthetic data are of high quality, \texttt{\gespi} yields smaller variance and therefore higher stability. This variance reduction is important in practice, as a practitioner has access to only a single dataset, and lower variance ensures that the holdout-data-conditional coverage is more likely to be close to the target level.
All in all, this experiment provides further evidence that \texttt{\gespi} is useful over using only the limited real data.

\begin{figure}[!h]
    \centering
    \includegraphics[width=0.4\linewidth]{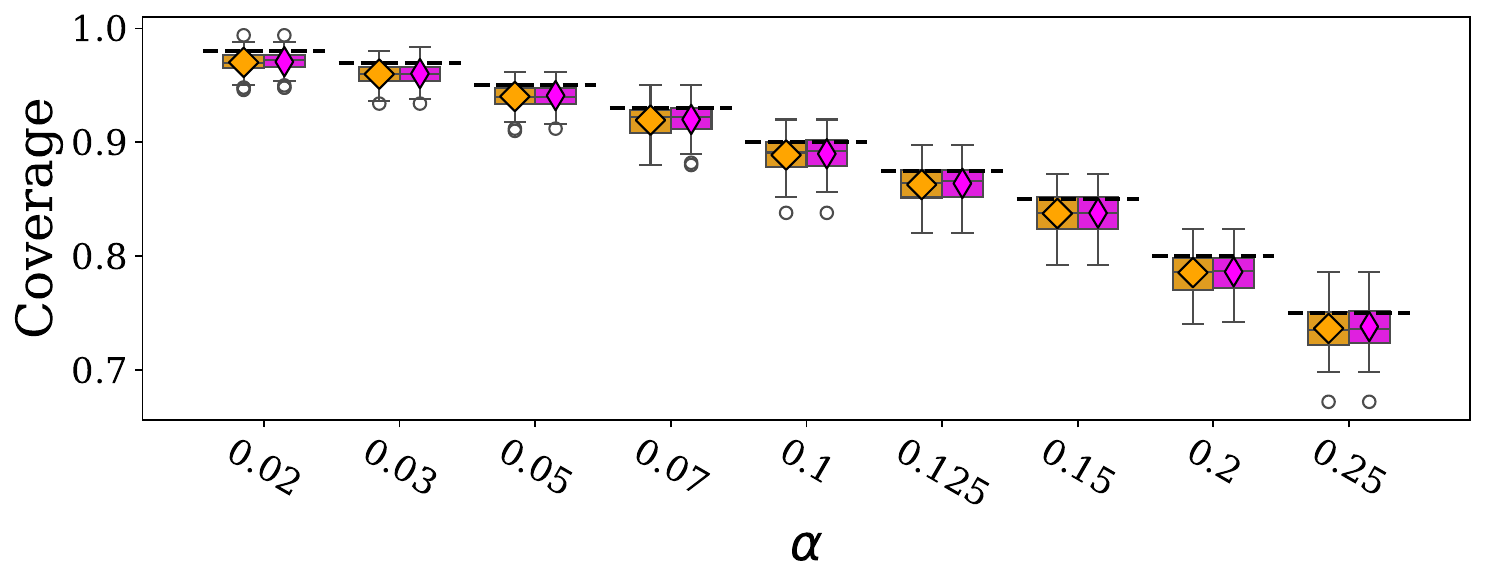}
    \includegraphics[width=0.4\linewidth]{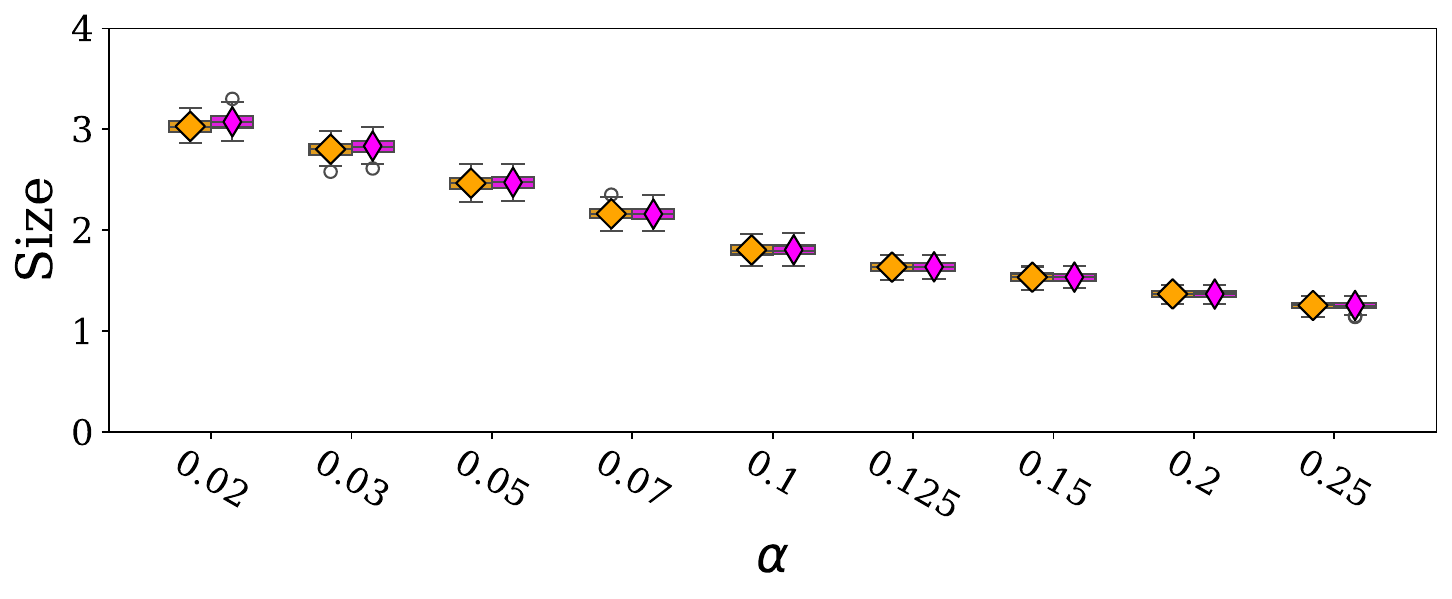}
    \includegraphics[width=0.15\textwidth]{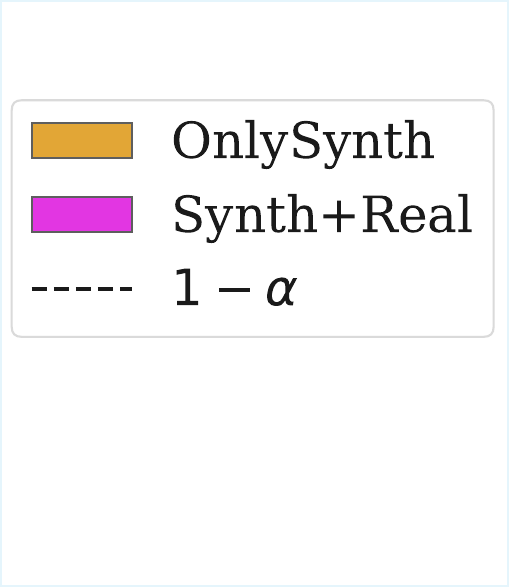}
    \caption{\textbf{Performance of OnlySynth and Synth+Real as a function of the target miscoverage rate $\alpha$ for image classification on ImageNet}. 
    Same experimental setup as \Cref{app-fig:cp-two-sided}. The dashed black lines indicate the target coverage level $1-\alpha$. These methods illustrate the quality of the synthetic data but do not provide any distribution-free guarantees.
    }
    \label{app-fig:cp-synth-coverage}
\end{figure}

Note that the synthetic data used in this experiment are not perfect, but are of high quality. To illustrate this, \Cref{app-fig:cp-synth-coverage} presents the performance of \texttt{OnlySynth} and \texttt{Synth+Real} across varying miscoverage levels $\alpha$. These methods are intended only to illustrate the quality of the synthetic data and do not provide any distribution-free risk control guarantees. As shown, both methods consistently achieve coverage slightly below the target $1-\alpha$, indicating that the synthetic data alone does not suffice here. Nevertheless, as evidenced by \Cref{app-fig:cp-two-sided,app-fig:cp-one-sided}, the synthetic data are still of high quality, as \texttt{\gespi} achieves lower variance than \texttt{OnlyReal}.

\FloatBarrier

\subsubsection{Comparison of \texttt{\gespi} and \texttt{SPI}}

In this section, we compare \texttt{\gespi} and \texttt{SPI} on an image classification task using ImageNet, following the experimental setup of \Cref{fig:imagenet}. In the conformal prediction setting, both methods are applicable. The main difference is that, unlike \texttt{SPI}, \texttt{\gespi} leverages the pooled real and synthetic data.

\begin{figure}[!h]
    \centering
    \includegraphics[width=0.28\linewidth, valign=t]{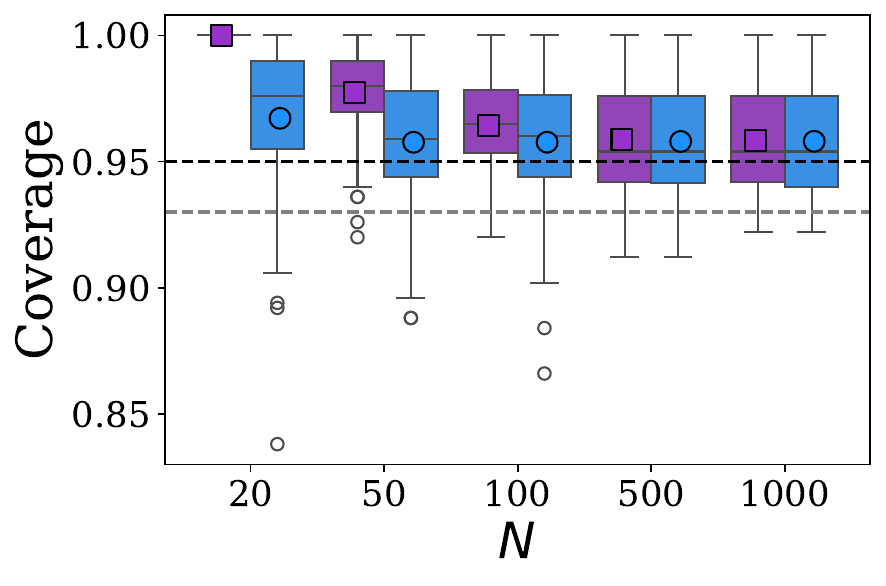}
    \includegraphics[width=0.28\linewidth, valign=t]{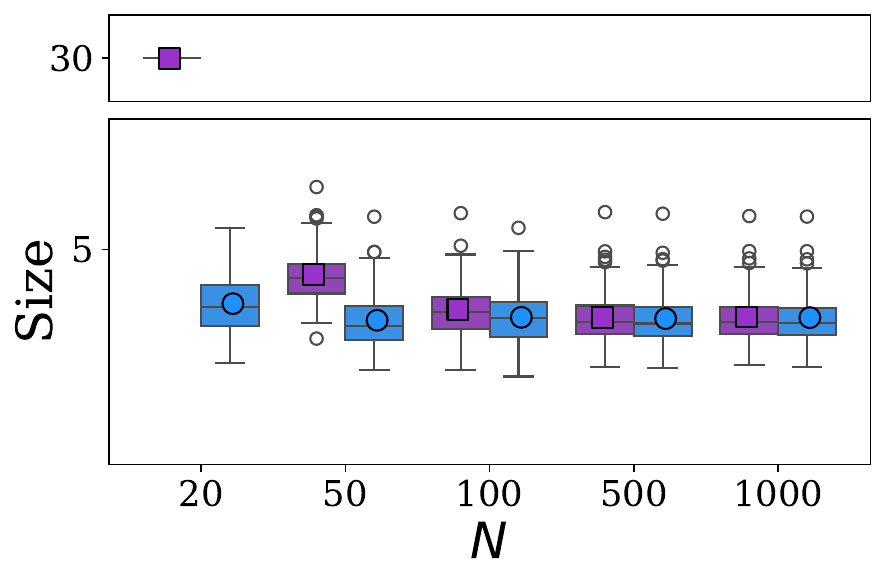}
    \includegraphics[width=0.12\linewidth, valign=t]{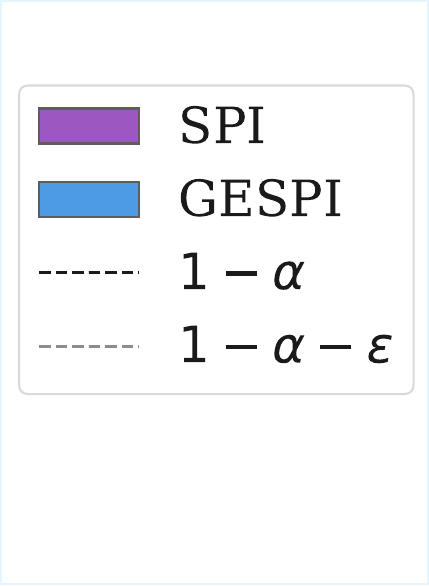}
    \caption{\textbf{Performance of \texttt{\gespi} and \texttt{SPI} as a function of the synthetic sample size $N$ for image classification on ImageNet}. Results shown for the \textit{Magpie} class. \texttt{SPI} is applied with $\beta = 0.4$, and all other details follow \Cref{fig:imagenet}.}
    \label{app-fig:gespi-vs-spi}
\end{figure}

This distinction can be practically relevant, as generating synthetic data can be costly, and one may wish to utilize them as efficiently as possible. To illustrate this point, \Cref{app-fig:gespi-vs-spi} compares the two methods as a function of the synthetic sample size $N$. For small values of $N$, \texttt{\gespi} is more sample-efficient, achieving coverage closer to $1-\alpha$ and producing smaller prediction sets. As $N$ increases, the performance of the two methods becomes comparable.

Beyond sample efficiency, \texttt{\gespi} offers greater methodological flexibility than \texttt{SPI}. In particular, \texttt{\gespi} allows the use of any inference procedure that takes both $(\tilde{\mathcal{D}}_N, \mathcal{D}_n)$ as inputs, while the guardrail bounds still hold. This flexibility allows the practitioners to filter, modify, or adapt the synthetic data using data-dependent choices without compromising validity. In contrast, \texttt{SPI} is restricted to applying only the split conformal prediction algorithm solely on the synthetic data.

\FloatBarrier

\subsection{Conformal Risk Control for Protein Structure Prediction}\label{app-sec:risk-control}

\begin{figure}[!h]
    \centering
    \begin{subfigure}[t]{\linewidth}
    \centering
    \includegraphics[width=0.8\linewidth]{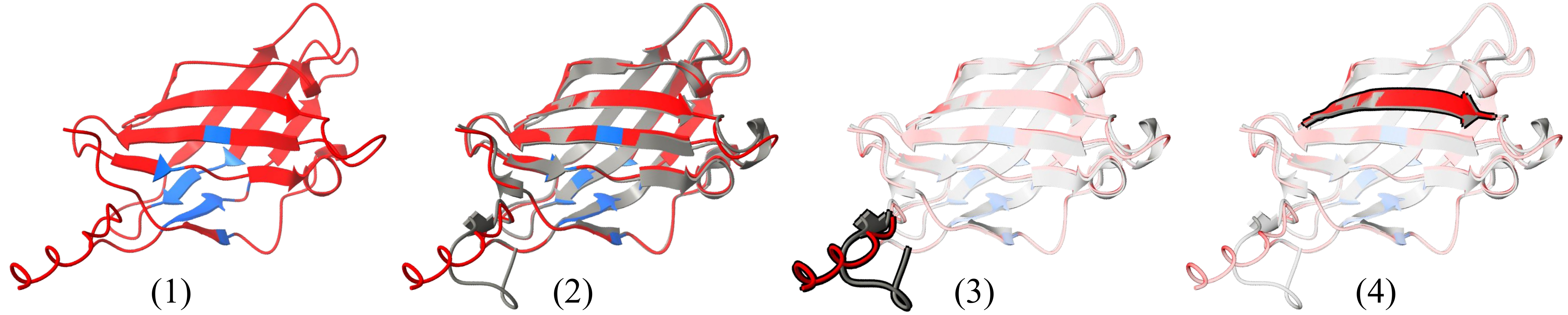}
    \caption{
    \texttt{OnlyReal}: Conformal risk control using only real data, applied at level $\alpha=5\%$}
    \label{app-fig:protein-illus-crc}
    \end{subfigure}
    \begin{subfigure}[t]{\linewidth}
    \centering
    \includegraphics[width=0.8\linewidth]{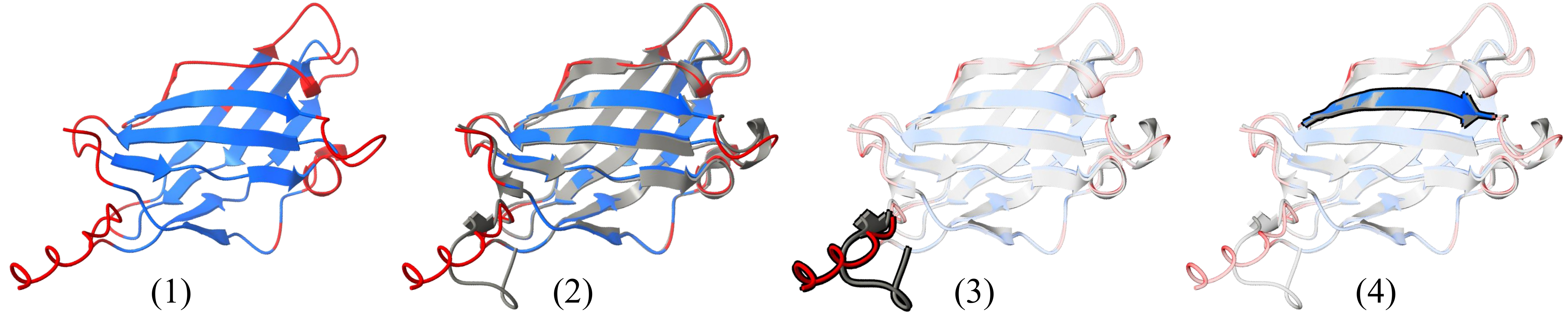}
    \caption{
    \texttt{\gespi}: our method applying conformal risk control at level $\alpha=5\%$ and guardrail $\alpha+\ep=10\%$}
    \label{app-fig:protein-illus-scrc}
    \end{subfigure}
    \caption{
    \textbf{Visualization of protein structure prediction with error rate control.} Panels show protein T1078 predictions with residues abstained on by (a) \texttt{OnlyReal} and (b) \texttt{\gespi} methods. {\color{red}Red}: residues abstained on; {\color{blue}Blue}: accepted residues. {\color{gray}Gray}: real experimental structure, aligned with AlphaFold2 predicted structure. Quantitative results \{abstention ratio, risk\}: \texttt{OnlyReal} -- \{$\approx91\%$,  $0\%$\}; \texttt{\gespi} -- \{$\approx48\%$,  $0\%$\}.}  
    \label{app-fig:protein-illus}
\end{figure}

\Cref{app-fig:protein-illus} visualizes protein T1078, for which AlphaFold produces a highly accurate predicted structure.
\Cref{app-fig:protein-illus-crc,app-fig:protein-illus-scrc} present the resulting prediction sets obtained by \texttt{OnlyReal} and \texttt{\gespi}, respectively. Panel (1) presents the predicted structure, with residues abstained on marked in red and accepted residues in blue. Both methods obtain risk equal to 0; however, \texttt{OnlyReal} conservatively abstains from most residues ($\sim91\%$), while \texttt{\gespi} abstains from significantly fewer ($\sim48\%$), demonstrating the benefit of leveraging synthetic data. Panel (2) shows the predicted structure (red and blue) aligned with the true structure (gray), illustrating that most of the protein is accurately predicted. 
Panel (3) highlights a small subset of residues whose predicted structure is inaccurate; both methods abstain. 
Lastly, panel (4) shows a well-predicted region where \texttt{OnlyReal} unnecessarily abstains, while \texttt{\gespi} correctly accepts the residues.

\begin{figure}[!h]
    \centering
    \begin{subfigure}[t]{0.2\linewidth}
    \centering
    \includegraphics[height=0.75\linewidth, valign=c]{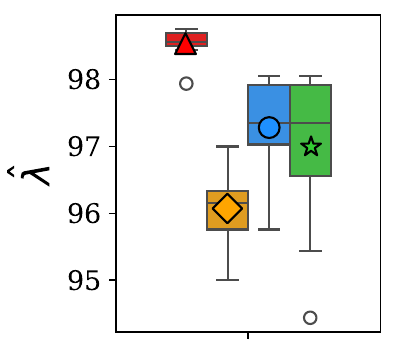}
        \caption{$\alpha = 5\%$}
    \end{subfigure}
    \begin{subfigure}[t]{0.2\linewidth}
    \centering
    \includegraphics[height=0.75\linewidth, valign=c]{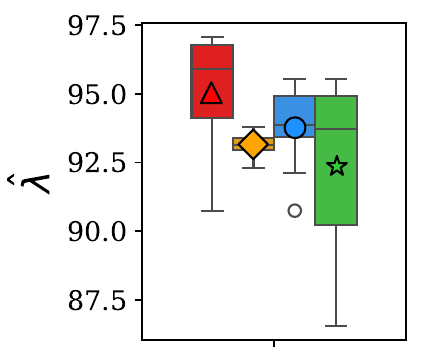}
        \caption{$\alpha=15\%$}
    \end{subfigure}
    \includegraphics[height=0.15\linewidth, valign=c]{figures/w_guardrail/protein_folding/legend.pdf}
    \caption{Chosen pLDDT thresholds, $\hat{\lambda}$, for $\alpha = 5\%$ and $15\%$, complementing the results shown in \Cref{fig:protein-alpha--0.05-0.15}.}
    \label{app-fig:protein-threshold}
\end{figure}

\Cref{app-fig:protein-threshold} complements \Cref{fig:protein-alpha--0.05-0.15} in the main manuscript by showing the chosen pLDDT thresholds, $\hat{\lambda}$, for $\alpha = 5\%$ and $15\%$. The figure demonstrates that the selected thresholds vary with $\alpha$ and differ across methods: \texttt{OnlyReal} conservatively chooses higher thresholds, while our \texttt{\gespi} procedure selects lower thresholds with lower variability than \texttt{Guardrail}. As a result, \texttt{\gespi} achieves risk closer to the nominal $\alpha$ level with lower variability.

\begin{figure}[!h]
    \centering
    \includegraphics[height=0.15\linewidth, valign=c]{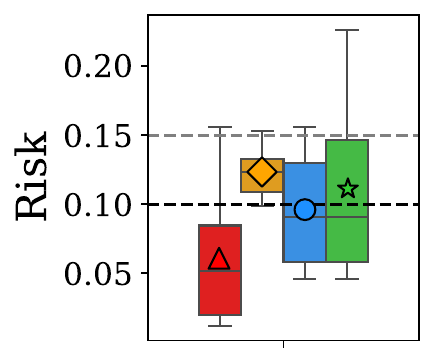}
    \includegraphics[height=0.15\linewidth, valign=c]{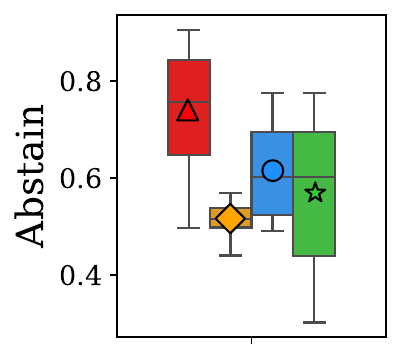}
    \includegraphics[height=0.15\linewidth, valign=c]{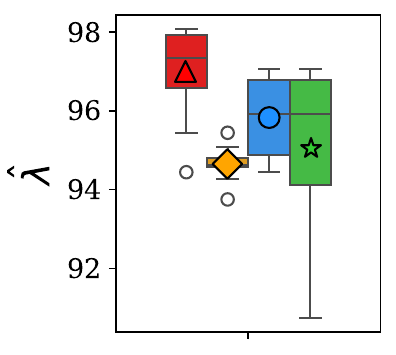}
    \includegraphics[height=0.15\linewidth, valign=c]{figures/w_guardrail/protein_folding/legend.pdf}    
    \label{app-fig:protein_alpha_0.15}
    \caption{
    \textbf{Performance comparisons for protein structure prediction with error rate control.}
    Conformal risk control methods applied at target level  $\alpha=10\%$. Left: average risk (fraction of residues with error $>3$\AA). Middle: average abstention rate. Right: selected pLDDT threshold $\hat{\lambda}$.}
    \label{app-fig:protein}
\end{figure}

\Cref{app-fig:protein} presents the performance of the conformal risk control methods at target level $\alpha=10\%$. The results show a similar trend to that in \Cref{fig:protein-alpha--0.05-0.15}. In particular, \texttt{OnlyReal} conservatively controls the risk and leads to a higher and more variable abstention rate. \texttt{OnlySynth}, which relies on approximate prediction errors, does not provide valid risk control guarantees. In contrast, our proposed method, \texttt{\gespi}, achieves risk close to the nominal $\alpha$ level with a lower abstention rate and lower variability compared to \texttt{Guardrail}.

\FloatBarrier

\subsection{Hypothesis Testing for Comparing Large Reasoning Models}\label{app-sec:exp-win-rate}

For brevity, we provide an overview of the experimental setup here and refer to~\Cref{app-sec:win-rate-exp-details} for full details.

Given two large language models (LLMs), we aim to test whether \texttt{model A} outperforms \texttt{model B} on a specific type of task. Formally, we consider the hypotheses
$
\mathcal{H}_0: p\leq0.5\text{ vs. }\mathcal{H}_1: p>0.5,
$
where $p$ denotes the win rate of \texttt{model A} over \texttt{model B}.

\textbf{Data.} We evaluate the win rate on the AIME25 dataset, which consists of 30 challenging reasoning math questions. 
The goal is to pin down the ranking of the models on AIME25 as closely as possible, which is hindered by the very small number of questions for this competition. 
For synthetic data, we use a subset of the OlympiadBench \citep{he2024olympiadbench} math competition questions, which resemble AIME problems but are drawn from a different distribution and therefore cannot be treated as real test data.\footnote{Since OlympiadBench was released in 2024, before AIME25, these two datasets are non-overlapping.}
For each question and model, we record whether the model’s answer is correct. Further details on the experimental setup are provided in \Cref{app-sec:win-rate-exp-details}.

\textbf{Experimental setup and metrics.} 
We randomly choose $n=15$ AIME25 math problems and $N=100$ synthetic problems to from $\Dn$ and $\tDn$; the choice of using half of the real dataset allows us to run multiple repetitions and estimate the power and Type I error.
In addition to this standard experiment, we evaluate the validity of our method by randomly shuffling the responses of the two models, which corresponds to the null hypothesis. This allows us to estimate the Type I error rate. 

We consider two experiments: the first is used for hyperparameter selection in the K-way majority vote, and the second compares the performance of different models.

\subsubsection{Hyperparameter Selection for K-Way Majority Vote}

Majority-vote scheme typically improves performance as $K$ increase, but at the cost of requiring $K$ forward passes per input. In this experiment, we compare the performance of the R1-Deepseek-distill-Qwen-7B model with an expensive majority vote using $K=64$ to cheaper alternatives with $K=1,4,8,$ and $16$.

\begin{figure}[!h]
    \centering
    \includegraphics[width=0.35\linewidth, valign=t]{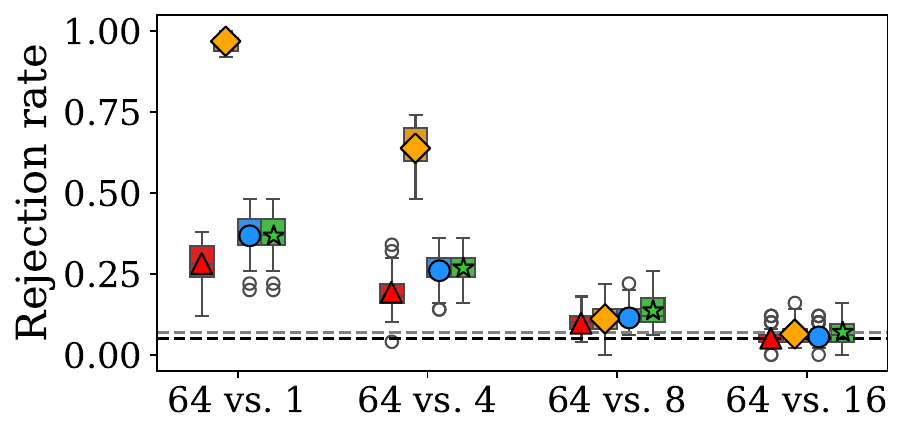}
    \includegraphics[width=0.35\linewidth, valign=t]{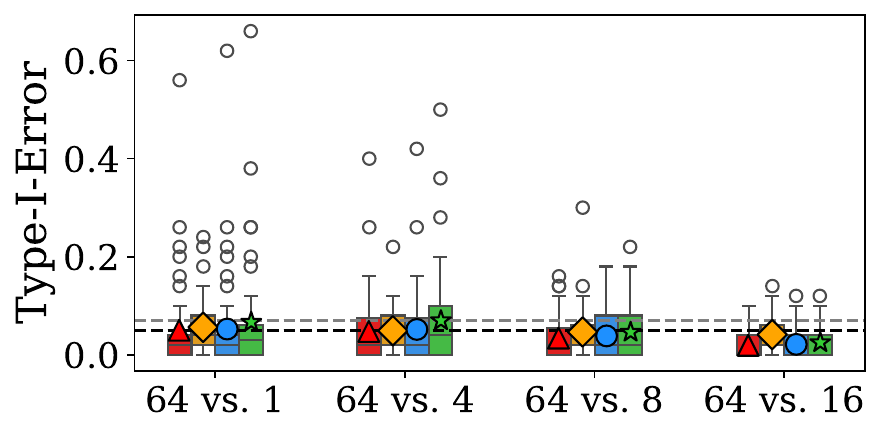}
    \includegraphics[width=0.13\linewidth, valign=t]{figures/w_guardrail/protein_folding/legend.pdf}
    \caption{\textbf{Performance comparisons for different K-way majority vote on the AIME25 dataset for the R1-Deepseek-distill-Qwen-7B model}. Hypothesis testing methods are applied at level $\alpha = 5\%$ and $\varepsilon = 2\%$. Comparisons are made between the expensive $K=64$ and cheaper $K=1,4,8,16$. Left: Rejection rate, measured under the standard setting. Right: Type I error, measured in the shuffled-response setting where the null hypothesis holds.}
    \label{app-fig:llm-k-majority}
\end{figure}

\Cref{app-fig:llm-k-majority} presents the performance of different hypothesis testing methods across these different $K$-way majority votes. All methods control the type I error close to the nominal level $\alpha$, except \texttt{Guardrail}, which always controls at the higher level $\alpha+\eps$ by design. Note that \texttt{OnlySynth} performs the test on a different dataset, which is intended to be similar to the real datset (AIME25) but may not follow the same distribution; therefore, it is not a competing method and is included for illustration only.

Following the left panel, we see that comparing $K = 64$ to $K \in \{1, 4, 8\}$ yields rejection rates above the nominal level, indicating that the alternative holds, with \texttt{\gespi} achieving higher power than \texttt{OnlyReal} and is comparable with \texttt{Guardrail}. Interestingly, there is no strong evidence that $K = 64$ outperforms $K = 16$ for this task, as all methods achieve rejection rates close to $\alpha$.

\subsection{Comparing Different Models}

\begin{figure}[!h]
    \centering
    \begin{subfigure}[t]{0.35\linewidth}
    \footnotesize
\texttt{Comp.~1:} DeepSeek-R1-Distill-Qwen-1.5B $>$ DeepSeek-R1-Distill-Qwen-7B (temp.$=0$)\\

\texttt{Comp.~2:} Qwen3 1.7B (temp. $=0$) $>$ DeepSeek-R1-Distill-Qwen-7B (temp. $=0$)
    \end{subfigure}%
    \hfill
    \begin{subfigure}[t]{0.6\linewidth}
    \includegraphics[width=0.35\linewidth, valign=t]{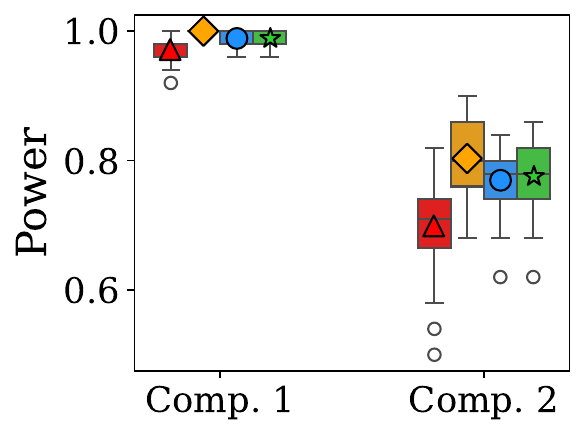}
    \includegraphics[width=0.35\linewidth, valign=t]{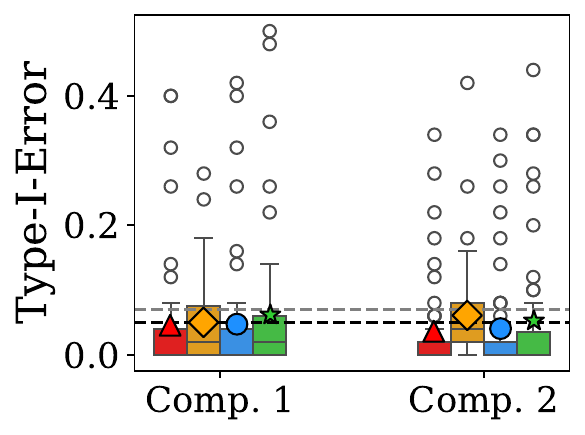}
    \includegraphics[width=0.2\linewidth, valign=t]{figures/w_guardrail/protein_folding/legend.pdf}
    \end{subfigure}
    \caption{
    \textbf{Performance comparisons for LLM win rate on AIME25 dataset}. Hypothesis testing methods applied at level $\alpha=5\%$ and $\ep=2\%$. Left: Description of model comparisons. Middle: Power, comparing the rejection rate under the standard setting. Right: Type I error, measured in the shuffled-response setting where the null holds. 
    }
    \label{app-fig:win-rate}
\end{figure}

\Cref{app-fig:win-rate} presents two model comparisons, each involving a distinct pair of LLMs. In both cases, the rejection rate (power) is well above the target level $\alpha$, indicating that the first model outperforms the second. Our proposed method, \texttt{\gespi}, achieves higher power than \texttt{OnlyReal} and is comparable with \texttt{Guardrail}. In the shuffled-answers setting, both \texttt{\gespi} and \texttt{OnlyReal} methods achieve Type I error at the target level $\alpha$, while \texttt{Guardrail} controls the error at the higher level $\alpha+\eps$.

\FloatBarrier

\subsection{Single and Multiple Hypothesis Testing for Outlier Detection}\label{app-sec:od}

We provide results for single and multiple hypothesis testing for outlier detection at specific $\alpha$ and $\eps$ levels, representing a real-world scenario, as well as ROC curves for different trimming proportions. We note that the ROC curves shown in the main manuscript require knowledge of the true outliers and inliers---information that is not available in practice---and are therefore included only for illustrative comparison.

\begin{figure*}[!tb]
    \centering
    \includegraphics[width=0.8\textwidth, valign=t]{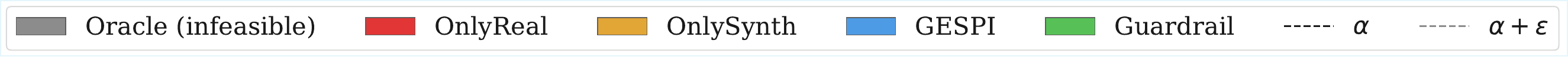}
    \begin{subfigure}[t]{0.45\linewidth}
    \includegraphics[height=0.35\linewidth, valign=t]{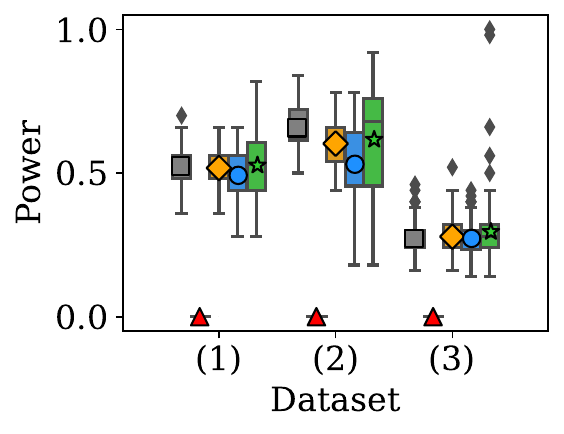}
    \includegraphics[height=0.35\linewidth, valign=t]{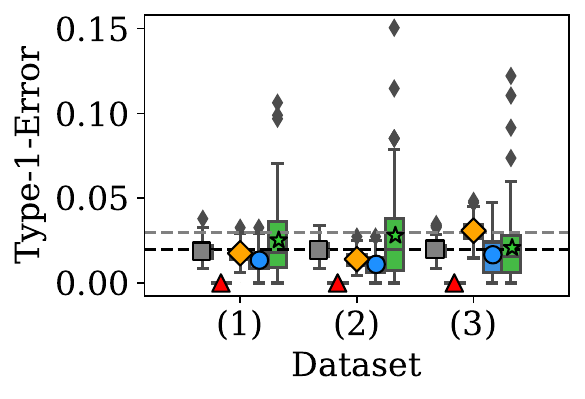}
    \end{subfigure}
    \begin{subfigure}[t]{0.45\linewidth}
    \includegraphics[height=0.35\linewidth, valign=t]{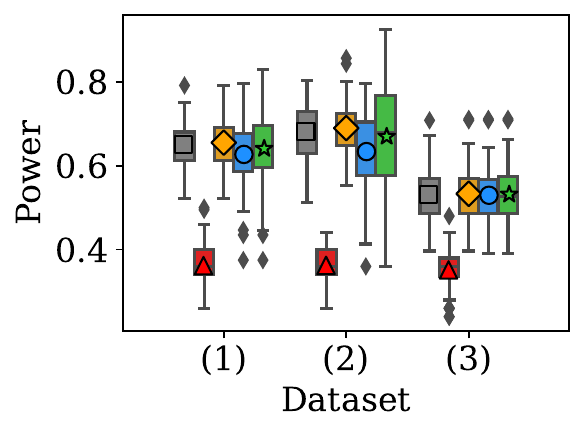}
    \includegraphics[height=0.33\linewidth, valign=t]{figures/w_guardrail/outlier-detection/initial_100_p_trim_0.025_fwer/FWER_no_legend.pdf}
    \end{subfigure}
    \caption{
    \textbf{Performance comparisons for outlier detection}.
    Evaluated on three datasets: (1) shuttle, (2) credit-card, (3) KDDCup99. Left two panels: single-outlier case ($\alpha=2\%$, $\ep=1\%$). Right two panels: multiple-outlier case ($\alpha=15\%$, $\ep=10\%$). The trimming proportion is $q=2.5\%$.}
    \label{fig:od-hypothesis-testing}
\end{figure*}

\begin{figure}[!h]
    \centering
    \includegraphics[width=0.9\textwidth, valign=t]{figures/w_guardrail/outlier-detection/legend_row.pdf}
    \begin{subfigure}[t]{0.45\linewidth}
    \includegraphics[height=0.35\linewidth, valign=t]{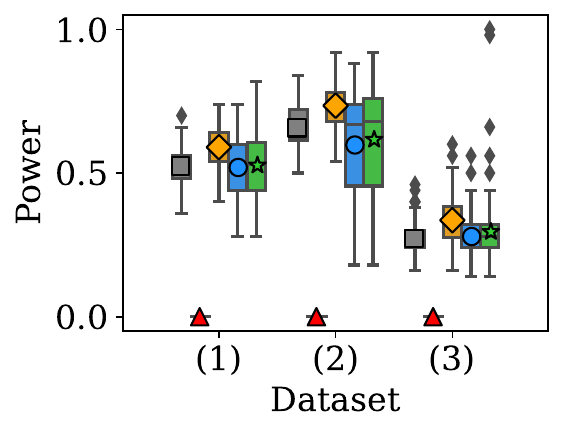}
    \includegraphics[height=0.35\linewidth, valign=t]{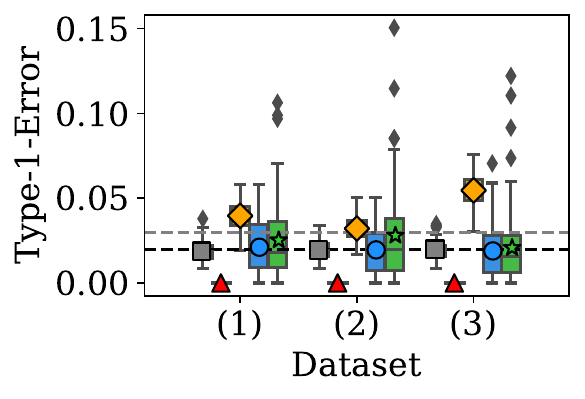}
    \end{subfigure}
    \begin{subfigure}[t]{0.45\linewidth}
    \includegraphics[height=0.35\linewidth, valign=t]{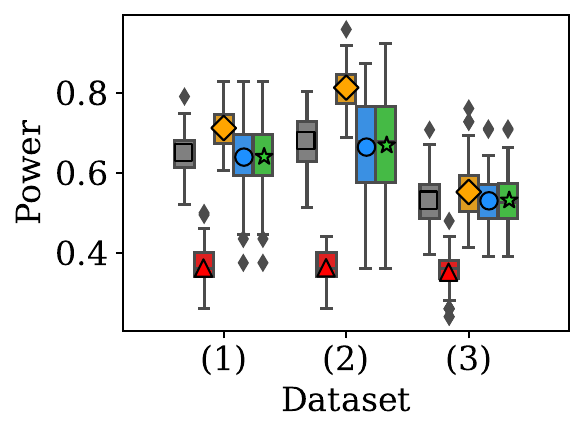}
    \includegraphics[height=0.33\linewidth, valign=t]{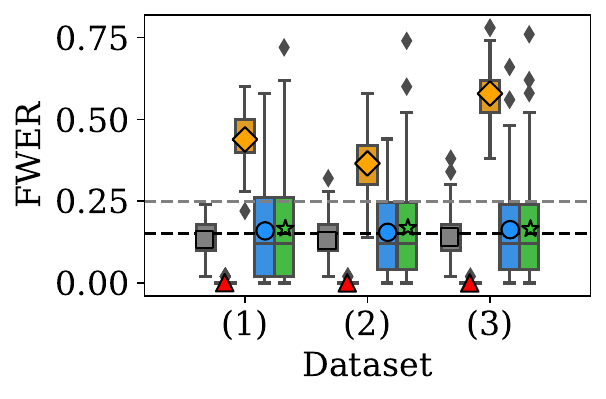}
    \end{subfigure}
    \caption{\textbf{Performance comparisons for outlier detection}.
    Evaluated on three datasets: (1) shuttle, (2) credit-card, (3) KDDCup99. Left two panels: 
    single-outlier case ($\alpha=2\%$, $\ep=1\%$). Right two panels: multiple-outlier case ($\alpha=15\%$, $\ep=10\%$). Same setup as in~\Cref{fig:od-hypothesis-testing}, 
    but with $q=5\%$ trimming proportion.
    }
    \label{app-fig:od_p_0.05}
\end{figure}

\Cref{fig:od-hypothesis-testing,app-fig:od_p_0.05} present the performance for single and multiple testing, for both $q=2.5\%$ and $5\%$, respectively. In both cases, we observe similar trends.
\texttt{OnlyReal} and \texttt{Guardrail} suffer from the small sample size: the former conservatively controls the error rate but obtains low power, whereas the latter controls the error rate at a higher level while still exhibiting high variability. The \texttt{OnlySynth} produces arbitrary error rates, while the \texttt{Oracle} method achieves error rate tightly regulated around the target level, as expected. In contrast, the error rate of \texttt{\gespi} is close to the nominal $\alpha$ level, while achieving substantially higher power than \texttt{OnlyReal}, lower variability than \texttt{Guardrail}, and performance approaching that of the \texttt{Oracle}.

\begin{figure}[!h]
    \centering
    \includegraphics[width=0.2\linewidth, valign=c]{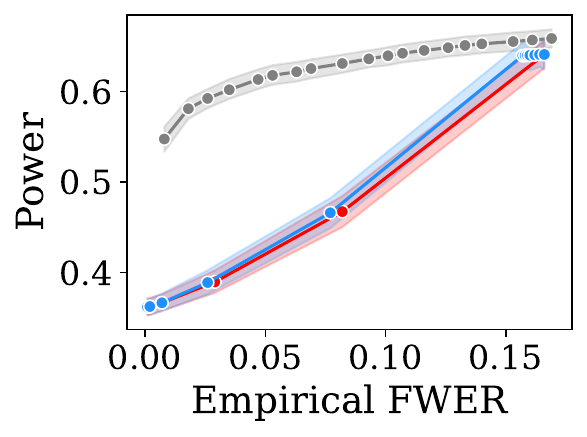}
    \includegraphics[width=0.15\linewidth, valign=c]{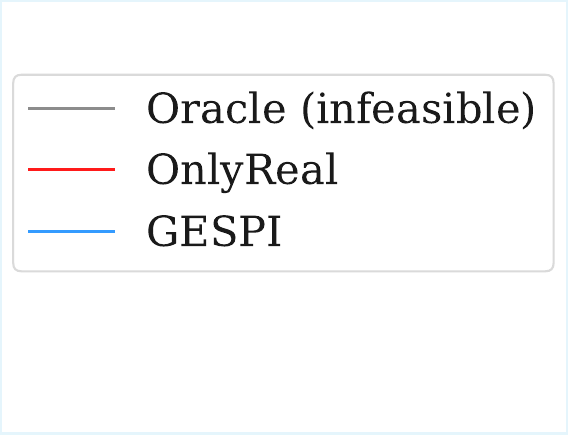}
        \caption{\textbf{Empirical power versus empirical FWER for outlier detection}. Results are shown for trimming proportion $5\%$, corresponding to a scenario with lower quality synthetic data. All other details follow in \Cref{fig:od-hypothesis-testing} for the FWER experiment.}
    \label{app-fig:od-fwer-roc}
\end{figure}

\Cref{app-fig:od-fwer-roc} presents the empirical power of multiple testing methods as a function of the empirical FWER for trimming proportion $q=5\%$, complementing the main manuscript results for $q=2.5\%$. For a trimming proportion of $5\%$, where the synthetic data are of lower quality, the ROC curve of \texttt{\gespi} closely follows that of \texttt{OnlyReal}. Notably, even in this regime, \texttt{\gespi} still exhibits lower variability in the error rate, as observed in \Cref{app-fig:od_p_0.05}.

\FloatBarrier

\subsection{Hypothesis Testing for Mechanistic Interpretability of a Vision Transformer Model}
\label{main:learn-rep-exp-details}

\paragraph{Overview.}
In this section, we consider a task related to the mechanistic interpretability of vision transformers. We aim to test whether a specific attention head exhibits a functional role---for example, detecting shapes, animals, or spatial locations. Concretely, we test whether the activation of a given head is stronger when the object it is hypothesized to detect is present in the image compared to when it is absent, which would provide evidence that the head indeed serves this role. As a case study, we examine the attention head at layer 22, head 6 (L22H6) in CLIP ViT-L/14 \citep{radford2021learning}, which has been previously reported to strongly respond to images of animals \citep{gandelsman2023interpreting}.

\paragraph{Datasets.}
We use ImageNet images as the real dataset, partitioned into \emph{animals} and\ \emph{non-animals} classes. For the synthetic dataset, we generate images using FLUX.1 \citep{flux2024}, also separated into \emph{animals} and \emph{non-animals}. Details for each dataset are as follows:
\begin{compactitem}
\item {\bf ImageNet (real)}  \citep{deng2009imagenet}: 
We use the training split, restricted to nine selected classes (listed below); all available training images from these classes are included. Grouping is inherited directly from the ImageNet taxonomy: classes depicting animals are assigned to the \emph{animals} group and the remainder to \emph{non-animals}. The exact class lists, any filtering rules, and per-class counts are provided below. 
\item {\bf FLUX.1 (synthetic)} \citep{flux2024}: For each class,  we generate $2{,}000$ images using the \texttt{FluxPipeline} from \texttt{diffusers} with mixed precision (\texttt{float16}) and $50$ inference steps. Prompts follow the CLIP-style template ``A photo of a \{class name\}'' \citep{radford2021learning}, where \{class name\} is the corresponding ImageNet label. The resulting synthetic images are labeled \emph{animals} / \emph{non-animals} based on the originating class.
\item \textbf{Class selection}: 
For the power experiment (animal vs.\ non-animal), we selected three animal categories---Labrador retriever (208), English springer spaniel (217), and kuvasz (222)---and three non-animal controls---lighter (626), tennis ball (852), and stinkhorn (994). For the Type I error experiment, both groups were drawn exclusively from non-animal categories.
We used gyromitra (993), coral fungus (991), tandem bicycle (444), tennis ball (852), stinkhorn (994), and lighter (626). The numbers in parentheses indicate the corresponding ImageNet class indices.
\end{compactitem}

\paragraph{Model and architecture.} 
We use the CLIP configuration of \citet{gandelsman2023interpreting} implemented via OpenCLIP with vision backbone \texttt{ViT-L-14} (24 transformer layers, 16 attention heads, patch size $14$) and the standard CLIP text transformer (context length $77$). 
We use the checkpoint \texttt{\textquotesingle laion2b\_s32b\_b82k\textquotesingle} of \texttt{\textquotesingle ViT-L-14\textquotesingle}, without any fine-tuning or architectural modification. Images are preprocessed using the model’s default pipeline (resize and center-crop to $224{\times}224$, CLIP normalization). 

\paragraph{Experimental procedure.} To evaluate whether attention head L22H6 detects animals, we quantify its response to images containing animals versus images without animals. Specifically, we compute a per-image ``activation score'' that summarizes how strongly each image patch (token) at this head aligns with the animal concept in CLIP’s embedding space, with larger scores indicating stronger evidence that the image contains an animal.

\begin{figure}[!ht]
    \centering
    \includegraphics[width=0.52\linewidth, valign=t]{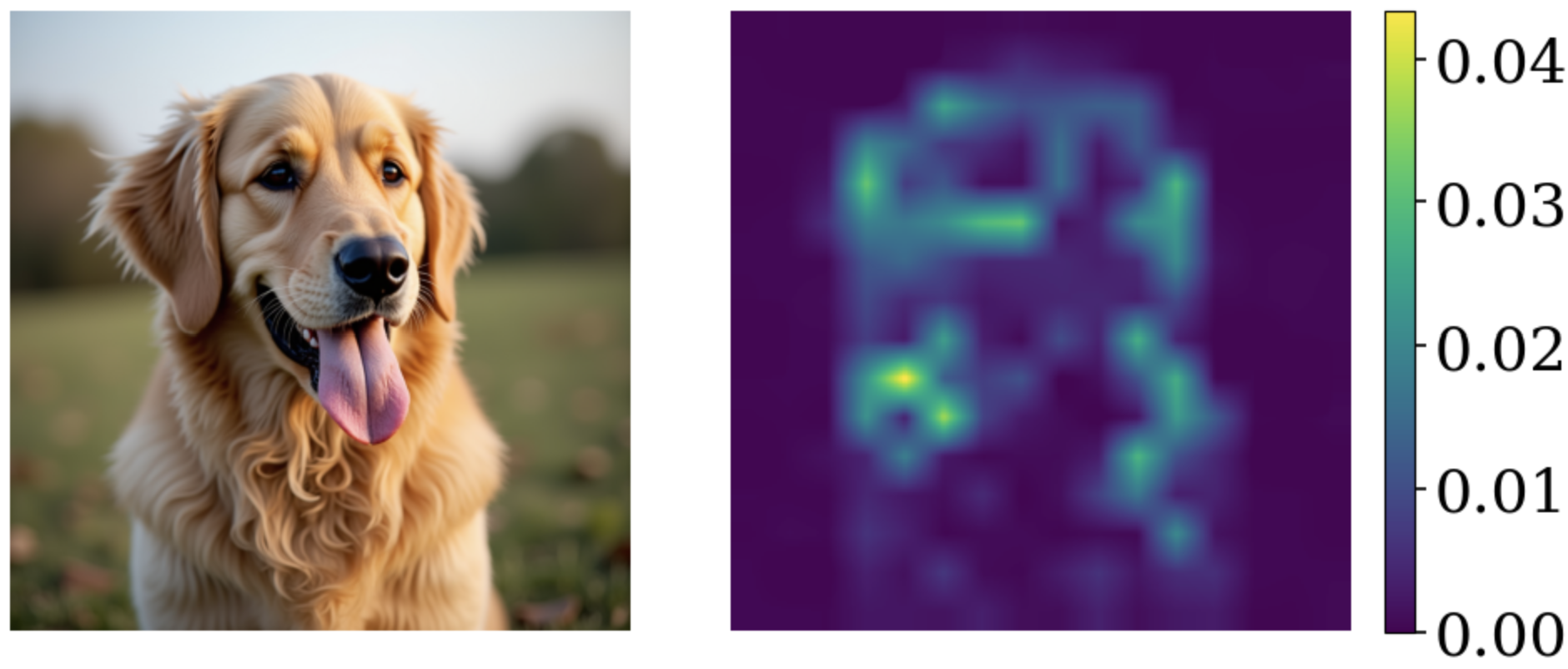}    
        \caption{Left: input image $X$ of a dog. Right: spatial heatmap $H^{22,6}(X)$ from layer 22, head 6, obtained by projecting token contributions onto $M_\text{text}(\text{``a photo of an animal''})$.}
\label{fig:clip-learned-rep-heatmap}
\end{figure}

We first encode the prompt $t^\star=\text{``a photo of an animal''}$ with the CLIP text encoder model $M_\text{text}$.
Then, for each input image $X$, we use
the heads-and-tokens decomposition of the multi-head self attention CLIP vision encoder output. Let $M^{l,h}_\text{Image}(X;i)$ denote the direct contribution of token $i$ at layer $l$ and head $h$ of the input $X$.
We define a per-head spatial heatmap by projecting each token’s contribution onto the text direction, as follows: 
$$
H^{l,h}_i(X)\;=\;\big\langle M^{l,h}_\text{Image}(X;i),\, M_\text{text}(t^\star)\big\rangle. 
$$
Intuitively, $H^{l,h}_i(X)$ measures how strongly token $i$ aligns with the animal text direction. 
See an example visualized as a spatial heatmap in Figure~\ref{fig:clip-learned-rep-heatmap}.

We summarize $H^{l,h}_i(X)$ across all tokens $i$
into a scalar ``activation'' score given by the mean absolute value over tokens:
\begin{equation}\label{app-eq:activation-score}
s^{l,h}(X)\;=\;\frac{1}{\text{\#{tokens}}}\sum_{\mathrm{tokens}\, i}^{}\big|H^{l,h}_i(X)\big|,
\end{equation}
where the number of tokens $i$ is a function of the image resolution and patch size.

\textbf{Formulation of the null hypothesis.}
We compute the activation score $s^{l,h}(X)$ \eqref{app-eq:activation-score} on two disjoint image sets, $\mathcal{D}_{\text{animals}}$ and $\mathcal{D}_{\text{non-animals}}$, and test whether head $(l,h)$ exhibits the same activation on animal and non-animal images.
Let $P^{(l,h)}_{\text{animals}}$ and $P^{(l,h)}_{\text{non-animals}}$ denote the
population distributions of activation scores of head $(l,h)$ over animals and non-animals, respectively.
We test the null hypothesis that these two distributions 
are equal:
$$
\mathcal{H}_0: P^{(l,h)}_{\text{animals}} = P^{(l,h)}_{\text{non-animals}}.
$$
As the test statistic, we use the standardized difference in mean activation scores: 
$$
\hat\Delta^{(l,h)} 
\;=\; \frac{\frac{1}{\big|\mathcal{D}_{\text{animals}}\big|}\sum_{X \in \mathcal{D}_{\text{animals}}} s^{l,h}(X)
-
\frac{1}{\big|\mathcal{D}_{\text{non-animals}}\big|}\sum_{X \in \mathcal{D}_{\text{non-animals}}} s^{l,h}(X).
}{\sqrt{\hat \sigma_{\text{animals}}^2/\big|\mathcal{D}_{\text{animals}}\big| + \hat \sigma_{\text{non-animals}}^2 / \big|\mathcal{D}_{\text{non-animals}}\big|)}}
$$

where $\hat \sigma^2_{\text{animals}}$ and $\hat \sigma^2_{\text{non-animals}}$ are the empirical variances of the activation scores of $\mathcal{D}_{\text{animals}}$ and $\mathcal{D}_{\text{non-animals}}$, respectively.

To set the critical value, we use a standard permutation approach, which is guaranteed to control the Type I error under the null hypothesis when the two distributions are equal \citep[see e.g.,][]{lehmann2005testing}. 
We use a one-sided test, assigning smaller $p$-values to larger values of $\hat\Delta^{(l,h)}$, corresponding to stronger activation on animal images relative to non-animal images.

\textbf{Base statistical test.} 
We use a permutation test with $10{,}000$ permutations per trial. For each setting (null for Type I error, alternative for power), we perform $100$ independent trials.
Within each trial, we compute the test statistic on the observed data 
and on the permuted datasets to obtain a $p$-value. Aggregating over the $100$ trials yields our estimates of Type I error and power.
We report the mean and standard deviation across trials to quantify variability.

\begin{figure}[!h]
    \centering
    \includegraphics[width=0.3\linewidth, valign=t]{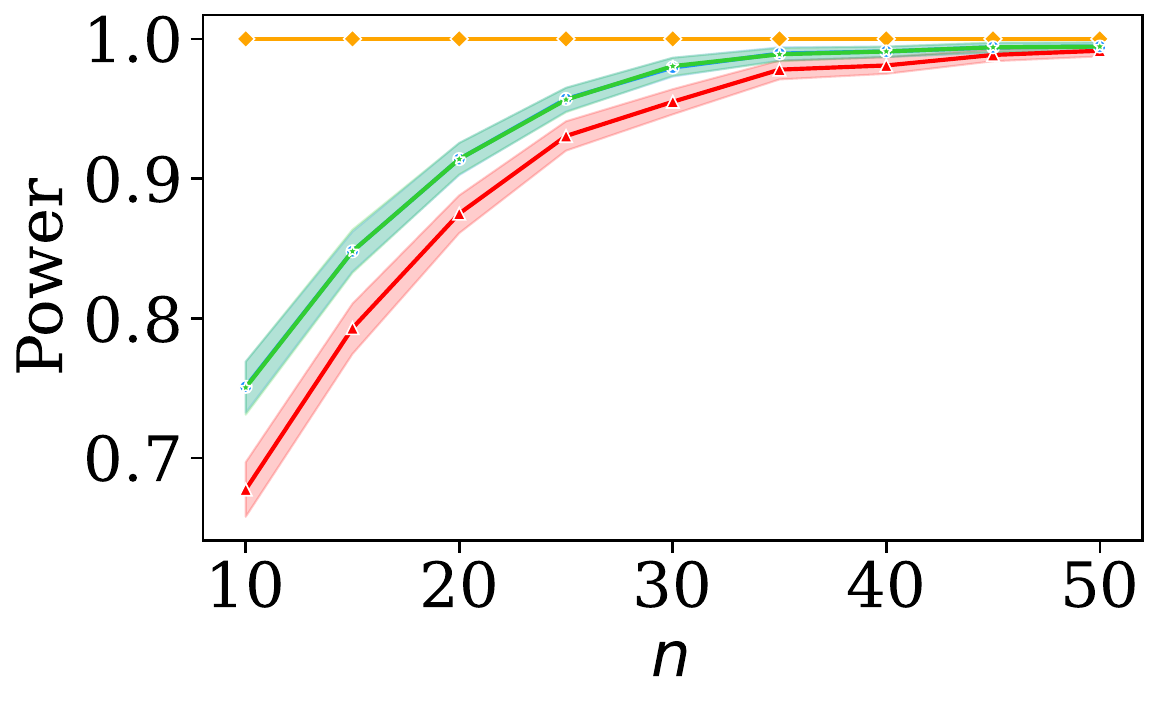}
    \includegraphics[height=0.1\linewidth, valign=t]{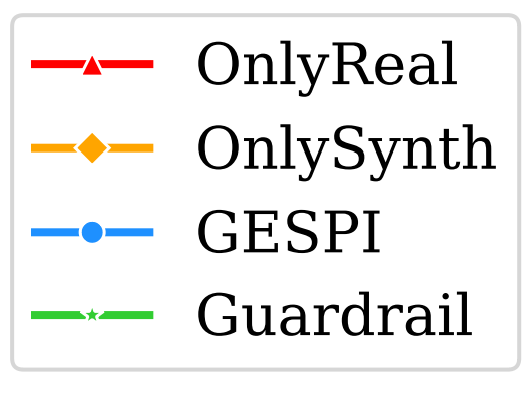}
    
    \caption{
    \textbf{Performance comparisons as a function of the real sample size $n$.}
    Hypothesis testing methods applied to animal versus non-animal groups, each containing three classes, at target level $\alpha=10\%$ and $\ep=5\%$.
    } 
\label{app-fig:clip-learned-rep-p}
\end{figure}

{\bf Results: Evaluating power.}
\Cref{app-fig:clip-learned-rep-p} presents the performance of different hypothesis testing methods as a function of the real sample size $n$, for $N=100$, $\alpha=10\%$, and $\ep= 5\%$. The rejection rate (power) is well above the target level $\alpha$, indicating that the mean activation score for images containing animals is indeed higher than for non-animals, supporting the conclusion that attention head L22H6 detects animals. 

Following that figure, we observe that \texttt{\gespi} consistently achieves higher power than the base method \texttt{OnlyReal} across all values of $n$, with power increasing as $n$ grows, illustrating the benefit of leveraging additional synthetic data. For sufficiently large $n$ ($n \geq 45$), both methods achieve power close to $1$ and obtain comparable performance. For all values of $n$, \texttt{\gespi} and \texttt{Guardrail} attain the same power. At the same time, \texttt{OnlySynth}, which relies only on synthetic images, achieves high power due to the large sample size; however, this approach does not provide valid error rate control, as the synthetic data may differ from the real data distribution.

\begin{figure}[!h]
    \centering
    \begin{subfigure}[t]{0.4\linewidth}
        \includegraphics[width=\linewidth, valign=t]{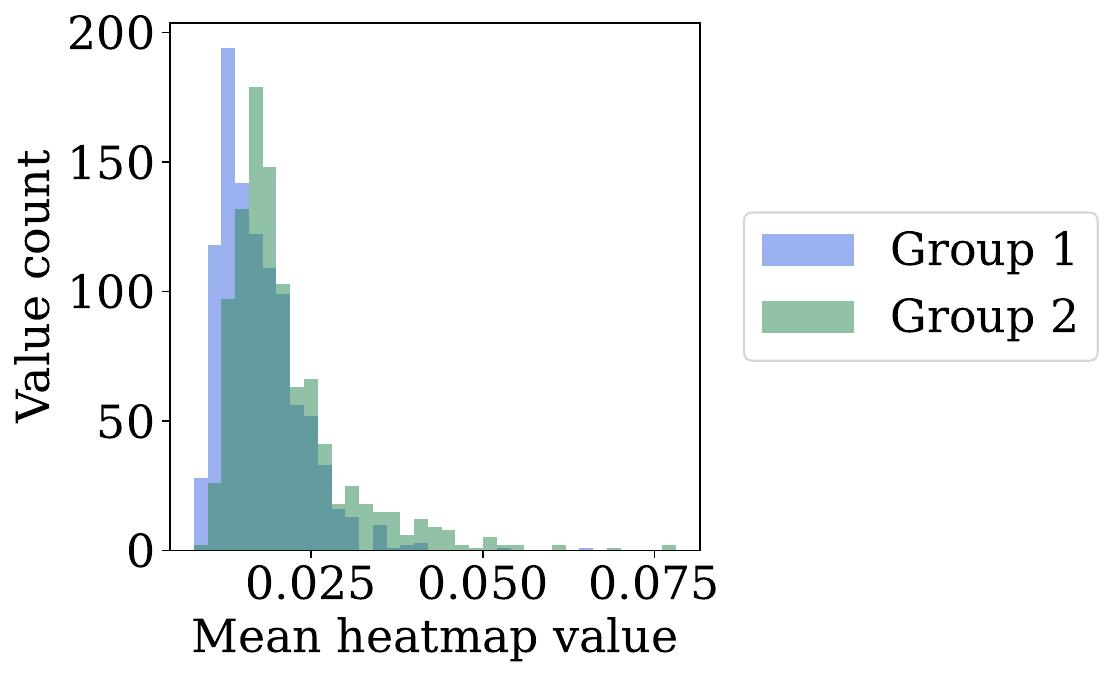}
        \caption{ImageNet (real data)}
        \label{app-fig:clip-learned-rep-hist-a}
    \end{subfigure}
    \begin{subfigure}[t]{0.4\linewidth}
        \includegraphics[width=\linewidth, valign=t]{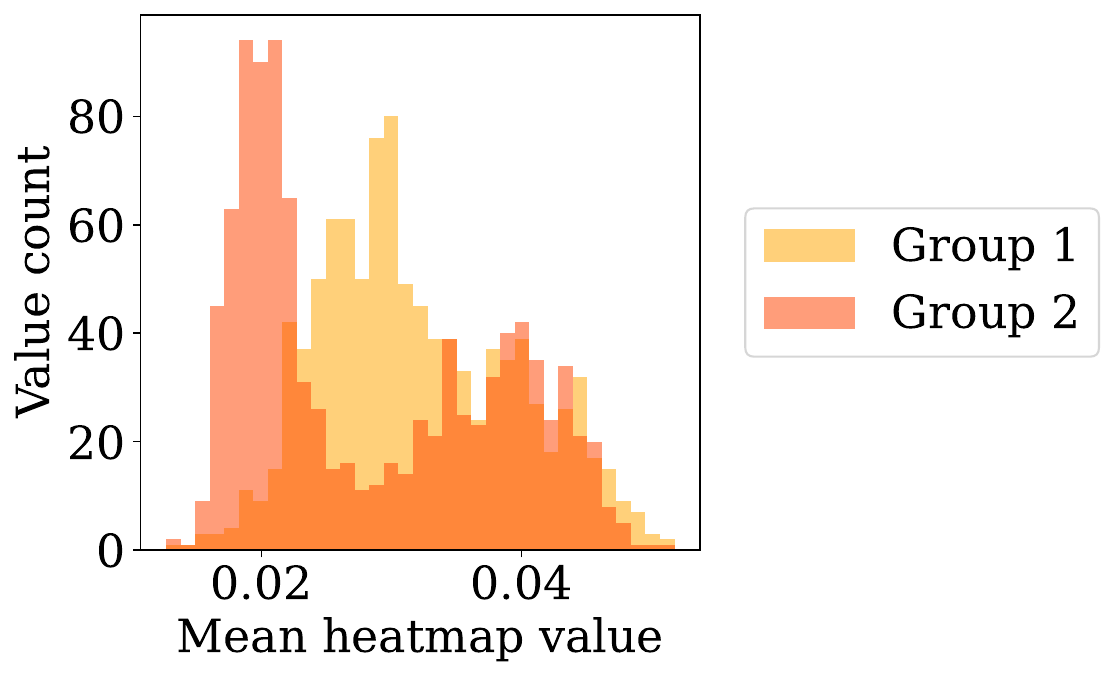}
        \caption{Flux-1 (synthetic data)}
        \label{app-fig:clip-learned-rep-hist-b}
    \end{subfigure}
    \caption{
    \textbf{Histograms of the per-image activation scores \(s^{l,h}(X)\) for two disjoin non-animal groups.}
    Each histogram is based on 1,000 samples drawn from the corresponding group; (a) shows results for real ImageNet images, and (b) for synthetic images generated with FLUX.1.
    }

    \label{app-fig:clip-learned-rep-hist}
\end{figure}

{\bf Results: Evaluating Type I error.} 
Next, we compare the performance of different hypothesis testing methods under the null hypothesis. To do so, we split the non-animal classes into two disjoint groups and repeat the experiment from~\Cref{app-fig:clip-learned-rep-p}. Under the null, the activation score distributions of the two groups are expected to be equal. Indeed, \Cref{app-fig:clip-learned-rep-hist-a} shows the histograms for the real data, where the two groups exhibit similar distributions. In contrast, \Cref{app-fig:clip-learned-rep-hist-b} shows the histograms for the synthetic data, where the two groups display noticeable differences, highlighting that the synthetic distribution does not perfectly match the real one.

\begin{figure}[!h]
    \centering
    \begin{subfigure}[t]{0.3\linewidth}
    \includegraphics[width=\linewidth, valign=t]{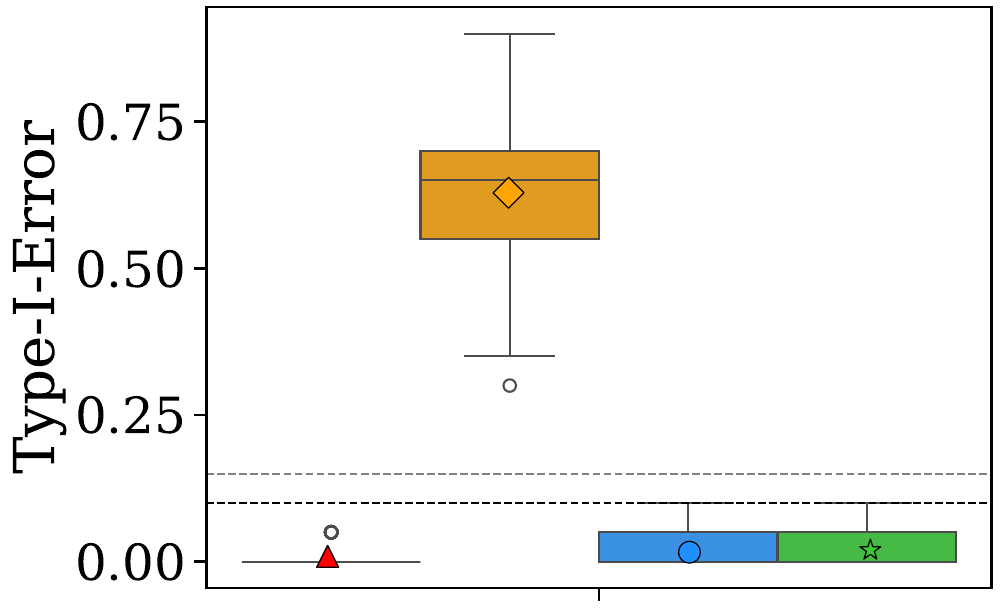}
    \caption{$n=10$}
    \end{subfigure}
    \begin{subfigure}[t]{0.3\linewidth}
    \includegraphics[width=\linewidth, valign=t]{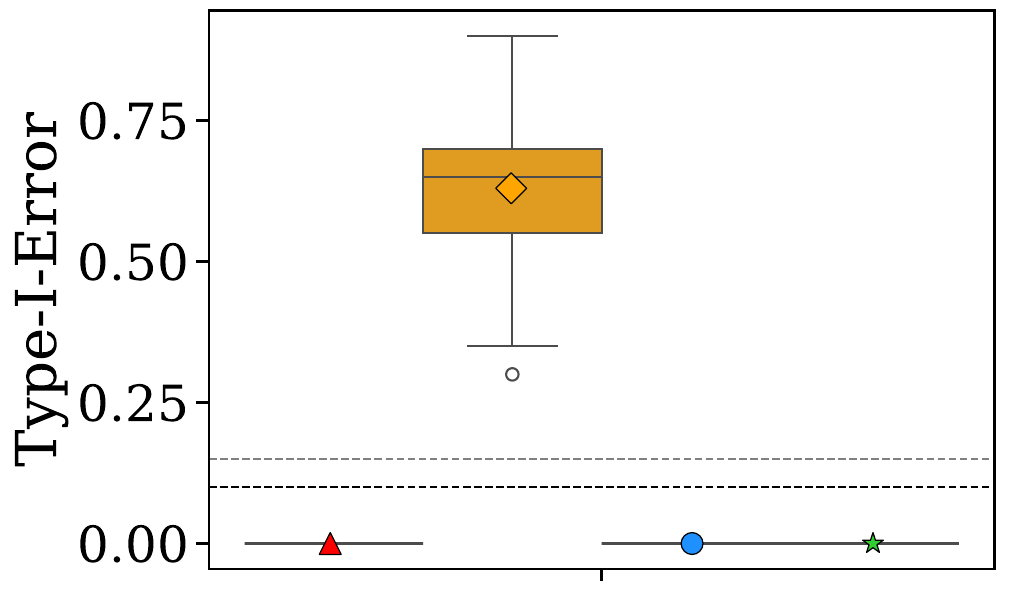}
    \caption{$n=100$}
    \end{subfigure}
    \includegraphics[height=0.15\linewidth, valign=t]{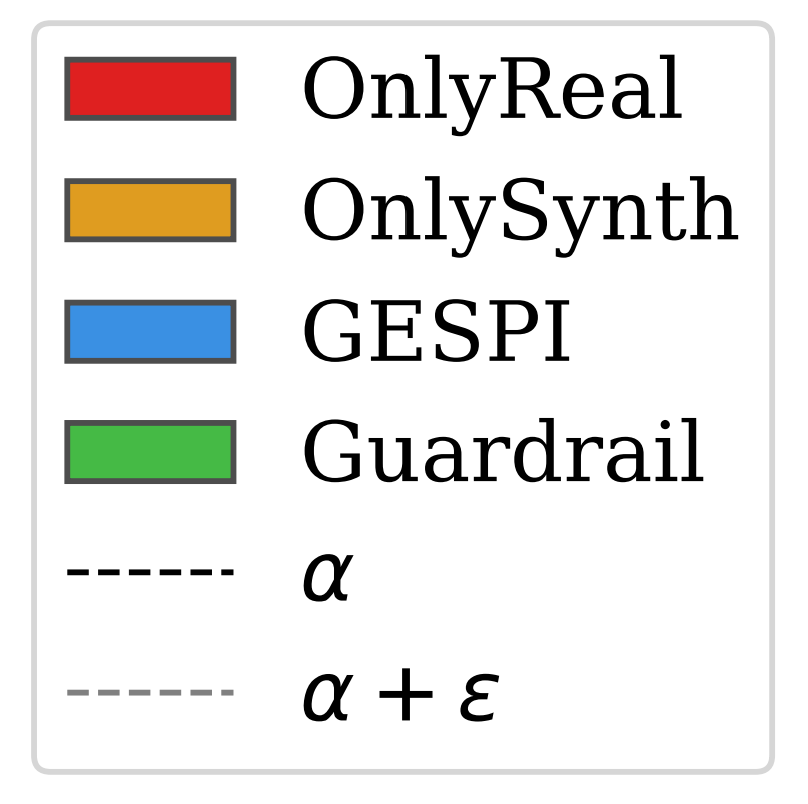}
    \caption{
    \textbf{Performance comparisons under the null hypothesis.} 
    Hypothesis testing methods applied to two disjoint groups of non-animal classes at target level $\alpha = 10\%$ and $\ep=5\%$; with (a) $n=10$ and (b) $n=100$.
    } 
\label{app-fig:clip-learned-rep-t1}
\end{figure}

\Cref{app-fig:clip-learned-rep-t1} shows the Type I error for $n=10$ and $100$; other details are as in \Cref{app-fig:clip-learned-rep-p}. \texttt{OnlyReal}, \texttt{Guardrail}, and \texttt{\gespi} all attain Type I error below the target level $\alpha$.
In contrast, \texttt{OnlySynth} obtains very high Type I error, illustrating that the synthetic data cannot be naively treated as if it were real. Since the synthetic data do not follow the null in this setting, as further illustrated in \Cref{app-fig:clip-learned-rep-hist}, \texttt{\gespi} attains a Type I error rate similar to \texttt{Guardrail}.

\section{Examples of Statistical Inference with Risk Control}\label{sec:examples}

In this section, we provide the details of how a number of classical statistical inference problems can be formulated in the form from~\eqref{eqn:target} required by our paper in order to apply the \gespi methodology. 
We aim to illustrate that the formulation~\eqref{eqn:target} covers a wide range of problems.

\begin{compactenum}
    \item \textit{Distribution-free predictive inference}\\
    Suppose we are given i.i.d. data $(X_1, Y_1), \ldots, (X_n, Y_n)$ and a new input $X_{n+1}$, with the goal of constructing a prediction set  for the unobserved outcome $Y_{n+1}$, while controlling its coverage rate. This corresponds to the following setting, where $\mathcal{B}(\Y)$ is the Borel sigma-algebra of $\Y$:
    \begin{align*}
        &\Z =\V = \X \times \Y,\,\, \A = \{g : \X \rightarrow \mathcal{B}(\Y), \textnormal{ measurable} \},\,\,\T(P) = P = P_X \times P_{Y \mid X},\\
        & V = (X_{n+1},Y_{n+1})\sim P,\\
        &\mathcal{P} =\Delta(\X \times \Y) =  \{\text{ set of all Borel probability distributions on $\X \times \Y$ }\},\\
        &\ell(g,(x,y)) = \One{y \notin g(x)}.
    \end{align*}
    
    Then the condition~\eqref{eqn:target} is equivalent to the following distribution-free marginal coverage guarantee:
    \[\Pp{(X_1,Y_1),\ldots,(X_n,Y_n),(X_{n+1},Y_{n+1}) \iidsim P}{Y_{n+1} \in \ch_n(X_{n+1})} \ge 1-\alpha,\qquad\text{ for all distributions $P$},\]
    where $\ch_n = \Alg(\Dn)$ denotes the output prediction set function from the algorithm $\Alg$.\\
    
    As a remark, if we instead set $\T(P) = Q_X \times P_{Y \mid X}$---assuming a known $Q_X$ or known likelihood ratio $dQ_X/dP_X$---then this setting corresponds to predictive inference under covariate shift.\\

    \item \textit{Hypothesis testing}\\
    Consider a hypothesis testing problem where one uses a sample $X_1, \ldots, X_n \iidsim P$ on $\X$ to test a null hypothesis of the form
    \[\text{$\mathcal{H}_0$ : $P \in \mathcal{P}_0$},\]
    for some set of (null) distributions $\mathcal{P}_0$,
    while controlling the Type I error. This problem can be viewed as a special case of the general target~\eqref{eqn:target} under the following setup:
    \begin{align*}
        &\Z = \X,\,\, \A = [0,1],\,\, \V = \{0\},\,\, \T(P) \equiv \delta_0 ,\,\, V = 0,\,\, \mathcal{P} = \mathcal{P}_0,\\
        &\ell(w, 0) = w.
    \end{align*}
    This results in the following Type I error/level control condition:
    \[\Ep{P}{\phi} \le \alpha,\qquad\text{ for all distributions $P \in \mathcal{P}_0$},\]
    where $\phi = \mathcal{A}(\Dn)$ denotes the output testing procedure from $\Alg$.\\

    \item \textit{Confidence intervals}\\
    Fix a space $\mathcal{P}$ of distributions and a parameter-functional $\theta : \mathcal{P} \to \R$---e.g., that maps a distribution to its mean or median.
    Consider constructing a confidence interval for $\theta(P)$ using the sample $X_1,\ldots,X_n \iidsim P$. 
    We let
    \begin{align*}
        &\Z = \X,\,\, \A = \mathcal{B}(\R),\,\, \V = \R,\,\, \T(P) = \delta_{\theta(P)},\,\, V = \theta(P)\\
        &\ell(I, v) = \One{v \notin I},
    \end{align*}
    which corresponds to the coverage guarantee
    \[\PP{\theta(P) \in \ch} \ge 1-\alpha, \text{ for all $P \in \mathcal{P}$ },\]
    where $\ch = \Alg(\Dn)$.\\

    \item \textit{Conformal risk control}\\
    Given sample $(X_1,Y_1),\ldots,(X_n,Y_n) \iidsim P$ on $\X \times \Y$ and $X_{n+1} \in \X$, suppose we now aim to construct a map $h : \X \rightarrow \Y'$, for some $\Y'$,
    with risk control, i.e.,
    \[\EE{\tilde{\ell}(h(X_{n+1}),Y_{n+1})} \le \alpha,\]
    for a loss function $\tilde{\ell} : \Y' \times \Y \rightarrow \R^+$, in the distribution-free sense. This corresponds to the following setup:
    \begin{align*}
        &\Z = \V = \X \times \Y,\,\, \A = \{h : \X \rightarrow \Y'\},\,\, \T(P) = P,\,\,         V = (X_{n+1},Y_{n+1})\sim P,\\
        &\mathcal{P} =\Delta(\X \times \Y) =  \{\text{ set of all Borel probability distributions on $\X \times \Y$ }\},\\
        &\ell(h,(x,y)) = \tilde{\ell}(h(x), y).
    \end{align*}
    \citet{angelopoulos2022conformal} introduces a procedure---conformal risk control---that achieves this guarantee.\\
    
    \item \textit{Simultaneous predictive inference on multiple test points}\\
    Consider the setting
    \begin{align*}
        &\Z = \X \times \Y,\,\, \A = \{g : \X \rightarrow \mathcal{B}(\Y)\},\,\, \V = \Z^m,\,\, \T(P)= P^m,\\
        &V = ((X_{n+1},Y_{n+1}),\ldots,(X_{n+m},Y_{n+m})) \iidsim P\\
        &\mathcal{P} =\Delta(\X \times \Y) =  \{\text{ set of all Borel probability distributions on $\X \times \Y$ }\},\\
        &\ell(g,(z_1,\ldots, z_m)) = \One{\frac{1}{m}\sum_{j=1}^m \One{y_j \in g(x_j)} < \delta}, \text{ where } 
        z_j=(x_j,y_j) \text{ for each } j \in [m].
    \end{align*}
    This setup leads to the following PAC-coverage guarantee
    \[\Pp{(X_1,Y_1),\ldots,(X_{n+m},Y_{n+m}) \iidsim P}{\frac{1}{m}\sum_{j=1}^m \One{Y_{n+j} \in \ch(X_{n+j})} \ge 1-\delta} \ge 1-\alpha, \text{ for all $P$},\]
    where $\ch = \Alg(\Dn)$. Methods for achieving this guarantee are discussed in~\citet{gazin2024transductive} and~\citet{lee2024batch}.

\end{compactenum}

\begin{table}[htbp]
\centering
\scalebox{0.8}{
\renewcommand{\arraystretch}{1.4}
\begin{tabular}{lcccccc}

\noalign{\hrule height 1pt}
\textbf{Example} & $\Z$ & $\A$ & $\V$ & $\T(P)$ & $\mathcal{P}$ & $\ell$ \\
\hline

{Predictive inference} & 
$\X \times \Y$ & 
$\{g: \X \to \mathcal{B}(\Y)\}$ & 
$\X \times \Y$ & 
$P$ & 
$\Delta(\X \times \Y)$ & 
$\One{y \notin g(x)}$ \\

{Hypothesis testing} & 
$\X$ & 
$[0,1]$ & 
$\{0\}$ & 
$\delta_0$ & 
$\mathcal{P}_0$ & 
$w$ \\
{Confidence intervals} & 
$\X$ & 
$\mathcal{B}(\R)$ & 
$\R$ & 
$\delta_{\theta(P)}$ & 
General $\mathcal{P}$ & 
$\One{v \notin I}$ \\
{Risk control} & 
$\X \times \Y$ & 
$\{h: \X \to \Y'\}$ & 
$\X \times \Y$ & 
$P$ & 
$\Delta(\X \times \Y)$ & 
$\tilde{\ell}(h(x), y)$ \\

{Simult. pred. inf.} & 
$\X \times \Y$ & 
$\{g: \X \to \mathcal{B}(\Y)\}$ & 
$(\X \times \Y)^m$ & 
$P^m$ & 
$\Delta(\X \times \Y)$ & 
$\mathbb{I}\{\frac{1}{m}\sum\limits_{j=1}^m \One{y_j \in g(x_j)} < \delta\}$ \\

\noalign{\hrule height 1pt}
\end{tabular}
}
\caption{
Summary of examples of our framework.}
\label{tab:examples}
\end{table}

\section{Impossibility Results}\label{sec:impossibility}

 One may consider whether it is possible to achieve a stronger target than what we aim for in the paper. 
 Namely, that for any target and any synthetic data distribution, the risk is controlled at the level $\alpha$ instead of $\alpha + \ep$. 
 Specifically, one may consider the following guarantees: 
\begin{equation}\label{eqn:target_synthetic}
\begin{split}
&\Ep{\Dn \iidsim P, \tDn \sim \bar{Q}, V \sim \T(P)}{\ell(\tAlg(\Dn, \tDn), V)} \le \alpha,\\
&\text{ for all } P \in \mathcal{P} \text{ and any distribution } \bar{Q} \text{ on } \Z^N, \text{ for all } n, N \in \N.
\end{split}
\end{equation}

However, it turns out that achieving the guarantee~\eqref{eqn:target_synthetic}---i.e., uniformly controlling the loss for all potential synthetic distributions $\bar{Q}$---is generally not a practical target.
 Indeed, as we show below, 
 if we can attain the above condition,
 then we can also do it by ignoring the synthetic data. 
\begin{proposition}\label{prop:impossibility}
    If an algorithm $\tAlg : Z^\infty \times Z^\infty \rightarrow \A$ satisfies the condition~\eqref{eqn:target_synthetic}, then it can be dominated by an algorithm that takes only the real datapoints $\Dn$ as input, in the following sense:
    For any arbitrarily small $\delta > 0$, there exists an algorithm $\Alg : Z^\infty \rightarrow \A$ such that
    \[\sup_{\substack{P\in \mathcal{P} \\ \bar{Q}}} \Ep{\substack{\Dn \iidsim P, \tDn \sim \bar{Q}\\ V \sim \T(P)}}{\ell(\tAlg(\Dn, \tDn), V)} \leq \sup_{P \in \mathcal{P}} \Ep{\Dn \iidsim P, V \sim \T(P)}{\ell(\Alg(\Dn), V)} \leq \alpha+\delta.\]
\end{proposition}
See Section \ref{pfi} for the proof. 
Proposition~\ref{prop:impossibility} states that for any procedure leveraging synthetic data, there exists an algorithm that uses only the real data and achieves the target level with equal or better tightness---potentially leading to higher power, narrower confidence intervals or prediction sets, etc. In other words, if the goal is to uniformly control the loss over all theoretically possible synthetic distributions, then ignoring the synthetic data may be the optimal choice.\footnote{Note, however, that the dominating procedure $\Alg$ for a specific $\tAlg$ in Proposition~\ref{prop:impossibility} may not be explicitly known in certain scenarios. Nevertheless, it is likely that well-known methods in various examples correspond to $\Alg$.}
However, in practical scenarios where some similarity between the true and synthetic distributions can reasonably be expected, ignoring the synthetic data based on the strong condition~\eqref{eqn:target_synthetic} would not be the preferred choice.

\section{Additional Theoretical Results for GESPI}\label{app-sec:general-theory}

We first provide a formal statement of the conditions required by our algorithm. 

\begin{condition}\label{con:target_order_formal}
The following conditions hold.
\begin{compactenum}
    \item The action space $\A$ is endowed with a partial order denoted by $\succeq$. For any pair of elements $a_1,a_2 \in \A$, there uniquely exists an element $a_1 \vee a_2$ in $\A$ such that the following hold:
\[\text{(i) } a_1 \vee a_2 \succeq a_1 \text{ and } a_1 \vee a_2 \succeq a_2.\;\; \text{ (ii) }\text{ If } a \succeq a_1\text{ and } a \succeq a_2, \text{ then } a \succeq a_1 \vee a_2.\]
    Similarly, there uniquely exists an element $a_1 \wedge a_2$ in $\A$ such that
\[\text{(i) } a_1 \wedge a_2 \preceq a_1 \text{ and } a_1 \wedge a_2 \preceq a_2.\;\; \text{ (ii) } \text{ If } a \preceq a_1\text{ and } a \preceq a_2, \text{ then } a \preceq a_1 \wedge a_2.\]

    \item For any $a_1,a_2,a_3 \in \A$, $(a_1 \wedge a_2) \vee a_3 = (a_1 \vee a_3) \wedge (a_2 \vee a_3)$.

    \item For any $\alpha_1 \leq \alpha_2$ and $z \in \Z^\infty$, $\Alg_{\alpha_1}(z) \preceq \Alg_{\alpha_2}(z)$.

    \item The loss function $\ell : \A \times \V \rightarrow \R^+$ satisfies the following.
    \begin{compactenum}
        \item \textit{boundedness}: There exists $c > 0$ such that $\ell(a,v) \le c$ for all $a \in \A$ and $v \in \V$.
        \item \textit{monotonicity}: $\text{ If } a_1 \preceq a_2, \text{ then } \ell(a_1,v) \le \ell(a_2 ,v) \text{ for any } v \in \V.$
    \end{compactenum}
\end{compactenum}
\end{condition}

The first two statements in Condition~\ref{con:target_order_formal} states that there exists a partial order on the action space with ``reasonable properties''---namely, that the minimum and maximum of any two elements are well-defined and that the distributive law holds.

 Table~\ref{table:order_examples}
 illustrates what the partial order means in our examples.
 The monotonicity of the algorithm and loss can be checked on a case-by-case basis. For instance, conformal prediction sets are nested by construction \citep{vovk2005algorithmic}.
 The requirements in
 Condition~\ref{con:target_order_formal}
 hold in a variety of statistical inference problems, including the examples in Section~\ref{sec:gespi_appl}. See also Table~\ref{table:order_examples} for simple examples. 

 \begin{table}[htbp]
 \centering
 \scalebox{0.8}{
 \renewcommand{\arraystretch}{1.4}
 \begin{tabular}{lcccc}
\noalign{\hrule height 1pt}
 \textbf{Example} & $\A$ & $a_1 \preceq a_2$ & $a_1 \vee a_2$ & $a_1 \wedge a_2$ \\
 \hline

 {Hypothesis testing} & $\{0,1\}$ & $a_1 \leq a_2$ &
 $a_1$ OR $a_2$ & $a_1$ AND $a_2$ \\

 {Confidence intervals} & $\mathcal{B}(\X)$ & $a_1 \supseteq a_2$ & 
 $a_1 \cap a_2$ & $a_1 \cup a_2$ \\

 {Predictive inference} & $\{g : \X \rightarrow \mathcal{B}(\Y), \textnormal{ measurable} \}$ & $a_1(x) \supseteq a_2(x)\; \forall x$ &
 $x \mapsto a_1(x) \cap a_2(x)$ & $x \mapsto a_1(x) \cup a_2(x)$ \\

\noalign{\hrule height 1pt}

 \end{tabular}
 }
 \caption{\footnotesize Examples of partially ordered action spaces for different inference problems. Here, $\mathcal{B}(\X)$ denotes the Borel sigma-algebra of $\X$.}
 \label{table:order_examples}
 \end{table}

We now present an extended theoretical result for the \texttt{\gespi} procedure with two-sided guardrail~\eqref{eqn:alg_spi_2}. Let us consider scenarios in which the algorithm $\Alg_{\alpha}$ additionally satisfies a tightness guarantee:

\begin{equation}\label{eqn:target_2}
\alpha - \delta_n \leq \Ep{\Dn \iidsim P, V \sim \T(P)}{\ell(\Alg_{\alpha}(\Dn),V)} \le \alpha, \text{ for all } P \in \mathcal{P} \text{ and } n \in \N.
\end{equation}
The term $\delta_n$ denotes the tightness level which may depend on the sample size $n$. Examples include:
\begin{compactenum}
    \item The confidence interval $\bar{X} \pm z_{\alpha/2} \frac{\sigma}{\sqrt{n}}$ for the mean of a normal distribution satisfies~\eqref{eqn:target_2} with $\delta_n = 0$.
    \item The split conformal prediction set attains a tightness level of $\delta_n = \frac{1}{n+1}$~\citep{vovk2005algorithmic}.
    \item Conformal risk control~\citep{angelopoulos2022conformal} attains a tightness level of $\delta_n = \frac{2B}{n+1}$, where $B$ denotes an upper bound on the loss.
\end{compactenum}

The following theorem extends the results in Theorem~\ref{thm:main_2}

\begin{theorem}\label{thm:main_2_extended}
Suppose that Condition~\ref{con:target_order_formal} holds. Then given $\alpha,\ep >0$, the algorithm $\tAlg$ defined in~\eqref{eqn:alg_spi_2} satisfies
\[\Alg_{\alpha}(\Dn) \preceq \tAlg(\Dn, \tDn) \preceq \Alg_{\alpha+\ep}(\Dn), \]
deterministically. If the algorithm $\Alg_{\alpha}$ satisfies~\eqref{eqn:target_2} for $\alpha$ and $\alpha+\ep$, then
   \[\alpha-\min\{\delta_n,c\tau+\delta_{N+n}+c\cdot\tdell(P,Q)\} \le \EE{\ell(\tAlg(\Dn, \tDn), V)} 
   \le \alpha + \min\{
   \ep,c\tau+c\cdot\tdell(P,Q)\}\]
   for all $P,Q \in \mathcal{P}$, where $\tdell(P,Q) =\dtv(\tilde{P}_{\ell,\Alg}(P,Q), \tilde{P}_{\ell,\Alg}(Q,Q))$, and $\tilde{P}_{\ell,\Alg}(P,Q)$ denotes the distribution of $\ell(\tAlg(\Dn, \tDn),V)$ under $\Dn \iidsim P,\tDn \iidsim Q$ and $\V \sim \T(P)$. The term $\tau$ is defined as
   \[\tau := 1-\Pp{\Dn \cup \tDn \iidsim Q}{\Alg_{\alpha}(\Dn) \preceq \Alg_{\alpha}(\Dn \cup \tDn) \preceq \Alg_{\alpha+\ep}(\Dn)}.\]
\end{theorem}

The first claim of Theorem~\ref{thm:main_2_extended} simply restates the result of Theorem~\ref{thm:main_2}. 
The key strength of this theorem lies in its generality. 
When it is particularized to specific examples,
 and sharpened by leveraging additional structure---such as exchangeability in conformal prediction---it can lead to more interpretable and tighter guarantees.  

The general bounds in the theorem imply that the risk lies between $\alpha - c\tau - \delta_{N+n}$ and $\alpha + c\tau$ when $P = Q$, i.e., under ``perfect” synthetic datapoints. The term $c\tau$, which appears in addition to the total variation distance, represents the potential bias introduced by the guardrail procedures. In other words, there is a tradeoff in using two-sided guardrails---between the bias in risk introduced by the guardrails in the ideal scenario of $P=Q$, and the range of worst-case scenarios controlled by them.\\

\begin{remark}
    The lower-guardrail procedure $\Alg_{\alpha}(\Dn)$ can be replaced with other choices, e.g., $\Alg_{\alpha-\ep'}(\Dn)$ for some $\ep' > 0$. However, we consider $\Alg_{\alpha}(\Dn)$ as the standard choice, since in most relevant settings, synthetic data is desired because the original procedure $\Alg_{\alpha}(\Dn)$ is not satisfactory.
    Thus, one typically wants the new procedure to have at least the power or at most the width of the standard method, respectively, in the worst case.\\
\end{remark}

\begin{remark}
    The first claim of Theorem~\ref{thm:main_2_extended} holds generally for any procedure of the form
    \[\tAlg(\Dn, \tDn) = (\Alg'(\Dn ; \tDn) \wedge \Alg_{\alpha+\ep}(\Dn)) \vee \Alg_{\alpha}(\Dn),\]
    where $\Alg'$ is any algorithm that takes the real and synthetic data as inputs. In other words, the guardrail bounds still hold regardless of how we use the real and synthetic data---e.g., data-dependent choices, even in the case of double dipping---and regardless of what joint distribution $\tDn$ follows and what guarantee $\Alg'$ satisfies, etc.
\end{remark}

\section{Extension: Multiple Testing}
    Consider a problem in which one is given a sample $X_1, \ldots, X_n \iidsim P$ on $\X$, and the goal is to test the following $m\ge 1$ hypotheses:
    \[\mathcal{H}_{0,j} : P \in \mathcal{P}_j,\qquad j \in [m],\]
    where $\mathcal{P}_1, \ldots, \mathcal{P}_m \subset \mathcal{P}_\text{all}$ are sets of of distributions belonging to a prespecified set $\mathcal{P}_\text{all}$ of distributions of interest.
    Consider controlling the family-wise error rate (FWER), or more generally, the $k$-FWER \citep{lehmann2005generalizations}:
    \[\PP{\sum_{j=1}^m \One{\text{$\mathcal{H}_{0,j}$ is true but rejected}} \geq k } \le \alpha,\]
    for some predetermined $k > 0$. 
    This is equivalent to requiring that condition~\eqref{eqn:target} holds under the following setup, for all $J \subset [m]$ (except for the case $J = \emptyset$):
    \begin{align*}
        &\Z = \X,\,\, \A = \{0,1\}^m,\,\, \V = \{0\},\,\, \T(P) \equiv \delta_0,\,\, V=0,\\
        &\mathcal{P} = \mathcal{P}_J := \bigcap_{j \in J} \mathcal{P}_j \cap \bigcap_{j \notin J} (\mathcal{P}_\all\backslash \mathcal{P}_j),\\
        &\ell_J((a_1,\ldots,a_m),0) = \One{\sum_{j \in J} a_j \geq k }.
    \end{align*}
    Specifically, this corresponds to the following problem of simultaneously controlling multiple risks:
    \begin{equation*}
    \mathcal{R}_\gamma(\Alg,P) = \Ep{\Dn \iidsim P, V \sim \T(P)}{\ell_\gamma(\Alg(\Dn),V)} \le \alpha, \text{ for all } P \in \mathcal{P}_\gamma \text{ and } n \in \N,\;\forall \gamma \in \Gamma
    \end{equation*}
    for a set $\Gamma$ of indices $\gamma$, each corresponding to a different distribution space $\mathcal{P}_\gamma$ and loss function $\ell_\gamma$.

   For the above multiple testing problem with FWER control, Condition~\ref{con:target_order_formal} is satisfied under the following ordering, and with all $\ell_\gamma$s:
\[(a_1,a_2,\ldots,a_m) \preceq (b_1,b_2,\ldots,b_m) \iff a_j \leq b_j \;\; \forall j \in [m].\]
Therefore, for the synthetic-powered procedure defined in~\eqref{eqn:alg_spi_2}, the first claim of Theorem~\ref{thm:main_2_extended} holds, showing that the guardrails also function properly in this setting.

\section{Applications of \gespi}\label{app-sec:applications}

In this section, we provide more detailed discussions of different applications of \gespi, along with more tailored theoretical results. For simplicity, we restrict our discussion to the \gespi procedure with a one-sided guardrail. Throughout this section, we use the notations $\mathcal{B}(\X)$ and $\Delta(\X)$, whose definitions follow those in Section~\ref{sec:examples}.

\subsection{Conformal Prediction}\label{sec:ex_conformal}
Consider a predictive inference problem where we have samples 
\begin{align*}
    &\Dn = ((X_1,Y_1),\ldots,(X_n,Y_n)) \iidsim P = P_X \times P_{Y \mid X},\\
    &\tDn = ((\tilde{X}_1,\tilde{Y}_1),\ldots,(\tilde{X}_N,\tilde{Y}_N)) \iidsim Q = Q_X \times Q_{Y \mid X},
\end{align*}
and the goal is, given a new input $X_{n+1} \sim P_X$, to construct a prediction set for $Y_{n+1}$, where $Y_{n+1} \mid X_{n+1} \sim P_{Y \mid X}$. For this problem, split conformal prediction~\citep{vovk2005algorithmic, papadopoulos2002inductive} provides the following algorithm:
\begin{align*}
    &\Alg_{\alpha} : (\X \times \Y)^\infty \rightarrow \A = \{g : \X \rightarrow \mathcal{B}(\Y), \textnormal{ measurable} \},\\ &\Alg_{\alpha}((x_i,y_i)_{i \in [n]}) = \left( x \mapsto \left\{y \in \Y : s(x,y) \le \hat{q}_\alpha((x_i,y_i)_{i \in [n]})\right\} \right), \text{ where } \\
    &\hat{q}_\alpha((x_1,y_1), \ldots, (x_n,y_n)) = (\,\textnormal{the $\lceil(1-\alpha)(n+1)\rceil$-th smallest element among $s(x_1,y_1),\ldots,s(x_n,y_n)$}\,).
\end{align*}
Here, $s : \X \times \Y \rightarrow \R$ is a prespecified nonconformity score function. This procedure attains the following marginal coverage guarantee:
\begin{align*}
    &\Ep{\Dn \iidsim P, (X_{n+1},Y_{n+1}) \sim P}{\ell(\Alg_{\alpha}(\Dn), (X_{n+1},Y_{n+1}))} = \PP{Y_{n+1} \notin \ch_n(X_{n+1})} \le \alpha, \text{ for all $P \in \mathcal{P}$},\\
    &\text{ where }\ell(g,(x,y)) = \One{y \notin g(x)}, \ch_n = \Alg(\Dn), \text{ and } \mathcal{P} =\Delta(\X \times \Y).
\end{align*}
Now the \gespi procedure for conformal prediction is constructed as
\begin{equation}\label{eqn:synthetic_conformal}
\begin{split}
& \tAlg(\Dn,\tDn) = \Alg_{\alpha}(\Dn \cup \tDn) \cup \Alg_{\alpha+\ep}(\Dn) = \tilde{C}_{n,N}, \text{ where }\\
&\tilde{C}_{n,N}(x) = \left\{y \in \Y : s(x,y) \le \max\{\hat{q}_{\alpha+\ep}(\Dn), \hat{q}_{\alpha}(\Dn \cup \tDn)\}\right\}.
\end{split}
\end{equation}
This extends the method of~\citet{bashari2025synthetic} by using the real samples $\Dn$ not only as a guardrail, but also as part of the main synthetic-boosted procedure. Suppose we predefine another level $\delta \in (0,1)$ and then set $\ep$ as $\ep = 1 - \alpha - r_\delta/(n+1)$, where
    \[r_\delta = \min\left\{r : \sum_{k=1}^{\lceil(1-\alpha)(N+n+1)\rceil} \frac{\binom{k-1}{r-1}\cdot\binom{N+n-k}{n-r}}{\binom{N+n}{n}} \ge 1-\delta\right\}.\]
Then we have the following guarantee.

\begin{theorem}\label{thm:conformal}
    Let $P_S$ and $Q_S$ denote the distributions of the score $s(Z)$ when $Z \sim P$ and $Z \sim Q$, respectively, and suppose that both $P_S$ and $Q_S$ are continuous. Then the procedure $\tilde{C}_{n,N} =  \tAlg(\Dn,\tDn)$ in~\eqref{eqn:synthetic_conformal} satisfies
    \[\PP{Y_{n+1} \in \tilde{C}_{n,N}(X_{n+1})} \ge 1-\alpha-\min\{\ep,d_{P,Q}\},\]
    where $d_{P,Q} = \frac{1}{n+1}\sum_{r=1}^{n+1} \dtv(P_{(r)}^{n+1}, Q_{(r)}^{n+1})$, and $P_{(r)}^{n+1}$ and $Q_{(r)}^{n+1}$ denote the distributions of the $r$-th order statistic from $n+1$ i.i.d. draws from $P_S$ and $Q_S$, respectively.
    
    Furthermore, if the scores are all distinct almost surely, then
    \[\PP{Y_{n+1} \in \tilde{C}_{n,N}(X_{n+1})} \le 1-\alpha+\frac{1}{N+n+1}+d_{P,Q}+\dtv(P_{(t_{\alpha+\ep})}^n, Q_{(t_{\alpha+\ep})}^n) + \delta,\]
    where $t_{\alpha+\ep} = \lceil(1-\alpha-\ep)(n+1)\rceil$. 
\end{theorem}

\subsection{Conformal Risk Control}
Consider the setting in the previous section~\ref{sec:ex_conformal}, and suppose now that we instead aim for risk control, i.e., we construct a map $h : \X \rightarrow \Y'$ with the guarantee
\[\Ep{(X_1,Y_1),\ldots,(X_n,Y_n),(X_{n+1},Y_{n+1}) \iidsim P}{\tilde{\ell}(h(X_{n+1}),Y_{n+1})} \le \alpha,\qquad \text{ for all distributions $P$}, \]
for some loss function $\tilde{\ell} : \Y' \times \Y \rightarrow [0,B]$. For this problem, \citet{angelopoulos2022conformal} introduces the following algorithm:
\begin{align*}
    &\Alg_{\alpha} : (\X \times \Y)^\infty \rightarrow \mathcal{H}_\Lambda = \{h_\lambda : \lambda \in \Lambda \} \subset \{g : \X \rightarrow \mathcal{B}(\Y), \textnormal{ measurable} \},\\
    &\Alg_{\alpha}((x_i,y_i)_{i \in [n]}) = h_{\hat{\lambda}_\alpha((x_i,y_i)_{i \in [n]})}, 
\end{align*}
where 
\begin{align*}
    &\hat{\lambda}_\alpha((x_i,y_i)_{i \in [n]}) = \inf\left\{\lambda \in \Lambda : \frac{\sum_{i=1}^n \tilde{\ell}(h_\lambda(X_i),Y_i) + B}{n+1} \le \alpha\right\}.
\end{align*}
Here, $\mathcal{H}_\Lambda$ is a set of measurable functions parameterized by $\lambda \in \Lambda \subset \R$, and the loss $\tilde{\ell}$ and the set $\mathcal{H}_\Lambda$ are chosen so that the function $\lambda \mapsto \tilde{\ell}(h_\lambda(x), y)$ is monotone increasing for any $(x, y) \in \X \times \Y$.

Note that for the set $\mathcal{H}_\Lambda$, the minimum operation $\wedge$ satisfies 
\[\tilde{\ell}(h_{\lambda_1 \wedge \lambda_2}(x),y) \le \min\{\tilde{\ell}(h_{\lambda_1}(x),y),\tilde{\ell}(h_{\lambda_2}(x),y)\}, \text{ for any $\lambda_1, \lambda_2 \in \Lambda$ and $x \in X, y \in \Y$ }.\]

Therefore, the \gespi risk control procedure is given as
\[\tAlg(\Dn,\tDn) = h_{\tilde{\lambda}}, \text{ where } \tilde{\lambda} = \hat{\lambda}_\alpha(\Dn \cup \tDn) \wedge \hat{\lambda}_{\alpha+\ep}(\Dn).\]

\subsection{Test for the Median}
Consider a testing problem where we use real-valued samples $\Dn = (X_1,\ldots,X_n) \iidsim P$ to test
\[\mathcal{H}_0 : Q_{1/2}(P) \le 0.\]
We consider a procedure $\Alg_{\alpha} : \R^n \rightarrow \{0,1\}$, defined as
\begin{equation}\label{eqn:alg_test_median}
    \Alg_{\alpha}((x_1,\ldots,x_n)) = \One{\sum_{i=1}^n \One{x_i > 0} > \hat{k}_{n,\alpha}}, \text{ where } \hat{k}_{n,\alpha} = Q_{1-\alpha}\left(\text{Binom}\left(n,\frac{1}{2}\right)\right).
\end{equation}

\begin{proposition}\label{prop:median}
    The algorithm $\Alg_{\alpha}$ in~\eqref{eqn:alg_test_median} satisfies
      \[\Ep{\Dn \iidsim P}{\Alg_{\alpha}(\Dn)} \le \alpha,\; \text{ for all } P \in \mathcal{P}_{\mathcal{H}_0} = \{ \textnormal{ all distributions on $\R$ with non-positive median }\}.\]
\end{proposition}

Now, given a synthetic data $\tDn = (\tilde{X}_1,\ldots,\tilde{X}_N)$, the \gespi procedure is given as
\begin{equation}\label{eqn:test_median}
    \tAlg(\Dn,\tDn) = \One{\sum_{i=1}^n \One{X_i > 0} > \hat{k}_{n,\alpha+\ep} \text{ and } \sum_{i=1}^n \One{X_i > 0} + \sum_{j=1}^N \One{\tilde{X}_j > 0} > \hat{k}_{N+n,\alpha}},
\end{equation}

and the following results hold.
\begin{theorem}\label{thm:median}
    Let $p = \Pp{X \sim P}{X > 0}$ and $q =  \Pp{X \sim Q}{X > 0}$. Then the testing procedure $\smash{\tAlg(\Dn,\tDn)}$ in~\eqref{eqn:test_median} satisfies
    \[\Ep{\Dn \iidsim P,\tDn \iidsim Q}{\tAlg(\Dn,\tDn)} \le \alpha + \min\{\ep, d_{P,Q}\}, \text{ where } d_{P,Q} = \sqrt{\frac{n}{2q(1-q)}}\cdot|p-q|,\]
    for all $P \in \mathcal{P}_{\mathcal{H}_0}^\ep = \left\{\text{ distributions $\bar{P}$ on $\R$} : \Pp{X \sim \bar{P}}{X > 0} \le \frac{1}{2}-\sqrt{\frac{1}{2n}}\ep\right\}$ and for any distribution $Q$.
\end{theorem}

\section{Proof of Theorems}\label{app-sec:proofs}

\subsection{Proof of Proposition~\ref{prop:impossibility}}
\label{pfi}
First observe that
\begin{align*}
    &\sup_{\substack{P\in \mathcal{P} \\ \bar{Q}}} \Ep{\Dn \iidsim P, \tDn \sim \bar{Q}, V \sim \T(P)}{\ell(\tAlg(\Dn, \tDn), V)}\\
    &= \sup_{\substack{P\in \mathcal{P} \\ \bar{Q}}} \Ep{\tDn \sim \bar{Q}}{\Epst{\Dn \iidsim P, V \sim \T(P)}{\ell(\tAlg(\Dn, \tDn), V)}{\tilde{D}_N}}\\
    &= \sup_{\bar{Q}} \Ep{\tDn \sim \bar{Q}}{\sup_{P\in \mathcal{P}}\Epst{\Dn \iidsim P, V \sim \T(P)}{\ell(\tAlg(\Dn, \tDn), V)}{\tilde{D}_N}} \leq \sup_{d_N \in \Z^N} h(d_N),
\end{align*}
where $h(d_N) = \sup_{P \in \mathcal{P}} \Ep{\Dn \iidsim P, V \sim \T(P)}{\ell(\tAlg(\Dn, d_N), V)}$. 
Next, for any $\delta>0$,
there exists $d_N^\delta$ such that $h(d_N^\delta) \ge \sup_{d_N \in \Z^N} h(d_N) - \delta$. Since condition~\eqref{eqn:target_synthetic} holds for any distribution $\bar{Q}$ on $\Z^N$, it also holds for $\bar{Q} = \delta_{d_N^\delta}$ (the point mass on $d_N^\delta$), which implies $h(d_N^\delta) \le \alpha$. Therefore, defining $\Alg(\Dn) = \tAlg(\Dn,d_N^\delta)$, we have that
\begin{multline*}
    \sup_{\substack{P\in \mathcal{P} \\ \bar{Q}}} \Ep{\Dn \iidsim P, \tDn \sim \bar{Q}, V \sim \T(P)}{\ell(\tAlg(\Dn, \tDn), V)} \leq h(d_N^\delta) + \delta\\
    = \sup_{P \in \mathcal{P}} \Ep{\Dn \iidsim P, V \sim \T(P)}{\ell(\Alg(\Dn), V)} +\delta \leq \alpha + \delta.
\end{multline*}
 This finishes the proof. 

\subsection{Proof of Theorem~\ref{thm:main}}

First observe that since 
\[\ell(a_1 \wedge a_2, v) \le \min\{\ell(a_1,v), \ell(a_2,v)\}, \text{ for any } a_1,a_2 \in \A \text{ and } v \in \V\]
holds by Condition~\ref{con:target_order_formal}, we have 
\begin{multline*}
    \Ep{\Dn \iidsim P, \tDn \iidsim Q, V \sim \T(P)}{\ell(\tAlg(\Dn, \tDn), V)} 
    \le \Ep{\Dn \iidsim P, \tDn \iidsim Q, V \sim \T(P)}{\ell(\Alg_{\alpha+\ep}(\Dn), V)}\\
    = \Ep{\Dn \iidsim P, V \sim \T(P)}{\ell(\Alg_{\alpha+\ep}(\Dn), V)} \le \alpha+\ep,
\end{multline*}
where the last inequality holds by the assumption that $\Alg_{\alpha}$ satisfies the guarantee~\eqref{eqn:target}. 
Similarly,
\begin{multline*}
    \Ep{\Dn \iidsim P, \tDn \iidsim Q, V \sim \T(P)}{\ell(\tAlg(\Dn, \tDn), V)} \le \Ep{\Dn \iidsim P, \tDn \iidsim Q, V \sim \T(P)}{\ell(\Alg_{\alpha}(\Dn \cup \tDn), V)}\\
    \le \Ep{\Dn \iidsim Q, \tDn \iidsim Q, V \sim \T(Q)}{\ell(\Alg_{\alpha}(\Dn \cup \tDn), V)} + c \cdot \dell(P,Q) \le \alpha+c \cdot \dell(P,Q).
\end{multline*}
Combining the two results above, we obtain the desired inequality.

\subsection{Proof of Theorem~\ref{thm:main_2}}
The proof of Theorem~\ref{thm:main_2} is covered in Section~\ref{sec:two_sided_proof}, which provides the proof for the extended statement, Theorem~\ref{thm:main_2_extended}.

\subsection{Proof of Theorem~\ref{thm:main_2_extended}}\label{sec:two_sided_proof}

First, the relation $\Alg_{\alpha}(\Dn) \preceq \tAlg(\Dn, \tDn)$ follows directly from the first part of Condition~\ref{con:target_order_formal}. Next, by the second and third condition in Condition~\ref{con:target_order_formal}, we have
\begin{align*}
    &\tAlg(\Dn, \tDn) = (\Alg_{\alpha}(\Dn \cup \tDn) \wedge \Alg_{\alpha+\ep}(\Dn)) \vee \Alg_{\alpha}(\Dn)\\
    &= (\Alg_{\alpha}(\Dn \cup \tDn) \vee \Alg_{\alpha}(\Dn)) \wedge (\Alg_{\alpha+\ep}(\Dn) \vee \Alg_{\alpha}(\Dn))\\
    &= (\Alg_{\alpha}(\Dn \cup \tDn) \vee \Alg_{\alpha}(\Dn)) \wedge \Alg_{\alpha+\ep}(\Dn),
\end{align*}
which implies $\tAlg(\Dn, \tDn) \preceq \Alg_{\alpha+\ep}(\Dn)$.

Now we prove the second claim. Observe that the first claim implies
\[\alpha-\delta_n \le \EE{\ell(\tAlg(\Dn, \tDn), V)} 
   \le \alpha + \ep,\]
and thus it is sufficient to show that
\[\alpha-c\tau-\delta_{N+n} \le \EE{\ell(\tAlg(\Dn, \tDn), V)} 
   \le \alpha +c\tau,\]
when $P=Q$. Then the final claim follows from the inequality
\[\left|\Ep{\substack{\Dn \iidsim P, \tDn \sim \bar{Q}\\ V \sim \T(P)}}{\ell(\tAlg(\Dn, \tDn), V)} - \Ep{\substack{\Dn \iidsim Q, \tDn \sim \bar{Q}\\ V \sim \T(P)}}{\ell(\tAlg(\Dn, \tDn), V)}\right| \leq c\cdot\tdell(P,Q),\]
which holds by the boundedness of $\ell$ and the properties of the total variation distance.

Let $E$ be the event that $\Alg_{\alpha}(\Dn) \preceq \Alg_{\alpha}(\Dn \cup \tDn) \preceq \Alg_{\alpha+\ep}(\Dn)$ holds. Observe that $\tAlg(\Dn, \tDn) = \Alg_{\alpha}(\Dn \cup \tDn)$ under the event $E$. Therefore, under $P=Q$,
\begin{align*}
    &\EE{\ell(\tAlg(\Dn, \tDn), V)} = \EE{\ell(\tAlg(\Dn, \tDn), V)\cdot\One{E}} + \EE{\ell(\tAlg(\Dn, \tDn), V)\cdot\One{E^c}}\\
    &\leq \EE{\ell(\Alg_{\alpha}(\Dn \cup \tDn),V)} + c\cdot\PP{E^c} \leq \alpha + c\tau.
\end{align*}
Moreover, we have
\begin{align*}
    &\EE{\ell(\tAlg(\Dn, \tDn), V)} \geq \EE{\ell(\tAlg(\Dn, \tDn), V)\cdot\One{E}}\\
&= \EE{\ell(\tAlg(\Dn, \tDn), V)} - \EE{\ell(\tAlg(\Dn, \tDn), V)\cdot\One{E^c}} \geq \alpha-\delta_{N+n} - c\tau,
\end{align*}
which completes the proof.

\subsection{Proof of Theorem~\ref{thm:conformal}}
Let $S_i = s(X_i,Y_i)$ for $i \in [n+1]$---including the test point---and let $\tilde{S}_j = s(\tilde{X}_i,\tilde{Y}_i)$ for $j \in [N]$. Since the distributions $P_S$ and $Q_S$ are continuous, there are no ties among the scores. Define $R$ as the rank of the test score $S_{n+1}$ among the scores $\{S_1,\ldots,S_{n+1}\}$:
\[R = \sum_{i=1}^{n+1} \One{S_i \le S_{n+1}}.\]
Then we have $S_{n+1} = S_{(R)}$, where $S_{(r)}$ denotes the $r$-th order statistics of $S_1,\ldots,S_{n+1}$. Moreover, by the exchangeability of $S_1, \ldots, S_{n+1}$, we have $R \sim \text{Unif}([n+1])$\footnote{For a finite set $A$, we write $\text{Unif}(A)$ to denote the uniform distribution on $A$.}

Now, we observe
\begin{align*}
    &\PP{Y_{n+1} \in \tilde{C}_{n,N}(X_{n+1})} = \PP{S_{n+1} \le \max\{\hat{q}_{\alpha+\ep}(\Dn), \hat{q}_{\alpha}(\Dn \cup \tDn)\}} \ge \PP{S_{n+1} \le \hat{q}_{\alpha}(\Dn \cup \tDn)}\\
    &= \PP{\text{(number of $S_i$'s and $\tilde{S}_j$'s smaller than $S_{n+1}$)} \le \lceil(1-\alpha)(N+n+1)\rceil}\\
    &= \PP{\sum_{i=1}^n \One{S_{i} < S_{(R)}} + \sum_{j=1}^N \One{\tilde{S}_j < S_{(R)}} \le \lceil(1-\alpha)(N+n+1)\rceil}\\
    &= \PP{R-1 + \sum_{j=1}^N \One{\tilde{S}_j \le S_{(R)}} \le \lceil(1-\alpha)(N+n+1)\rceil}\\
    &=\EE{\PPst{R-1 + \sum_{j=1}^N \One{\tilde{S}_j \le S_{(R)}} \le \lceil(1-\alpha)(N+n+1)\rceil}{R}}\\
    &= \frac{1}{n+1}\sum_{r=1}^{n+1} \PPst{r-1 + \sum_{j=1}^N \One{\tilde{S}_j \le S_{(r)}} \le \lceil(1-\alpha)(N+n+1)\rceil}{R=r}\\
    &= \frac{1}{n+1}\sum_{r=1}^{n+1} \PP{r-1 + \sum_{j=1}^N \One{\tilde{S}_j \le S_{(r)}} \le \lceil(1-\alpha)(N+n+1)\rceil},
\end{align*}
where the last equality holds since $R$ is independent of $S_{(1)}, \ldots, S_{(n+1)}$,\footnote{This follows from $S_{n+1} \mid (S_{(1)}, \ldots, S_{(n+1)}) \sim \frac{1}{n+1}\sum_{i=1}^{n+1} \delta_{S_{(i)}}$, which implies $R \mid (S_{(1)}, \ldots, S_{(n+1)})$ $\sim \text{Unif}([n+1])$.} 
as well as the synthetic scores $(\tilde{S}_j)_{j \in [N]}$. Now denote the event inside the probability by $E_r$, and observe that the probability inside the summation is taken with respect to
\[\tilde{S}_1,\ldots,\tilde{S}_N \iidsim Q_S, S_{(r)} \sim P_{(r)}^{n+1}, \text{ with } (\tilde{S}_j)_{j \in [N]} \indep S_{(r)}\]
Therefore, putting everything together, we have
\begin{align*}
    &\Pp{(X_i,Y_i)_{i \in [n+1]} \iidsim P, (\tilde{X}_j,\tilde{Y}_j)_{j \in [N]}\iidsim Q}{Y_{n+1} \in \tilde{C}_{n,N}(X_{n+1})} \ge \frac{1}{n+1}\sum_{r=1}^{n+1} \Pp{\tilde{S}_1,\ldots,\tilde{S}_N \iidsim Q_S, S_{(r)} \sim P_{(r)}^{n+1}}{E_r}\\
    &\ge \frac{1}{n+1}\sum_{r=1}^{n+1} \left[\Pp{\tilde{S}_1,\ldots,\tilde{S}_N \iidsim Q_S, S_{(r)} \sim P_{(r)}^{n+1}}{E_r} - \dtv(P_{(r)}^{n+1}, Q_{(r)}^{n+1})\right]\\
    &= \frac{1}{n+1}\sum_{r=1}^{n+1} \Pp{\tilde{S}_1,\ldots,\tilde{S}_N \iidsim Q_S, S_{(r)} \sim Q_{(r)}^{n+1}}{E_r} - d_{P,Q}\\
    &=\Pp{(X_i,Y_i)_{i \in [n+1]} \iidsim Q, (\tilde{X}_j,\tilde{Y}_j)_{j \in [N]}\iidsim Q}{S_{n+1} \le \hat{q}_{\alpha}(\Dn \cup \tDn)} - d_{P,Q}\\
    &\ge 1-\alpha - d_{P,Q},
\end{align*}
where the last equality holds by the same steps as derived previously—the distribution $P$ is simply replaced by $Q$. This proves the first claim, since $\PP{Y_{n+1} \in \tilde{C}_{n,N}(X_{n+1})} \ge \PP{S_{n+1} \le \hat{q}_{\alpha+\ep}(\Dn)} \ge 1-\alpha-\ep$.

We now prove the second claim. Observe that
\begin{multline*}
    \PP{Y_{n+1} \in \tilde{C}_{n,N}(X_{n+1})} = \PP{S_{n+1} \le \max\{\hat{q}_{\alpha+\ep}(\Dn), \hat{q}_{\alpha}(\Dn \cup \tDn)\}}\\
    \le \PP{S_{n+1} \le \hat{q}_{\alpha}(\Dn \cup \tDn)} + \PP{\hat{q}_{\alpha+\ep}(\Dn) >  \hat{q}_{\alpha}(\Dn \cup \tDn)}.
\end{multline*}
applying arguments similar to the one used previously, we have
\begin{multline*}
    \Pp{(X_i,Y_i)_{i \in [n+1]} \iidsim P, (\tilde{X}_j,\tilde{Y}_j)_{j \in [N]}\iidsim Q}{S_{n+1} \le \hat{q}_{\alpha}(\Dn \cup \tDn)}\\
\le \Pp{(X_i,Y_i)_{i \in [n+1]} \iidsim Q, (\tilde{X}_j,\tilde{Y}_j)_{j \in [N]}\iidsim Q}{S_{n+1} \le \hat{q}_{\alpha}(\Dn \cup \tDn)} + d_{P,Q}\\
\le 1-\alpha+\frac{1}{N+n+1}+d_{P,Q},
\end{multline*}
where the second inequality applies the result of~\citet{vovk2005algorithmic}.

Next, let $r_{\alpha+\ep} = \lceil(1-\alpha-\ep)(n+1)\rceil$, and let $\bar{S}_{(1)}, \ldots, \bar{S}_{(n)}$ be the order statistics of $S_1,\ldots,S_n$---note that these differ from $S_{(1)},\ldots,S_{(n+1)}$. We let $\bar{S}_{(n+1)} = +\infty$. We then compute
\begin{align*}
    &\PP{\hat{q}_{\alpha+\ep}(\Dn) >  \hat{q}_{\alpha}(\Dn \cup \tDn)} = \PP{\bar{S}_{(r_{\alpha+\ep})} > \hat{q}_{\alpha}(\Dn \cup \tDn)}\\
    &= \PP{\sum_{i=1}^n \One{S_i < \bar{S}_{(r_{\alpha+\ep})}} + 1 + \sum_{j=1}^N \One{\tilde{S}_j \le \bar{S}_{(r_{\alpha+\ep})}} \ge \lceil(1-\alpha)(N+n+1))\rceil}\\
    &= \PP{r_{\alpha+\ep} + \sum_{j=1}^N \One{\tilde{S}_j \le \bar{S}_{(r_{\alpha+\ep})}} \ge \lceil(1-\alpha)(N+n+1))\rceil}.
\end{align*}
Since the event inside the probability depends on $(S_i)_{i \in [n]}$ only through $\bar{S}_{(r_{\alpha+\ep})}$, we have
\begin{multline*}
    \Pp{(X_i,Y_i)_{i \in [n]} \iidsim P, (\tilde{X}_j,\tilde{Y}_j)_{j \in [N]}\iidsim Q}{\hat{q}_{\alpha+\ep}(\Dn) >  \hat{q}_{\alpha}(\Dn \cup \tDn)}\\
\le \Pp{(X_i,Y_i)_{i \in [n]} \iidsim Q, (\tilde{X}_j,\tilde{Y}_j)_{j \in [N]}\iidsim Q}{\hat{q}_{\alpha+\ep}(\Dn) >  \hat{q}_{\alpha}(\Dn \cup \tDn)} + \dtv(P_{(r_{\alpha+\ep})}^n, Q_{(r_{\alpha+\ep})}^n).
\end{multline*}
Now, let $R_r$ be the rank of $\bar{S}_{(r)}$ in $(S_i)_{i \in [n]} \cup (\tilde{S}_j)_{j \in [N]}$. Then under $\Dn \cup \tDn \iidsim Q$,
\begin{multline*}
    \PP{\bar{S}_{(r_{\alpha+\ep})} > \hat{q}_{\alpha}(\Dn \cup \tDn)} = \PP{R_{r_{\alpha+\ep}} > \lceil(1-\alpha)(N+n+1)\rceil}\\
    = \sum_{k=\lceil(1-\alpha)(N+n+1)\rceil+1}^{N+n} \PP{R_{r_{\alpha+\ep}} = k} = \sum_{k=\lceil(1-\alpha)(N+n+1)\rceil+1}^{N+n} \frac{\binom{k-1}{r_{\alpha+\ep}-1}\cdot\binom{N+n-k}{n-r_{\alpha+\ep}}}{\binom{N+n}{n}},
\end{multline*}
which is bounded by $\delta$ under the assumed choice of $\ep$. This completes the proof.

\subsection{Proof of Proposition~\ref{prop:median}}
Fix any $P \in \mathcal{P}$, and let $p = \PP{X > 0}$. Since $P \in \mathcal{P}$, we have $p < 1/2$. Now let $\tilde{k}_{n,\alpha} = Q_{1-\alpha}\left(\text{Binom}\left(n,p\right)\right)$, and observe that $\tilde{k}_{n,\alpha} \le \hat{k}_{n,\alpha}$. Then
\begin{align*}
    \Ep{\Dn \iidsim P}{\Alg_{\alpha}(\Dn)} = \PP{\sum_{i=1}^n \One{X_i > 0} > \hat{k}_{n,\alpha}} \le \PP{\sum_{i=1}^n \One{X_i > 0} > \tilde{k}_{n,\alpha}} \le \alpha,
\end{align*}
since $\sum_{i=1}^n \One{X_i > 0} \sim \text{Binom}(n,p)$.

\subsection{Proof of Theorem~\ref{thm:median}}
Fix any $P \in \mathcal{P}_{\mathcal{H}_0}^\ep$ and $Q$. We first have
\[\Ep{\Dn \iidsim P,\tDn \iidsim Q}{\tAlg(\Dn,\tDn)} \le \alpha+\ep\]
by Theorem~\ref{thm:main}. The claim holds trivially if $Q \notin \mathcal{P}_{\mathcal{H}_0}$, since
\[d_{P,Q} = \sqrt{\frac{n}{2q(1-q)}}\cdot(q-p) \ge \sqrt{2n}\cdot\left(\frac{1}{2}-\left(\frac{1}{2}-\sqrt{\frac{1}{2n}}\ep\right)\right) = \ep.\]
in that case, since $q(1-q) \leq 1/4$. Thus, we now focus on the case $Q \in \mathcal{P}_{\mathcal{H}_0}$.

Since the bound $\alpha + \ep$ follows directly from the guardrail component $\Alg(\Dn)$, it suffices to show that $\EE{\tAlg(\Dn,\tDn)} \leq \alpha+d_{P,Q}$ holds. Let $W_n = \sum_{i=1}^n \One{X_i} > 0$ and $\widetilde{W}_N = \sum_{j=1}^N \One{\tilde{X}_j > 0}$. Since the rejection event depends only on $W_n \sim \text{Binom}(n,p)$ and $\widetilde{W}_N \sim \text{Binom}(N,q)$, 
\begin{align*}
    &\Ep{\Dn \iidsim P,\tDn \iidsim Q}{\tAlg(\Dn,\tDn)} \leq \Ep{W_n \sim \text{Binom}(n,p),\widetilde{W}_N \sim \text{Binom}(N,q)}{W_n + \widetilde{W}_N > \hat{k}_{N+n,\alpha}}\\
    &\leq \Ep{W_n \sim \text{Binom}(n,q),\widetilde{W}_N \sim \text{Binom}(N,q)}{W_n + \widetilde{W}_N > \hat{k}_{N+n,\alpha}} + \dtv(\text{Binom}(n,p),\text{Binom}(n,q))\\
    &\leq \alpha+\dtv(\text{Binom}(n,p),\text{Binom}(n,q)),
\end{align*}
where the second inequality holds since we are assuming $Q \in \mathcal{P}_{\mathcal{H}_0}$. Then observe
\begin{align*}
    &\dtv(\text{Binom}(n,p),\text{Binom}(n,q)) \leq \sqrt{\frac{1}{2}\dkl(\text{Binom}(n,p),\text{Binom}(n,q))}\\
    &=\sqrt{\frac{n}{2}\left(p \log \frac{p}{q} + (1-p)\log\frac{1-p}{1-q}\right)} \leq \sqrt{\frac{n}{2}\left(p \cdot \left(\frac{p}{q}-1\right) + (1-p)\cdot\left(\frac{1-p}{1-q}-1\right)\right)}\\
    &=\sqrt{\frac{n}{2q(1-q)}} \cdot |p-q|.
\end{align*}
This completes the proof.

\clearpage

\section{Full Experimental Results (including \texttt{Synth+Real})}\label{app-sec:all-exp-full}

In this section, we present all experiments from the main manuscript and appendix, including an additional baseline, \texttt{Synth+Real}, which applies the base inference method on the pooled real and synthetic data at level~$\alpha$. Like \texttt{OnlySynth}, this baseline does not provide any error rate control guarantees and is included solely to illustrate the unknown quality of the pooled data.
Across all the experiments below, the performance of \texttt{Synth+Real} closely match that of \texttt{OnlySynth}, as expected when $N\gg n$.

\subsection{Conformal Risk Control for Protein Structure Prediction}
\begin{figure}[!h]
    \centering
\begin{subfigure}[t]{\linewidth}
\hspace{6em}
    \includegraphics[height=0.17\linewidth]{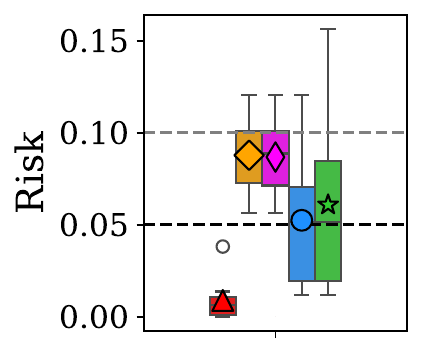}
    \includegraphics[height=0.17\linewidth]{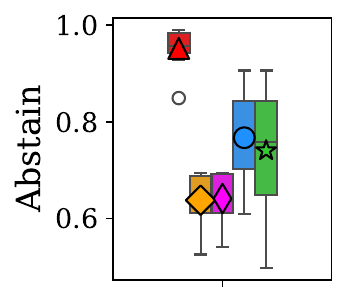}
    \includegraphics[height=0.17\linewidth]{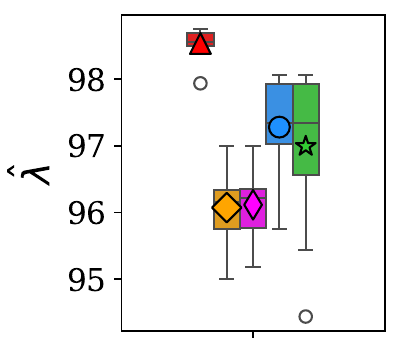}
    \caption{$\alpha=5\%$}
    \end{subfigure}
    
\begin{subfigure}[t]{\linewidth}
\hspace{6em}
    \includegraphics[height=0.17\linewidth]{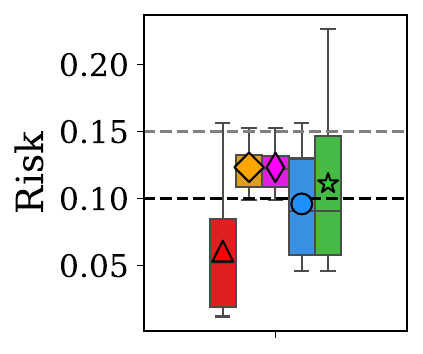}
    \includegraphics[height=0.17\linewidth]{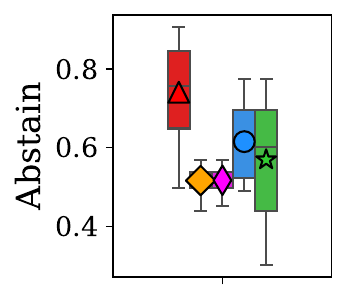}
    \includegraphics[height=0.17\linewidth]{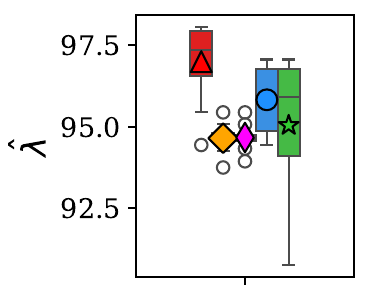}
    \caption{$\alpha=10\%$}
    \end{subfigure}
    
    \begin{subfigure}[t]{\linewidth}
\hspace{6em}
    \includegraphics[height=0.17\linewidth]{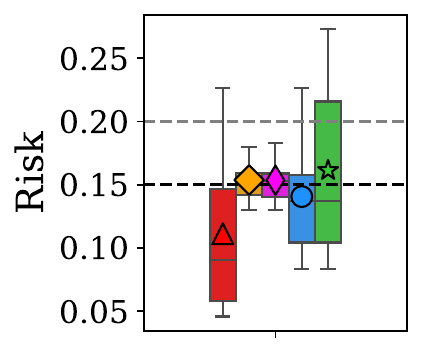}
    \includegraphics[height=0.17\linewidth]{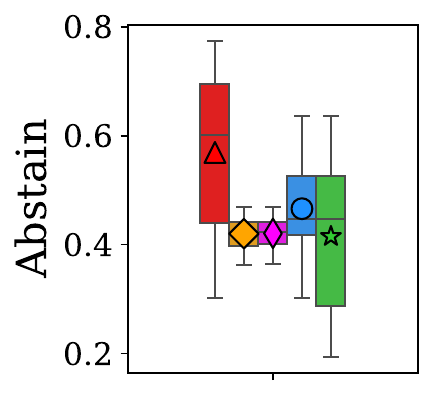}
    \includegraphics[height=0.17\linewidth]{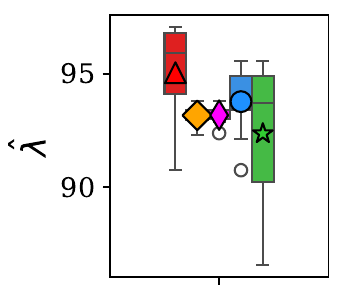}
    \includegraphics[height=0.17\linewidth]{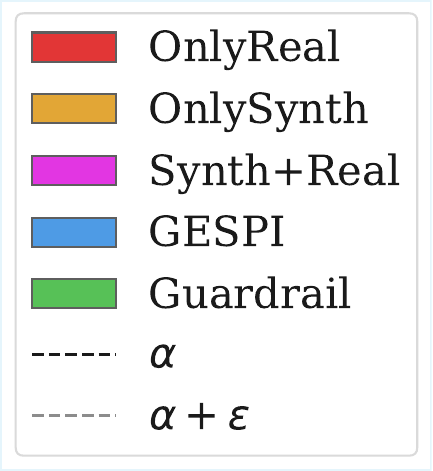}
    \caption{$\alpha=15\%$}
    \end{subfigure}
    \caption{\textbf{Performance comparisons for protein structure prediction with error rate control}.
Conformal risk control methods applied at target level $\alpha$ = 5\% (a), 10\% (b), and 15\% (c). Left: average risk (fraction of residues with error $>$ 3\AA). Middle: average abstention rate. Right: selected pLDDT threshold $\hat{\lambda}$.}
    \label{app-fig:protein-all}
\end{figure}
\FloatBarrier

\subsection{Conformal Prediction for Image Classification}

\begin{figure}[!h]
    \centering
    \includegraphics[width=0.4\linewidth, valign=t]{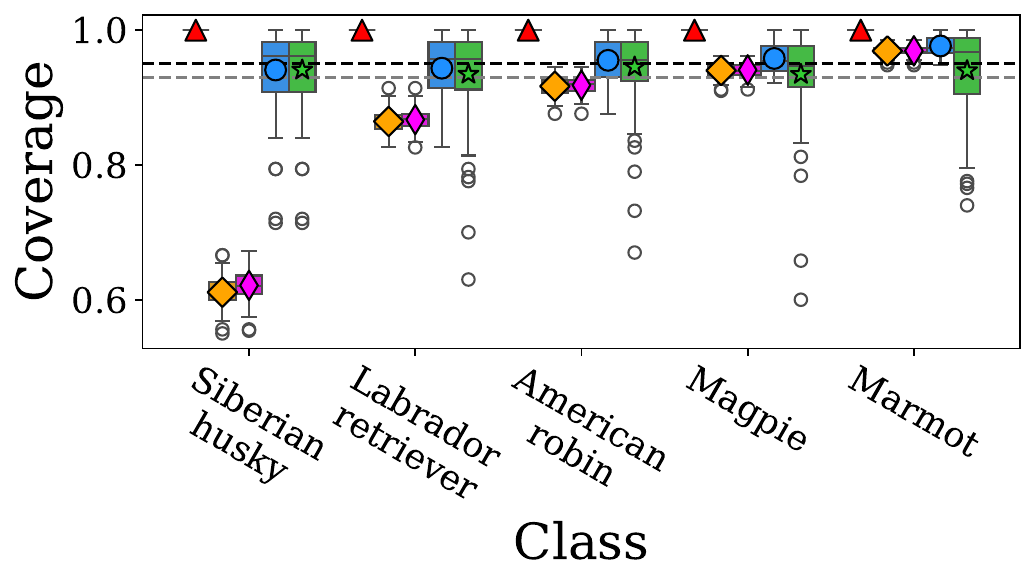}
    \includegraphics[width=0.4\linewidth, valign=t]{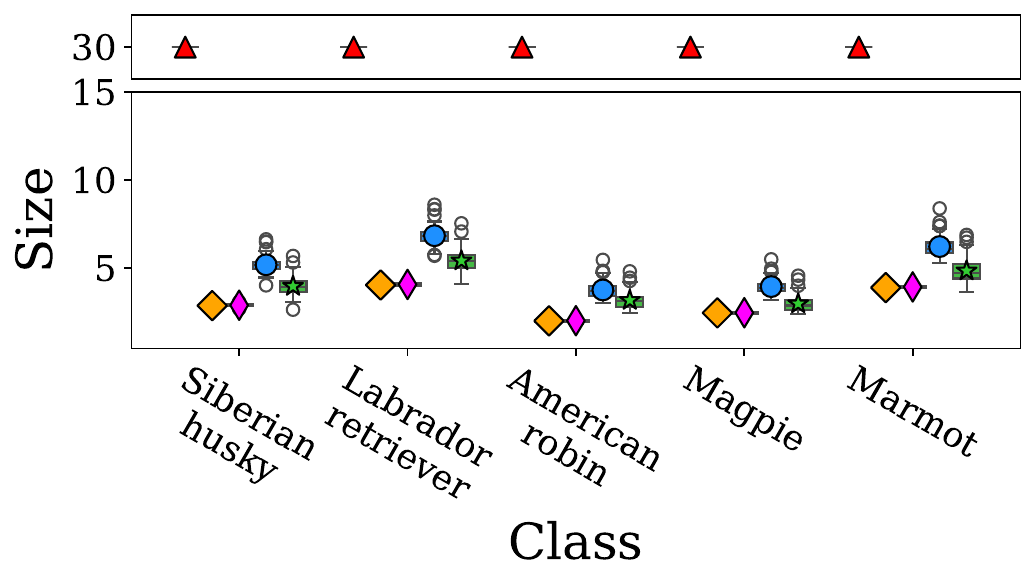}
    \includegraphics[width=0.15\linewidth, valign=t]{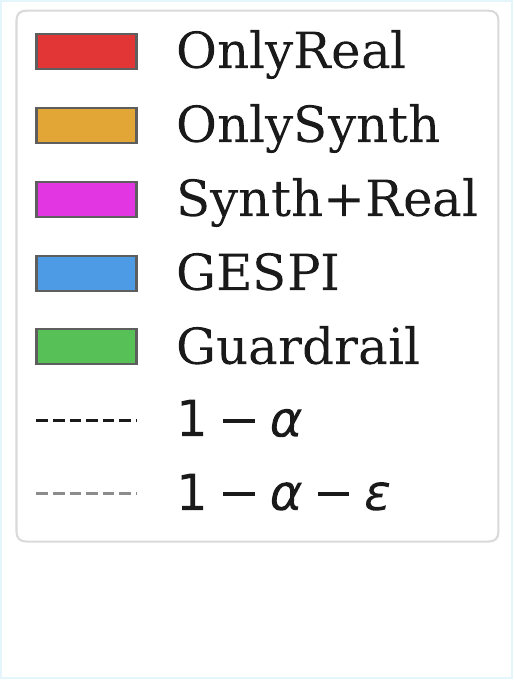}
    \caption{\textbf{Performance comparisons for image classification with class-conditional coverage on ImageNet}. Conformal prediction methods applied at level $\alpha = 5\%$ and $\varepsilon=2\%$. FLUX-generated~\citep{flux2024} images serve as the synthetic data. Results are shown for selected classes; see \Cref{app-tab:img-coverage,app-tab:img-size} for results across all classes.}
    \label{app-fig:imagenet-all}
\end{figure}

\begin{table}[!h]
\caption{Per-class conditional coverage (in \%) for each method, computed over 100 trials using FLUX-generated synthetic data. The target coverage level is $1 - \alpha = 0.95$ and the guardrail worst-case bound is $1-\alpha-\varepsilon= 0.93$. Standard errors are shown in parentheses.}
\label{app-tab:img-coverage}
\centering
\begin{tabular}{llllll}
\toprule
& \multicolumn{5}{c}{Coverage (\%)} \\
\toprule

Class &            \texttt{Only Real} &   \texttt{Only Synth} & \texttt{Synth+Real} &        \texttt{GESPI} &         \texttt{Guardrail}\\
\midrule
            Admiral & 100 (± 0) &  0.2 (± 0) &     0.2 (± 0) & 93.6 (± 0.6) & 93.6 (± 0.6)     \\
    American robin  & 100 (± 0) & 91.7 (± 0.1) &    91.8 (± 0.1) & 95.5 (± 0.3) & 94.5 (± 0.5) \\
         Barracouta & 100 (± 0) & 99.9 (± 0) &    99.9 (± 0) & 99.9 (± 0) & 95.3 (± 0.4)       \\
             Beaver & 100 (± 0) & 81.0 (± 0.2) &    81.4 (± 0.2) & 94.1 (± 0.5) & 94.0 (± 0.6) \\
            Bicycle & 100 (± 0) & 98.9 (± 0) &    98.9 (± 0) & 99.0 (± 0) & 94.2 (± 0.5)       \\
             Bulbul & 100 (± 0) & 93.8 (± 0.1) &    93.9 (± 0.1) & 96.1 (± 0.2) & 93.8 (± 0.6) \\
       Coral fungus & 100 (± 0) & 99.4 (± 0) &    99.4 (± 0) & 99.4 (± 0) & 93.5 (± 0.6)       \\
   English springer & 100 (± 0) & 95.1 (± 0.1) &    95.1 (± 0.1) & 96.8 (± 0.2) & 93.9 (± 0.6) \\
            Garfish & 100 (± 0) & 84.7 (± 0.2) &    85.0 (± 0.2) & 93.9 (± 0.5) & 93.5 (± 0.6) \\
   Golden retriever & 100 (± 0) & 89.9 (± 0.1) &    90.0 (± 0.1) & 95.0 (± 0.3) & 94.1 (± 0.6) \\
          Gyromitra & 100 (± 0) & 46.6 (± 0.2) &    47.8 (± 0.2) & 95.0 (± 0.6) & 95.0 (± 0.6) \\
                Jay & 100 (± 0) & 31.7 (± 0.2) &    35.4 (± 0.3) & 93.3 (± 0.7) & 93.3 (± 0.7) \\
    Junco, snowbird & 100 (± 0) & 97.9 (± 0.1) &    98.0 (± 0.1) & 98.3 (± 0.1) & 94.2 (± 0.5) \\
             Kuvasz & 100 (± 0) & 99.2 (± 0) &    99.3 (± 0) & 99.3 (± 0) & 93.4 (± 0.6)       \\
Labrador retriever  & 100 (± 0) & 86.5 (± 0.2) &    86.7 (± 0.2) & 94.4 (± 0.5) & 93.5 (± 0.7) \\
     Lighter, Light & 100 (± 0) & 67.0 (± 0.2) &    68.1 (± 0.2) & 94.2 (± 0.6) & 94.2 (± 0.6) \\
 Lycaenid butterfly & 100 (± 0) & 88.5 (± 0.1) &    88.8 (± 0.1) & 94.6 (± 0.4) & 94.0 (± 0.5) \\
             Magpie & 100 (± 0) & 94.0 (± 0.1) &    94.1 (± 0.1) & 95.8 (± 0.2) & 93.5 (± 0.6) \\
             Marmot & 100 (± 0) & 96.9 (± 0.1) &    96.9 (± 0.1) & 97.7 (± 0.1) & 94.0 (± 0.6) \\
             Muzzle & 100 (± 0) & 94.2 (± 0.1) &    94.3 (± 0.1) & 96.1 (± 0.2) & 93.5 (± 0.6) \\
           Papillon & 100 (± 0) & 99.9 (± 0) &    99.9 (± 0) & 99.9 (± 0) & 93.8 (± 0.6)       \\
        Rock beauty & 100 (± 0) & 87.9 (± 0.1) &    88.1 (± 0.1) & 94.6 (± 0.4) & 94.2 (± 0.5) \\
    Siberian husky  & 100 (± 0) & 61.1 (± 0.2) &    62.1 (± 0.2) & 94.1 (± 0.6) & 94.1 (± 0.6) \\
          Stinkhorn & 100 (± 0) & 97.3 (± 0.1) &    97.3 (± 0.1) & 97.7 (± 0.1) & 93.4 (± 0.6) \\
        Tennis ball & 100 (± 0) & 84.4 (± 0.2) &    84.8 (± 0.2) & 93.7 (± 0.5) & 93.5 (± 0.6) \\
        Tinca tinca & 100 (± 0) & 98.7 (± 0) &    98.7 (± 0) & 98.8 (± 0.1) & 93.2 (± 0.6)     \\
              Torch & 100 (± 0) & 86.9 (± 0.2) &    87.1 (± 0.2) & 95.2 (± 0.4) & 94.8 (± 0.5) \\
           Unicycle & 100 (± 0) & 99.8 (± 0) &    99.8 (± 0) & 99.8 (± 0) & 93.3 (± 0.6)       \\
        Water ouzel & 100 (± 0) & 98.9 (± 0) &    98.9 (± 0) & 99.0 (± 0) & 94.3 (± 0.5)       \\
         White wolf & 100 (± 0) & 78.5 (± 0.2) &    78.9 (± 0.2) & 93.9 (± 0.6) & 93.9 (± 0.6) \\
\bottomrule
\end{tabular}
\end{table}

\begin{table}[!h]
\caption{Per-class prediction set size for each method, computed over 100 trials using FLUX-generated synthetic data. The target coverage level is $1 - \alpha = 0.95$ and the guardrail worst-case bound is $1-\alpha-\varepsilon= 0.93$. Standard errors are shown in parentheses.}
\label{app-tab:img-size}
\centering
\begin{tabular}{llllll}
\toprule
& \multicolumn{5}{c}{Size} \\
\toprule
Class &            \texttt{Only Real} &   \texttt{Only Synth} & \texttt{Synth+Real} &        \texttt{GESPI} &         \texttt{Guardrail}\\

\midrule
            Admiral & 30 (± 0) & 4.4 (± 0) &     4.4 (± 0) & 6.2 (± 0) & 5.6 (± 0) \\
    American robin  & 30 (± 0) & 2.0 (± 0) &     2.0 (± 0) & 3.8 (± 0) & 3.2 (± 0) \\
         Barracouta & 30 (± 0) & 4.8 (± 0) &     4.8 (± 0) & 7.7 (± 0.1) & 6.6 (± 0.1) \\
             Beaver & 30 (± 0) & 3.1 (± 0) &     3.1 (± 0) & 4.8 (± 0) & 3.6 (± 0.1) \\
            Bicycle & 30 (± 0) & 3.1 (± 0) &     3.1 (± 0) & 4.2 (± 0) & 3.2 (± 0) \\
             Bulbul & 30 (± 0) & 2.6 (± 0) &     2.6 (± 0) & 4.3 (± 0) & 3.3 (± 0) \\
       Coral fungus & 30 (± 0) & 2.4 (± 0) &     2.5 (± 0) & 3.6 (± 0) & 2.9 (± 0) \\
   English springer & 30 (± 0) & 3.0 (± 0) &     3.0 (± 0) & 4.9 (± 0) & 3.6 (± 0) \\
            Garfish & 30 (± 0) & 4.0 (± 0) &     4.0 (± 0) & 6.3 (± 0) & 5.4 (± 0.1) \\
   Golden retriever & 30 (± 0) & 3.8 (± 0) &     3.9 (± 0) & 6.5 (± 0.1) & 5.0 (± 0.1) \\
          Gyromitra & 30 (± 0) & 3.0 (± 0) &     3.0 (± 0) & 4.1 (± 0) & 3.3 (± 0) \\
                Jay & 30 (± 0) & 5.7 (± 0) &     5.8 (± 0) & 8.5 (± 0.1) & 6.8 (± 0.1) \\
    Junco, snowbird & 30 (± 0) & 2.0 (± 0) &     2.0 (± 0) & 3.3 (± 0) & 2.7 (± 0) \\
             Kuvasz & 30 (± 0) & 2.3 (± 0) &     2.3 (± 0) & 4.3 (± 0) & 3.5 (± 0) \\
Labrador retriever  & 30 (± 0) & 4.1 (± 0) &     4.1 (± 0) & 6.8 (± 0.1) & 5.4 (± 0.1) \\
     Lighter, Light & 30 (± 0) & 3.9 (± 0) &     4.0 (± 0) & 6.6 (± 0) & 5.4 (± 0.1) \\
 Lycaenid butterfly & 30 (± 0) & 3.3 (± 0) &     3.4 (± 0) & 4.6 (± 0) & 3.9 (± 0.1) \\
             Magpie & 30 (± 0) & 2.5 (± 0) &     2.5 (± 0) & 4.0 (± 0) & 3.0 (± 0) \\
             Marmot & 30 (± 0) & 3.9 (± 0) &     3.9 (± 0) & 6.2 (± 0.1) & 4.9 (± 0.1) \\
             Muzzle & 30 (± 0) & 3.2 (± 0) &     3.2 (± 0) & 5.5 (± 0) & 4.3 (± 0) \\
           Papillon & 30 (± 0) & 2.2 (± 0) &     2.2 (± 0) & 4.3 (± 0) & 3.8 (± 0.1) \\
        Rock beauty & 30 (± 0) & 4.4 (± 0) &     4.4 (± 0) & 6.8 (± 0) & 5.1 (± 0.1) \\
    Siberian husky  & 30 (± 0) & 2.9 (± 0) &     2.9 (± 0) & 5.2 (± 0) & 4.0 (± 0) \\
          Stinkhorn & 30 (± 0) & 4.4 (± 0) &     4.4 (± 0) & 6.0 (± 0) & 4.5 (± 0.1) \\
        Tennis ball & 30 (± 0) & 2.8 (± 0) &     2.8 (± 0) & 4.4 (± 0) & 3.5 (± 0) \\
        Tinca tinca & 30 (± 0) & 2.7 (± 0) &     2.7 (± 0) & 4.6 (± 0) & 3.8 (± 0) \\
              Torch & 30 (± 0) & 5.2 (± 0) &     5.3 (± 0) & 8.0 (± 0) & 6.2 (± 0.1) \\
           Unicycle & 30 (± 0) & 4.0 (± 0) &     4.0 (± 0) & 6.0 (± 0) & 4.8 (± 0) \\
        Water ouzel & 30 (± 0) & 2.0 (± 0) &     2.0 (± 0) & 3.5 (± 0) & 3.0 (± 0) \\
         White wolf & 30 (± 0) & 2.8 (± 0) &     2.9 (± 0) & 4.7 (± 0) & 3.4 (± 0) \\
\bottomrule
\end{tabular}
\end{table}

\FloatBarrier
\clearpage

\subsection{Hypothesis Testing for Comparing Large Reasoning Models}
\subsubsection{Hyperparameter Selection for K-Way Majority Vote}
\begin{figure}[!h]
    \centering
    \includegraphics[width=0.4\linewidth, valign=t]{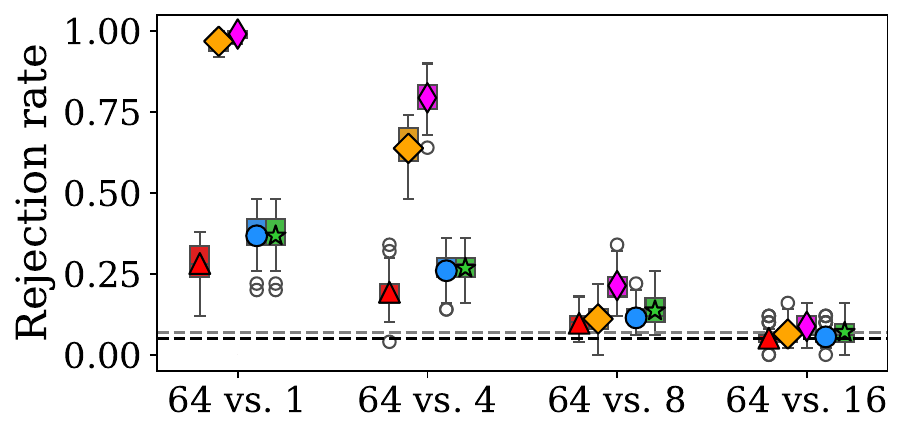}
    \includegraphics[width=0.4\linewidth, valign=t]{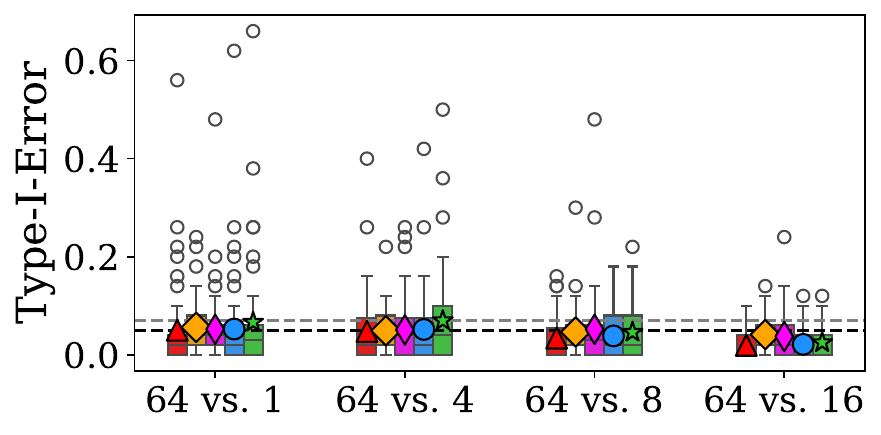}
    \includegraphics[width=0.15\linewidth, valign=t]{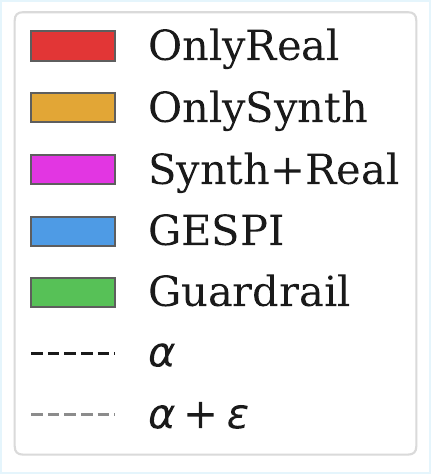}
    \caption{\textbf{Performance comparisons for different K-way majority vote on the AIME25 dataset for the R1-Deepseek-distill-Qwen-7B model}. Hypothesis testing methods are applied at level $\alpha = 5\%$ and $\varepsilon = 2\%$. Comparisons are made between the expensive $K=64$ and cheaper $K=1,4,8,16$. Left: Rejection rate, measured under the standard setting. Right: Type I error, measured in the shuffled-response setting where the null hypothesis holds.}
    \label{app-fig:llm-k-majority-all}
\end{figure}
\FloatBarrier

\subsubsection{Comparing Different Models}
\begin{figure}[!h]
    \centering
    \begin{subfigure}[t]{0.3\textwidth}
    \footnotesize
\texttt{Comp.~1:} DeepSeek-R1-Distill-Qwen-1.5B $>$ DeepSeek-R1-Distill-Qwen-7B (temp.$=0$)\\

\texttt{Comp.~2:} Qwen3 1.7B (temp. $=0$) $>$ DeepSeek-R1-Distill-Qwen-7B (temp. $=0$)
    \end{subfigure}%
    \hfill
    \begin{subfigure}[t]{0.67\textwidth}
    \includegraphics[width=0.36\linewidth, valign=t]{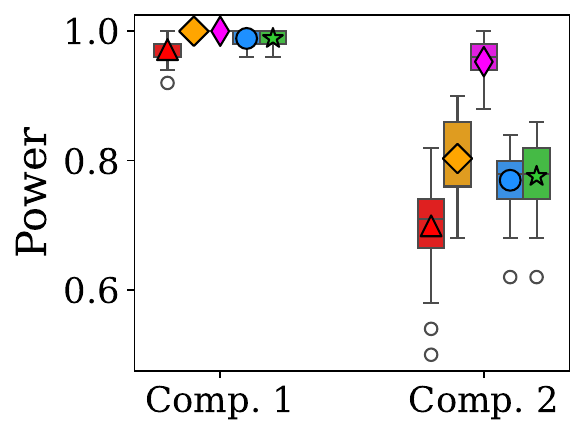}
    \includegraphics[width=0.36\linewidth, valign=t]{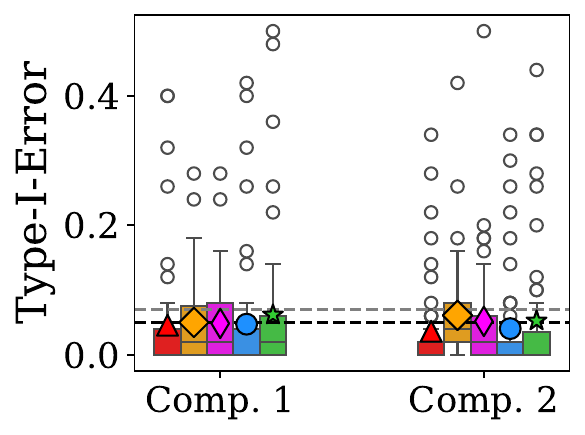}
    \includegraphics[width=0.25\linewidth, valign=t]{figures/more/color/llm_eval/legend.pdf}
    \end{subfigure}
    \caption{
    \textbf{Performance comparisons for LLM win rate on AIME25 dataset}. Hypothesis testing methods applied at level $\alpha=5\%$ and $\varepsilon=2\%$. Left: Description of model comparisons. Middle: Power, comparing the rejection rate under the standard setting. Right: Type I error, measured in the shuffled-response setting where the null holds. 
    }
    \label{app-fig:win-rate-all}
\end{figure}
\FloatBarrier

\clearpage

\subsection{Single and Multiple Hypothesis Testing for Outlier Detection}

\begin{figure}[!h]
    \centering
\begin{subfigure}[t]{\linewidth}
\hspace{5em}
    \includegraphics[width=0.33\linewidth, valign=t]{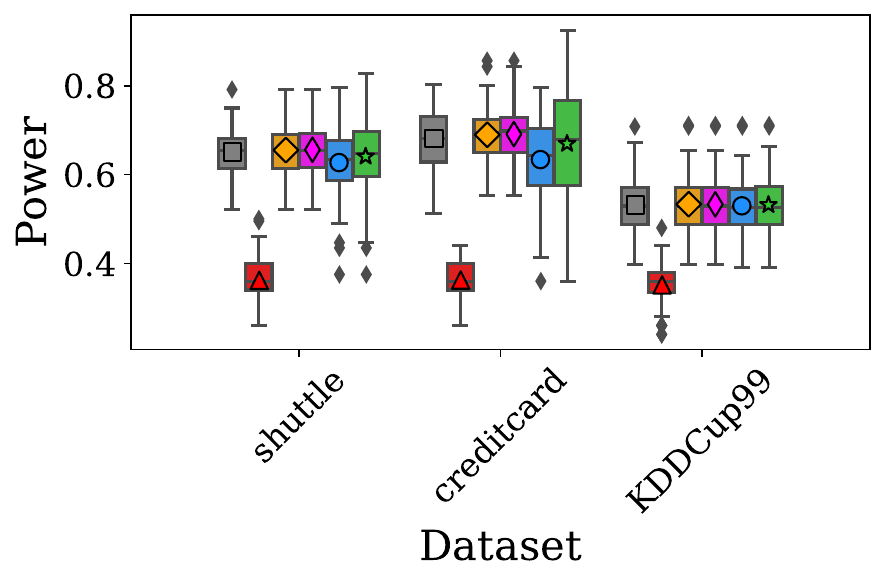}
    \includegraphics[width=0.33\linewidth, valign=t]{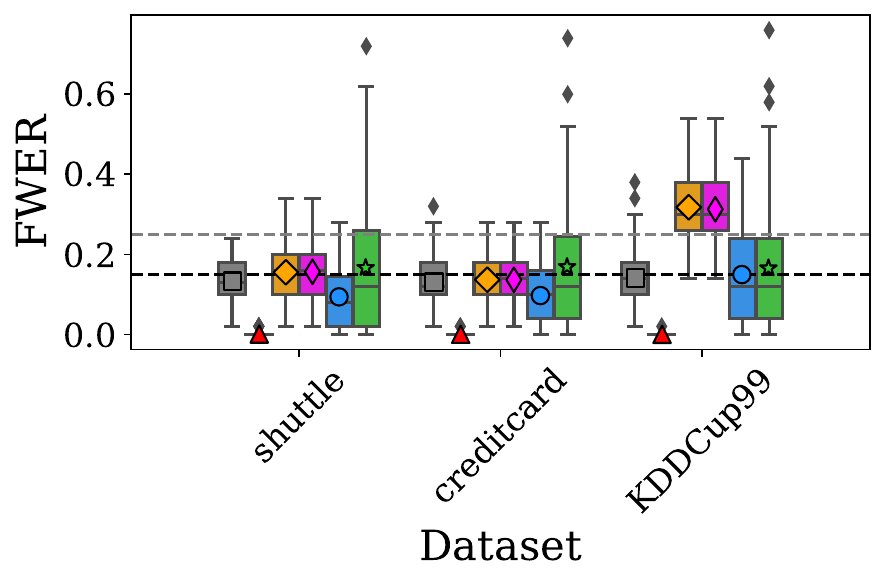}
        \caption{$q=2.5\%$}

    \end{subfigure}
    
\begin{subfigure}[t]{\linewidth}
\hspace{5em}
    \includegraphics[width=0.33\linewidth, valign=t]{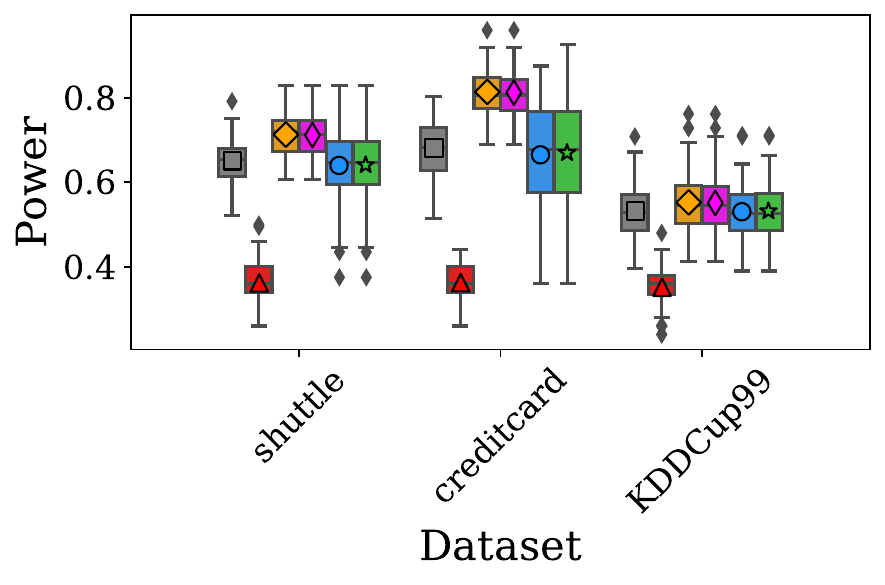}
    \includegraphics[width=0.33\linewidth, valign=t]{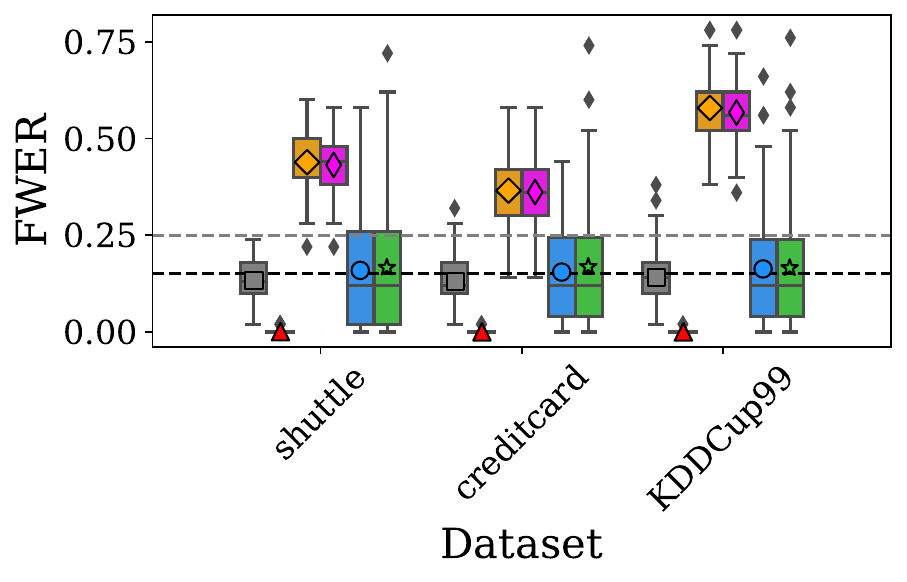}
    \includegraphics[width=0.2\linewidth, valign=t]{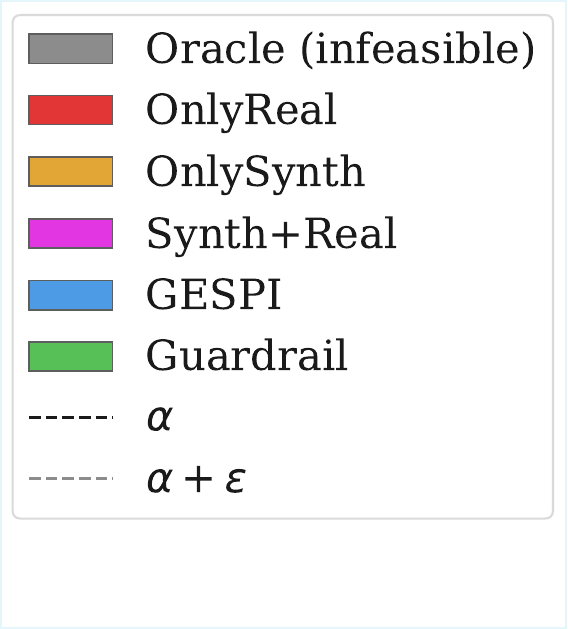}
        \caption{$q=5\%$}
\end{subfigure}
    \caption{\textbf{Performance comparisons for outlier detection with FWER control}. Evaluated on three datasets: shuttle, credit-card, KDDCup99, for $\alpha=15\%$ and $\varepsilon=10\%$, and trimming proportion $q=2.5\%$ (a) and $5\%$ (b).}
    \label{app-fig:od-fwer-all}
\end{figure}

\begin{figure}[!h]
    \centering
    
\begin{subfigure}[t]{\linewidth}
\hspace{5em}
    \includegraphics[width=0.33\linewidth, valign=t]{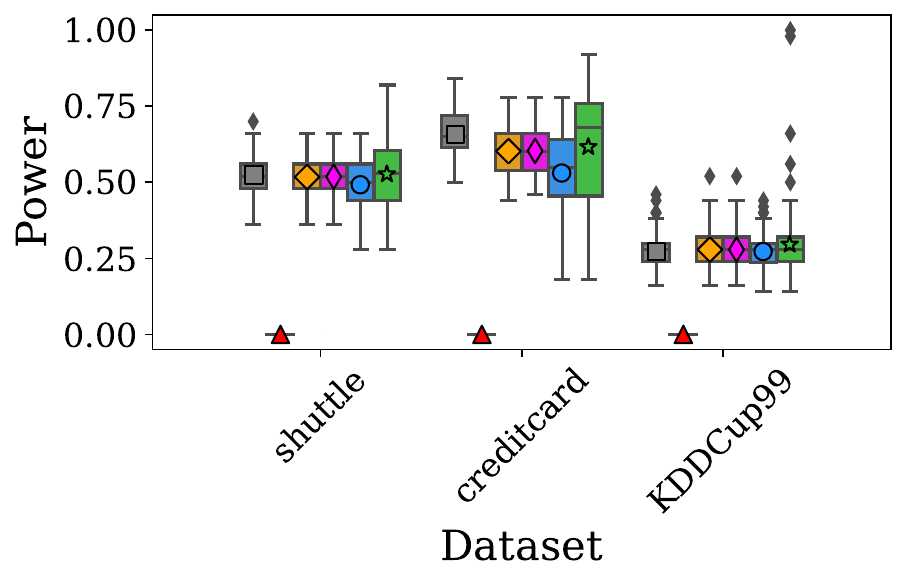}
    \includegraphics[width=0.33\linewidth, valign=t]{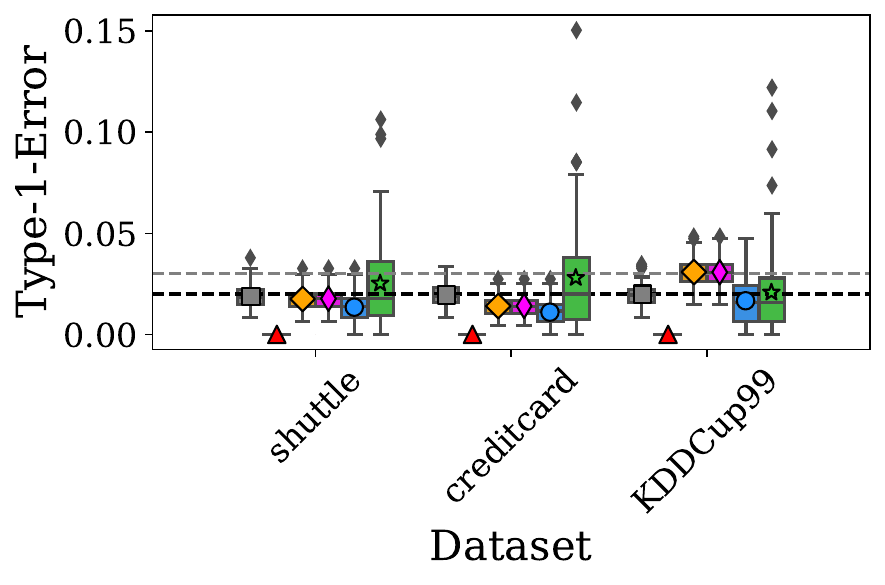}
        \caption{$q=2.5\%$}
    \end{subfigure}
    
\begin{subfigure}[t]{\linewidth}
\hspace{5em}
    \includegraphics[width=0.33\linewidth, valign=t]{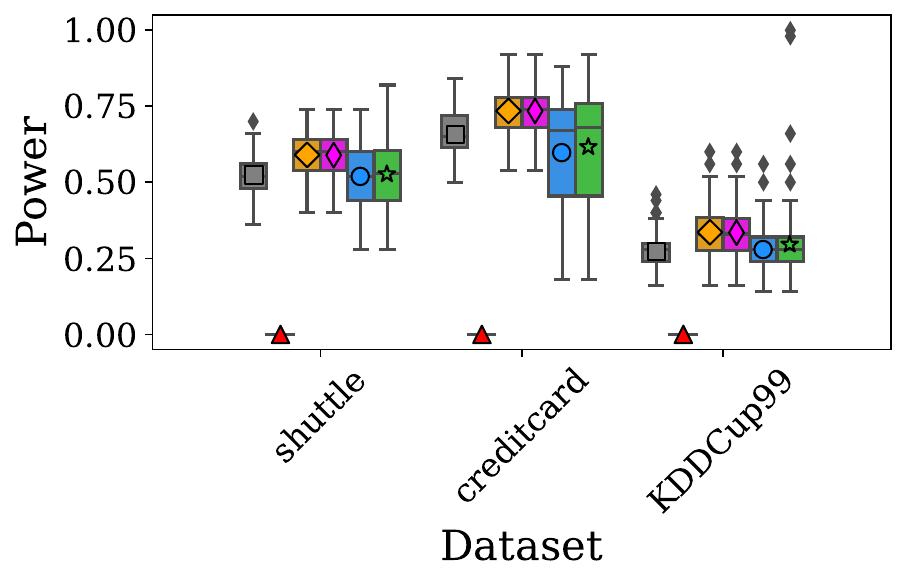}
    \includegraphics[width=0.33\linewidth, valign=t]{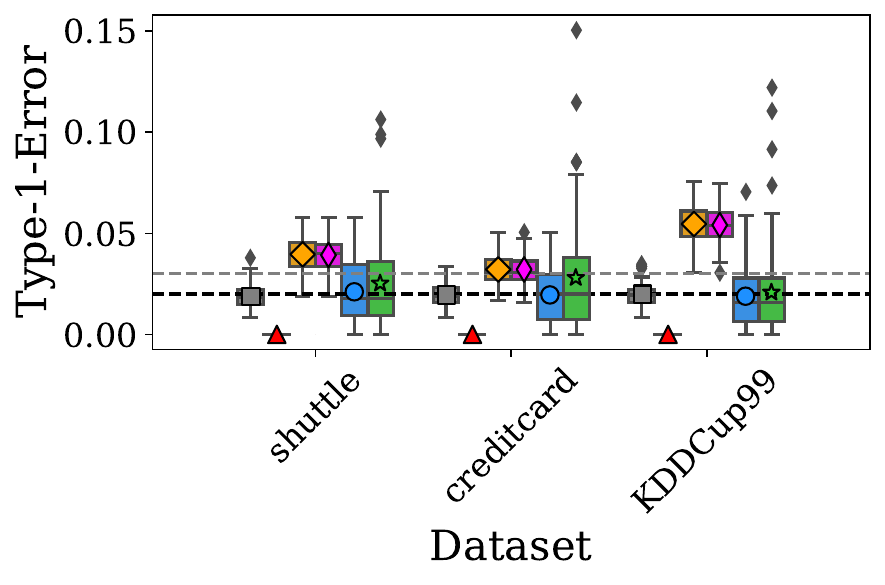}
    \includegraphics[width=0.2\linewidth, valign=t]{figures/more/color/outlier_detection/legend.pdf}
    \caption{$q=5\%$}
\end{subfigure}
    \caption{\textbf{Performance comparisons for outlier detection with Type I error rate control}. Evaluated on three datasets: shuttle, credit-card, KDDCup99, for $\alpha=2\%$ and $\varepsilon=1\%$, and trimming proportion $q=2.5\%$ (a) and $5\%$ (b).
}
    \label{app-fig:od-all}
\end{figure}
\FloatBarrier

\clearpage
\subsection{Hypothesis Testing for Mechanistic Interpretability of a Vision Transformer Model}

\begin{figure}[!h]
    \centering
    \includegraphics[width=0.3\linewidth, valign=t]{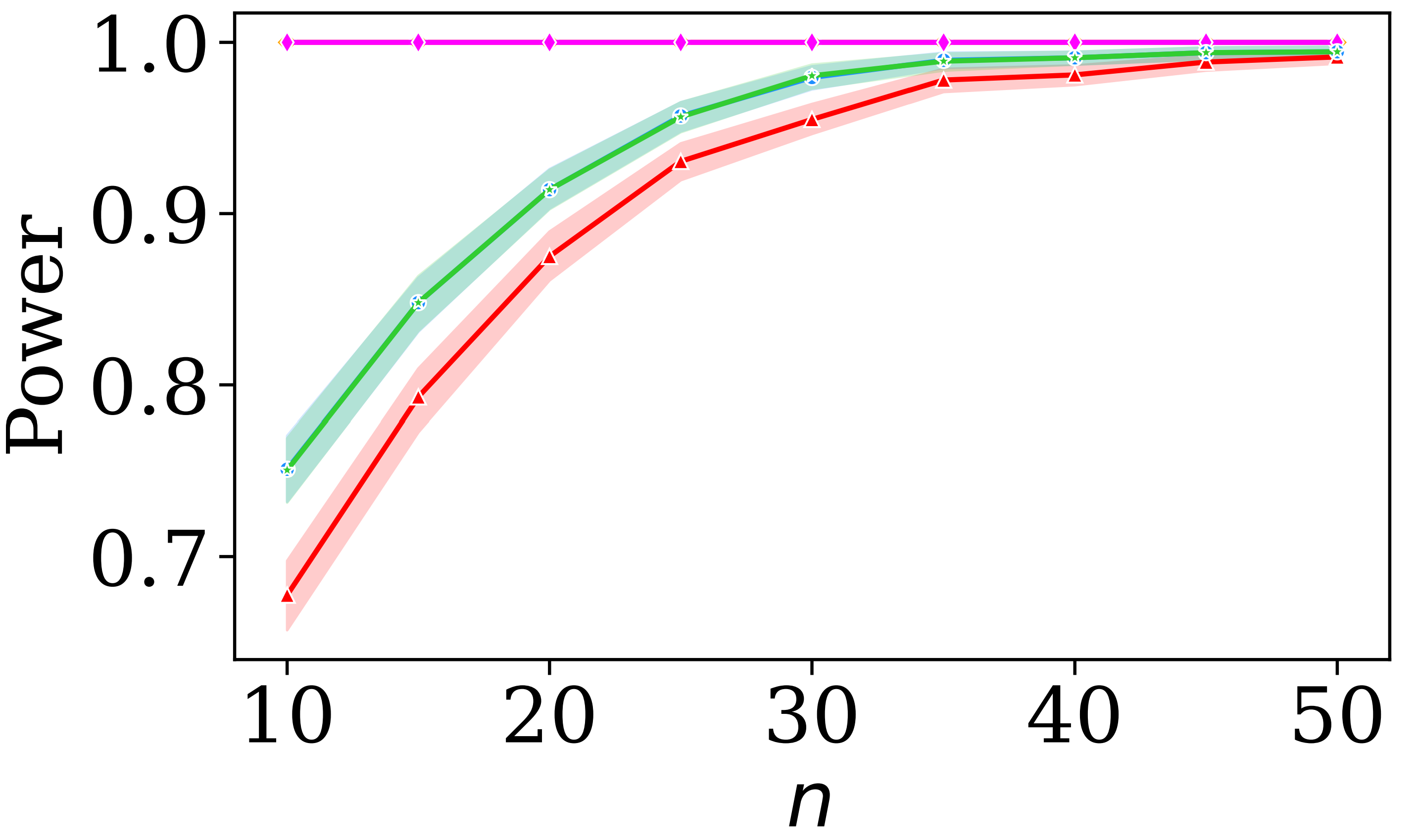}
    \includegraphics[height=0.1\linewidth, valign=t]{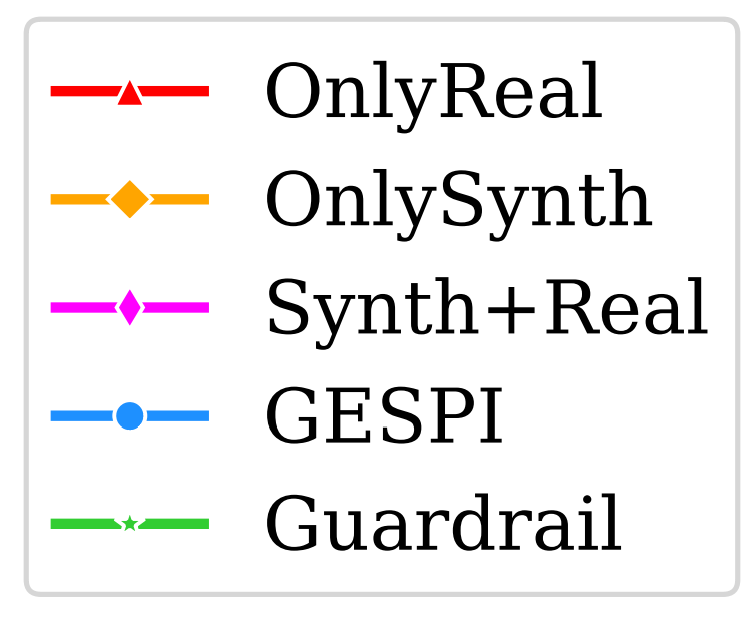}\vspace{-0.5em}
    
    \caption{
    \textbf{Performance comparisons as a function of the real sample size $n$.}
    Hypothesis testing methods applied to animal versus non-animal groups, each containing three classes, at target level $\alpha=10\%$ and $\varepsilon=5\%$.
    } 
\label{app-fig:clip-learned-rep-p-all}
\end{figure}

\begin{figure}[!h]
    \centering
    \begin{subfigure}[t]{0.3\linewidth}
    \includegraphics[width=\linewidth, valign=t]{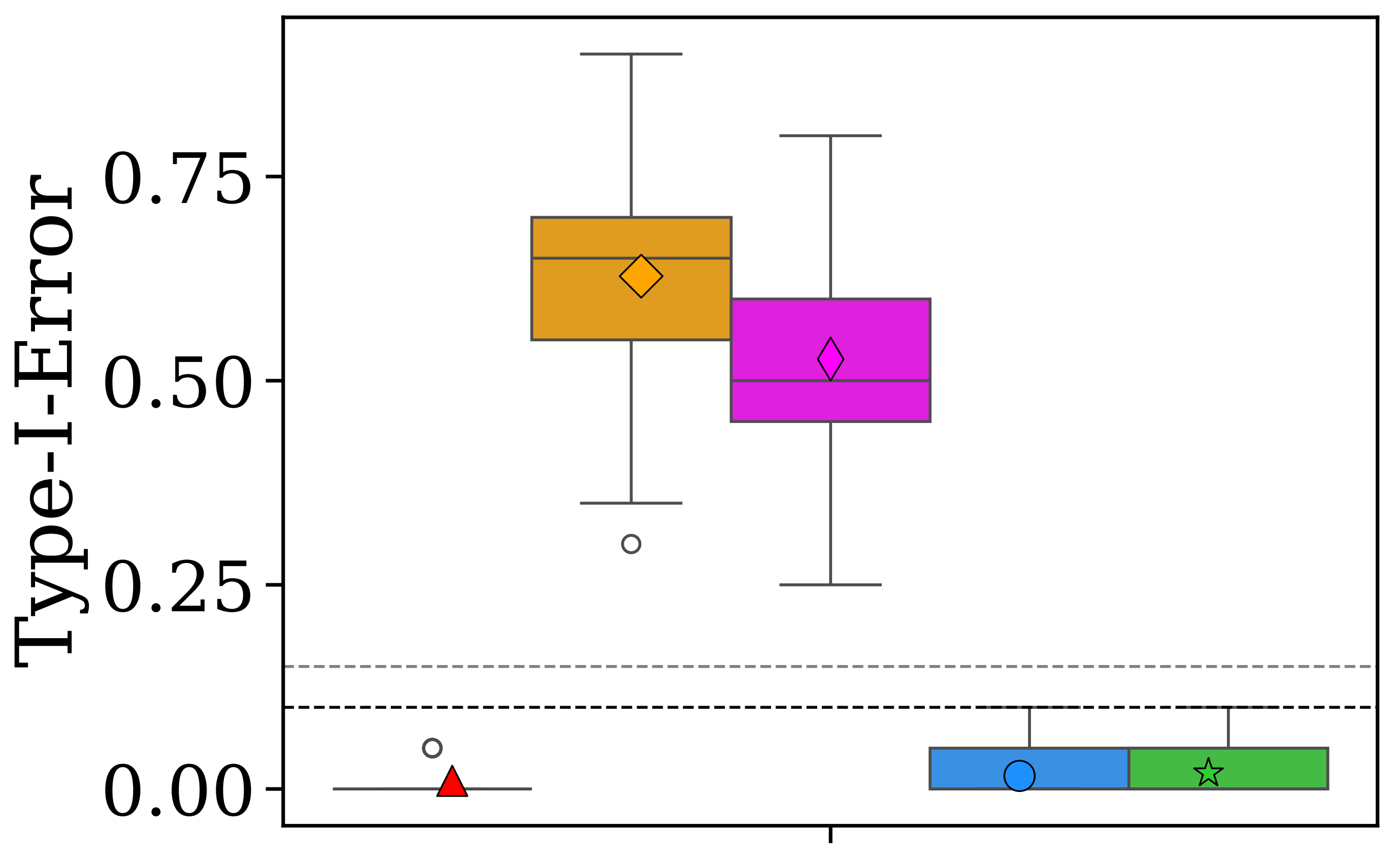}
    \caption{$n=10$}
    \end{subfigure}
    \begin{subfigure}[t]{0.3\linewidth}
    \includegraphics[width=\linewidth, valign=t]{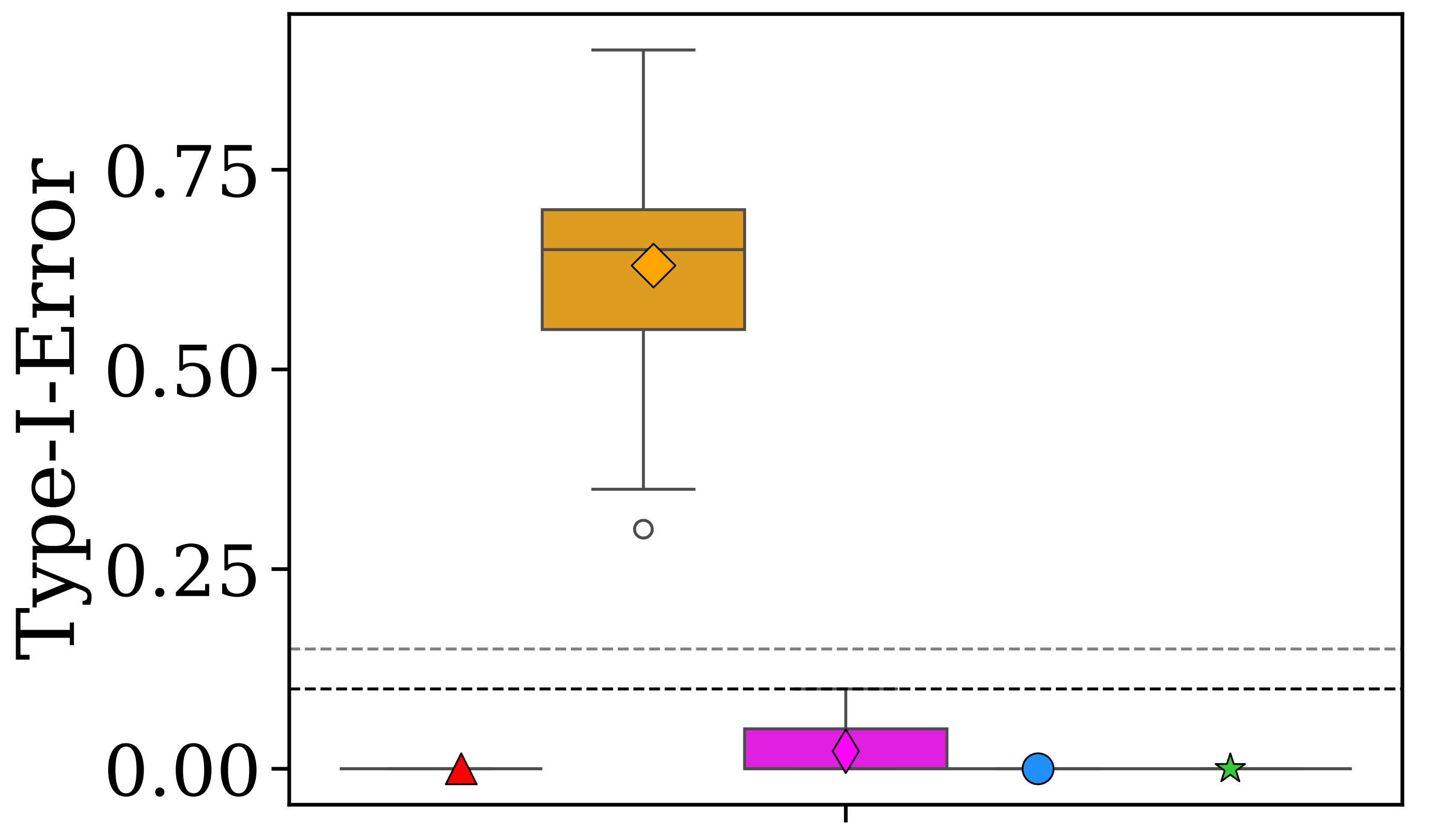}
    \caption{$n=100$}
    \end{subfigure}
    \includegraphics[height=0.15\linewidth, valign=t]{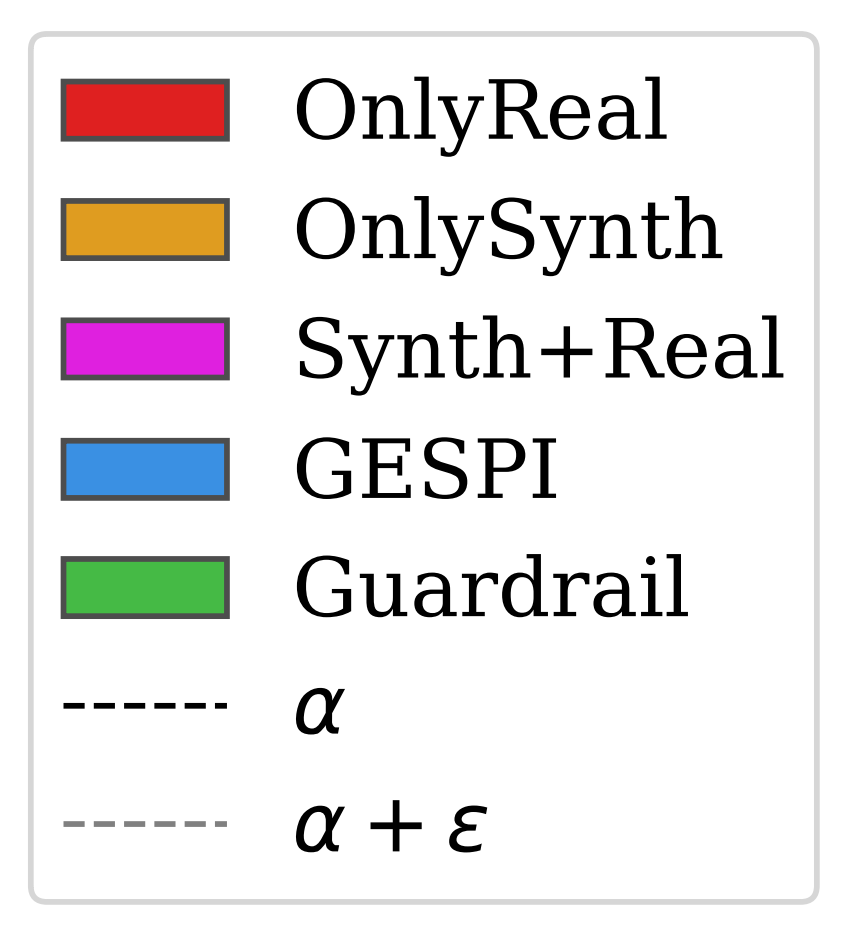}
    \caption{
    \textbf{Performance comparisons under the null hypothesis.} 
    Hypothesis testing methods applied to two disjoint groups of non-animal classes at target level $\alpha = 10\%$ and $\varepsilon=5\%$; with $n=10$ (a) and $n=100$ (b).
    } 
\label{app-fig:clip-learned-rep-t1-all}
\end{figure}
\FloatBarrier

\subsection{Hypothesis Testing with Simulated Data}
\begin{figure}[!h]
    \centering
    \hspace{-7em}
    \begin{subfigure}[t]{0.3\linewidth}
    \centering
    \includegraphics[width=\linewidth, valign=t]{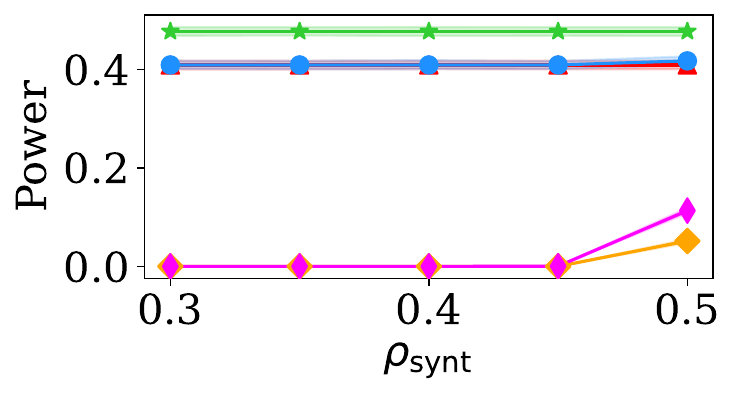}
    
    \caption{
    real under the alternative,\\ synthetic under the null}
    \end{subfigure}
    \begin{subfigure}[t]{0.3\linewidth}
    \centering
    \includegraphics[width=\linewidth, valign=t]{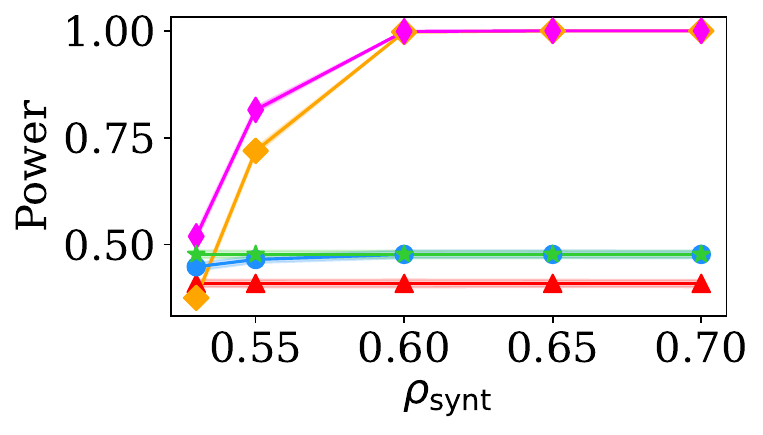}
    \caption{
    both real and synthetic under the alternative}
    \end{subfigure}\\
    \begin{subfigure}[t]{0.3\linewidth}
    \centering
    \includegraphics[width=\linewidth, valign=t]{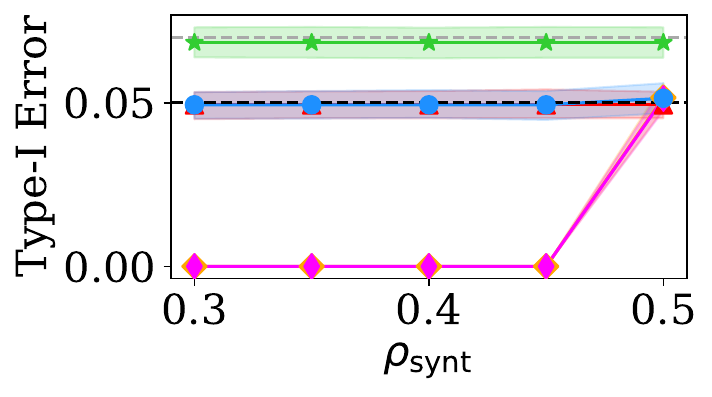}
    \caption{
    both under the null}
    \end{subfigure}
    \begin{subfigure}[t]{0.3\linewidth}
    \centering
    \includegraphics[width=\linewidth, valign=t]{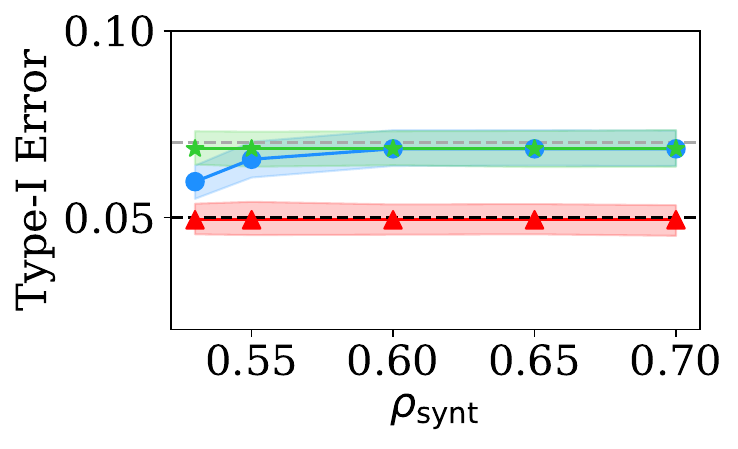}
    \caption{
    real under the null, synthetic under the alternative}
    \end{subfigure}
    \includegraphics[width=0.15\linewidth, valign=t]{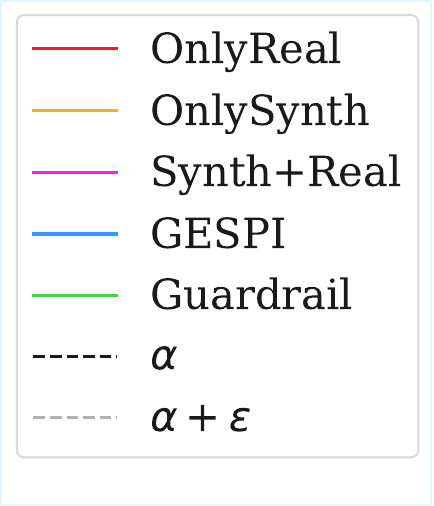}
    \caption{
    \textbf{Performance comparison as a function of $\rho_{\text{synt}}$}.
    Hypothesis testing methods across different values of $\rho$ applied at level $\alpha=5\%$ and $\varepsilon=2\%$. Top row: $\rho = 0.6$ (alternative). Bottom row: $\rho = 0.5$ (null).}
    \label{app-fig:synt-p_synth-all}
\end{figure}

\begin{figure}[!h]
    \centering
    \hspace{-7em}
    \begin{subfigure}[t]{0.3\linewidth}
    \includegraphics[width=\linewidth, valign=t]{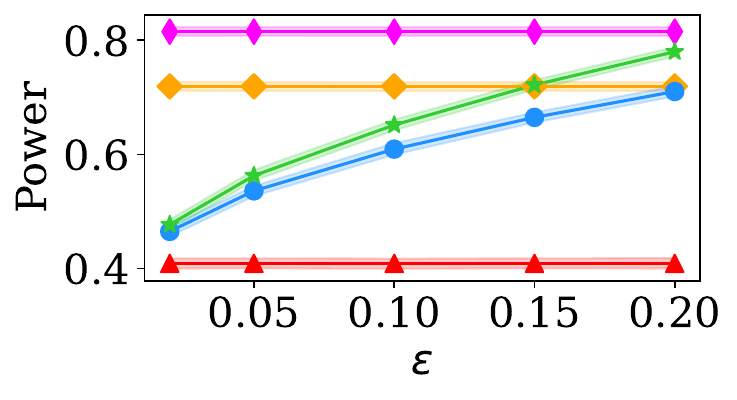}
    \caption{
    $\rho=0.6, \rho_{\text{synt}}=0.55$}
    \end{subfigure}
    \begin{subfigure}[t]{0.3\linewidth}
    \includegraphics[width=\linewidth, valign=t]{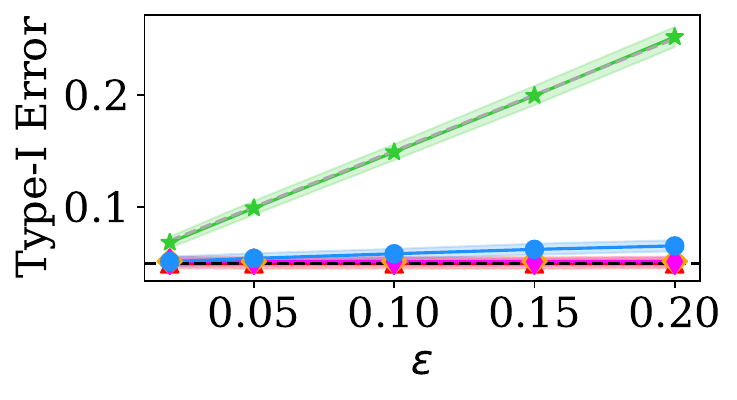}
    \caption{
    $\rho=0.5, \rho_{\text{synt}}=0.5$}
    \end{subfigure}
    \\
    \begin{subfigure}[t]{0.3\linewidth}
    \includegraphics[width=\linewidth, valign=t]{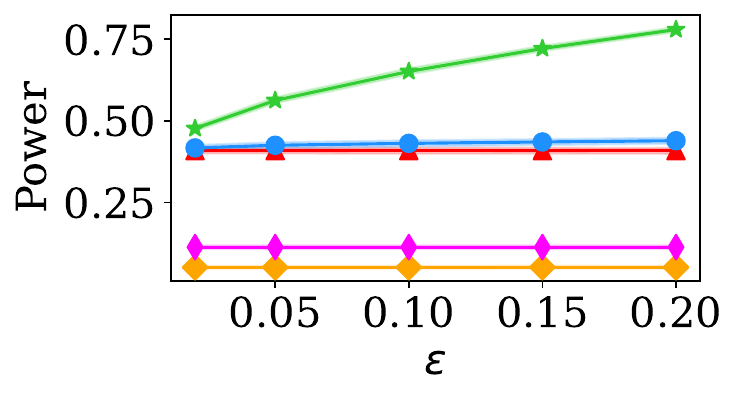}
    \caption{
    $\rho=0.6, \rho_{\text{synt}}=0.5$}
    \end{subfigure}
    \begin{subfigure}[t]{0.3\linewidth}
    \includegraphics[width=\linewidth, valign=t]{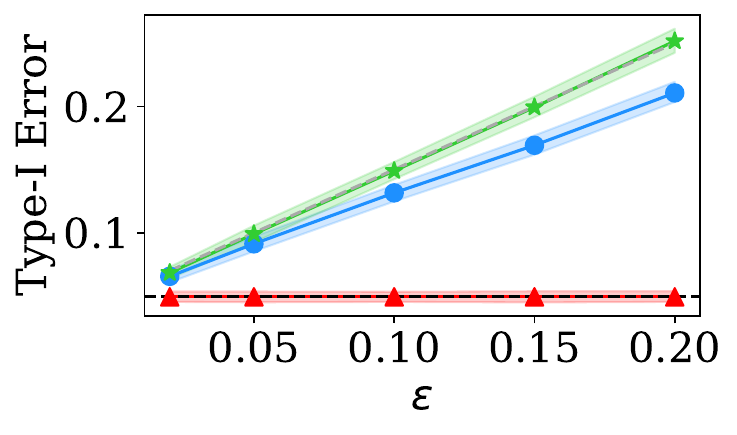}
    \caption{
    $\rho=0.5, \rho_{\text{synt}}=0.55$}
    \end{subfigure}
    \includegraphics[width=0.15\linewidth, valign=t]{figures/more/color/simulated/Type-I-Error_legend.pdf}
    \caption{
    \textbf{Performance comparison as a function of $\varepsilon$}.
    Hypothesis testing methods across different values of $\rho$ and $\rho_\text{synt}$ applied at level $\alpha=5\%$.}
    \label{app-fig:synt-epsilon-all}
\end{figure}

\begin{figure}[!h]
    \centering
    \includegraphics[width=0.3\linewidth, valign=t]{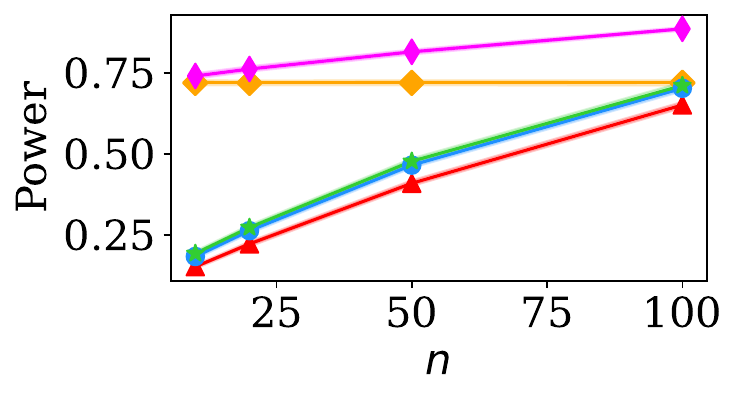}
    \includegraphics[width=0.3\linewidth, valign=t]{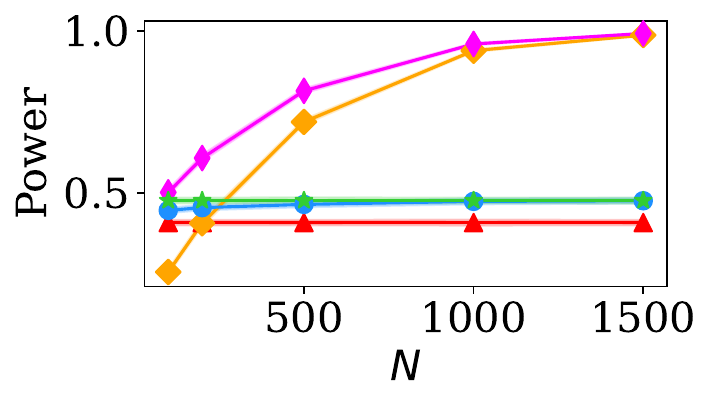}
    \includegraphics[width=0.15\linewidth, valign=t]{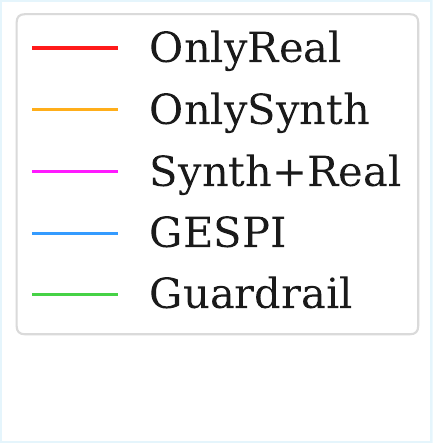}
    \caption{
    \textbf{Performance comparison as a function of the real dataset size $n$ and the synthetic dataset size $N$}.
    Hypothesis testing methods under the alternative ($\rho=0.6$ and $\rho_\text{synt}=0.55$) applied at level $\alpha=5\%$ and $\varepsilon=2\%$.}
    \label{app-fig:synt-n-all}
\end{figure}

\begin{figure}[!h]
    \centering
    \hspace{-7em}
    \begin{subfigure}[t]{0.3\linewidth}
    \includegraphics[width=\linewidth, valign=t]{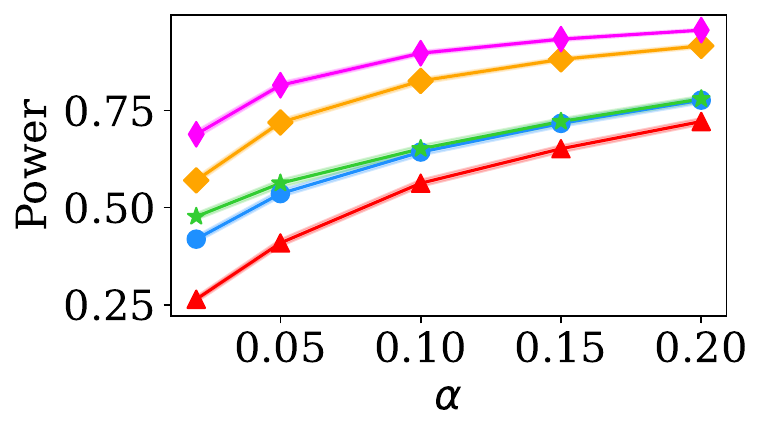}
    \caption{
    $\rho=0.6, \rho_{\text{synt}}=0.55$}
    \end{subfigure}
    \begin{subfigure}[t]{0.3\linewidth}
    \includegraphics[width=\linewidth, valign=t]{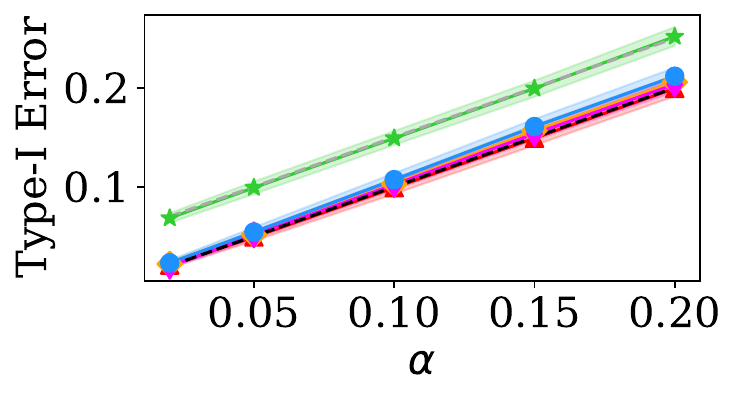}
    \caption{
    $\rho=0.5, \rho_{\text{synt}}=0.5$}
    \end{subfigure}
    \\
    \begin{subfigure}[t]{0.3\linewidth}
    \includegraphics[width=\linewidth, valign=t]{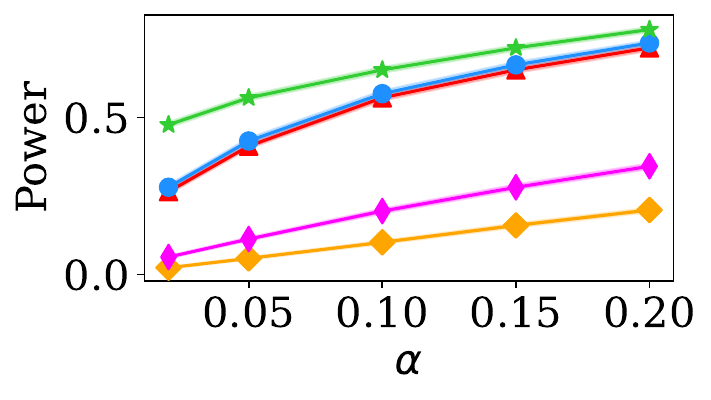}
    \caption{
    $\rho=0.6, \rho_{\text{synt}}=0.5$}
    \end{subfigure}
    \begin{subfigure}[t]{0.3\linewidth}
    \includegraphics[width=\linewidth, valign=t]{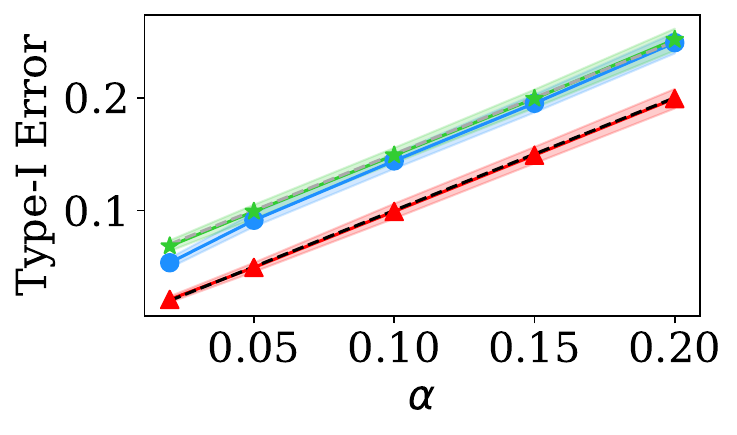}
    \caption{
    $\rho=0.5, \rho_{\text{synt}}=0.55$}
    \end{subfigure}
    \includegraphics[width=0.15\linewidth, valign=t]{figures/more/color/simulated/Type-I-Error_legend.pdf}
    \caption{
    \textbf{Performance comparison as a function of the target Type I error level $\alpha$}.
    Hypothesis testing methods across different values of $\rho$ and $\rho_\text{synt}$. GESPI applied with $\varepsilon=5\%$.}
    \label{app-fig:synt-alpha-eps-05-all}
\end{figure}

\begin{figure}[!h]
    \centering
    \hspace{-7em}
    \begin{subfigure}[t]{0.3\linewidth}
    \includegraphics[width=\linewidth, valign=t]{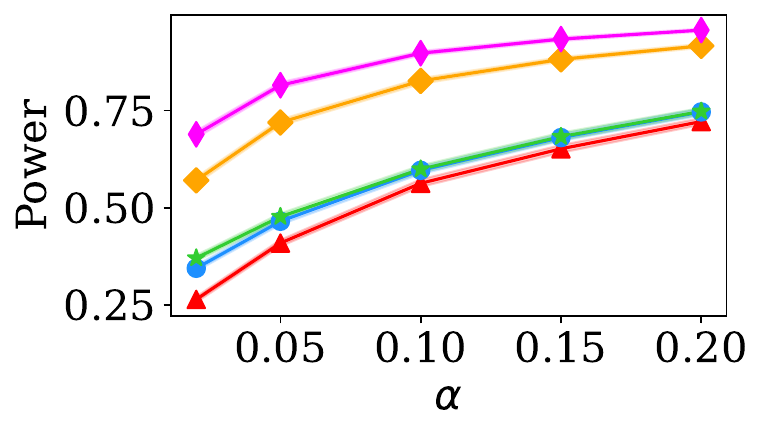}
    \caption{
    $\rho=0.6, \rho_{\text{synt}}=0.55$}
    \end{subfigure}
    \begin{subfigure}[t]{0.3\linewidth}
    \includegraphics[width=\linewidth, valign=t]{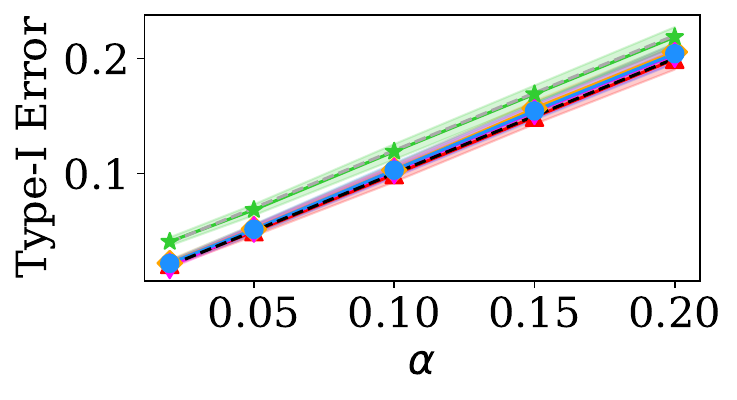}
    \caption{
    $\rho=0.5, \rho_{\text{synt}}=0.5$}
    \end{subfigure}
    \\
    \begin{subfigure}[t]{0.3\linewidth}
    \includegraphics[width=\linewidth, valign=t]{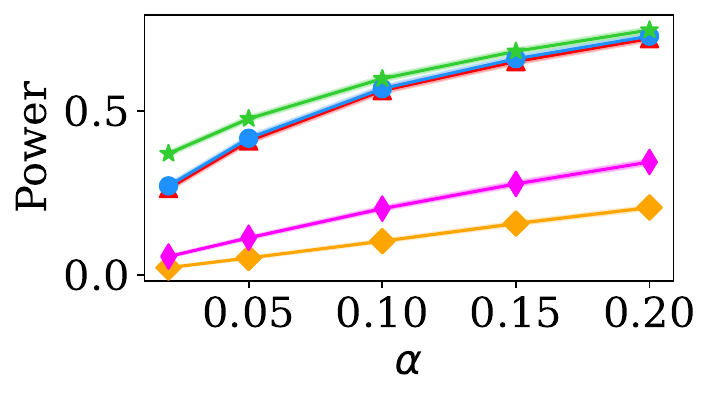}
    \caption{
    $\rho=0.6, \rho_{\text{synt}}=0.5$}
    \end{subfigure}
    \begin{subfigure}[t]{0.3\linewidth}
    \includegraphics[width=\linewidth, valign=t]{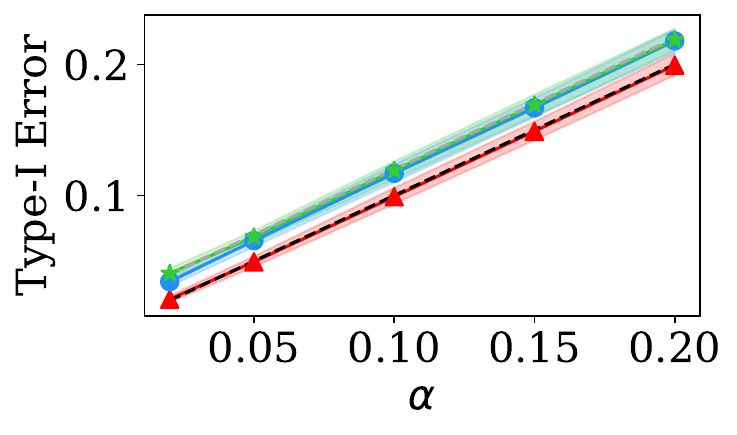}
    \caption{
    $\rho=0.5, \rho_{\text{synt}}=0.55$}
    \end{subfigure}
    \includegraphics[width=0.15\linewidth, valign=t]{figures/more/color/simulated/Type-I-Error_legend.pdf}
    \caption{
    \textbf{Performance comparison as a function of the target Type I error level $\alpha$}.
    Hypothesis testing methods across different values of $\rho$ and $\rho_\text{synt}$. GESPI applied with $\varepsilon=2\%$.}
    \label{app-fig:synt-alpha-all}
\end{figure}

\FloatBarrier

\end{document}